\documentclass[longauth]{aa} 
\usepackage[colorlinks=true,linkcolor=blue,citecolor=blue]{hyperref}
\usepackage[dvipsnames]{xcolor}
\usepackage{graphicx}
\graphicspath{{graphics}}
\usepackage{txfonts}
\usepackage[separate-uncertainty=true, list-units=repeat]{siunitx}
\usepackage{subcaption}
\usepackage{hyperref}
\usepackage[inline]{enumitem}
\usepackage{threeparttable} 
\usepackage{physics} 	
\usepackage{multirow}
\usepackage{bm}

\newcommand{\Lephare}{\textsc{LePHARE~}}

\newcommand{\hst}{\textit{HST}}
\newcommand{\JWST}{\textit{JWST}}

\newcommand{\UVISTA}{UltraVISTA}

\newcommand{\NIRCAM}{NIRCam}
\newcommand{\MIRI}{MIRI}

\DeclareSIUnit{\Msun}{M_\odot}
\DeclareSIUnit{\year}{yr}
\DeclareSIUnit{\pc}{pc}
\DeclareSIUnit{\mag}{mag}
\DeclareSIUnit{\mas}{mas}
\DeclareSIUnit{\dex}{dex}
\DeclareSIUnit{\jansky}{Jy}
\DeclareSIUnit{\kpc}{kpc}

\usepackage{graphicx}
\usepackage{txfonts}
\begin{document}

   \title{COSMOS-Web: The emergence of the Hubble Sequence}

   \author{M. Huertas-Company      \inst{1,2,3}\thanks{mhuertas@iac.es}
          \and
          M. Shuntov \inst{4,5}
          \and
          Y. Dong\inst{6}
           \and
          M. Walmsley\inst{7}
           \and
          O. Ilbert\inst{8}
          \and
          H.J. McCracken\inst{9}
          \and
          H.B. Akins\inst{10}
          \and
          N. Allen\inst{4,5}
           \and
          C.M. Casey\inst{10,27,4}
           \and 
          L. Costantin\inst{11}
          \and
          E. Daddi\inst{13}
          \and 
          A. Dekel\inst{12}
           \and
          M. Franco\inst{10,13}
          \and
          I. L. Garland\inst{30}
          \and
          T. G\'{e}ron\inst{7}
          \and
          G. Gozaliasl\inst{28,29}
          \and
          M. Hirschmann\inst{25}
          \and
          J.S. Kartaltepe\inst{15}
          \and
          A.M. Koekemoer\inst{16}
           \and 
          C. Lintott\inst{17}
           \and
          D. Liu\inst{18}
           \and
          R. Lucas\inst{16}  
          \and
          K. Masters\inst{26}
          \and 
          F. Pacucci\inst{19,20}
          \and
          L. Paquereau \inst{9}
            \and 
          P.G. P\'erez-Gonz\'alez\inst{11}
          \and 
          J.D. Rhodes\inst{21} 
           \and
          B.E. Robertson\inst{22}      
         \and
          B. Simmons\inst{23}
          \and
          R. Smethurst\inst{17}         
          \and
          S. Toft\inst{4,5}
           \and
          L. Yang\inst{24} 
          }
   \institute{Instituto de Astrof´ısica de Canarias, La Laguna, Tenerife, Spain\\
              \email{mhuertas@iac.es}
         \and
             Universidad de la Laguna, La Laguna, Tenerife, Spain
             \and 
             Universit\'e de Paris, LERMA - Observatoire de Paris, PSL, Paris, France
             \and
             Cosmic Dawn Center (DAWN), Denmark
             \and
             Niels Bohr Institute, University of Copenhagen, Jagtvej 128, 2200
Copenhagen, Denmark
\and
ETH, Switzerland
\and 
Dunlap Institute for Astronomy \& Astrophysics, University of Toronto, Toronto, Canada 
\and 
Aix Marseille Univ, CNRS, CNES, LAM, Marseille, France
\and 
Institut d’Astrophysique de Paris, UMR 7095, CNRS, Sorbonne
Université, 98 bis boulevard Arago, F-75014 Paris, France
\and 
 The University of Texas at Austin, 2515 Speedway Blvd Stop
C1400, Austin, TX 78712, USA
\and 
Centro de Astrobiología (CAB), CSIC-INTA, Ctra. de Ajalvir km 4, Torrejón de Ardoz, E-28850, Madrid, Spain
\and 
Racah Institute of Physics, The Hebrew University, Jerusalem 91904 Israel
\and
Université Paris-Saclay, Université Paris Cité, CEA, CNRS, AIM,
91191, Gif-sur-Yvette, France
\and
Physics Department, Lancaster University, Lancaster, LA1 4YB, UK
\and
Laboratory for Multiwavelength Astrophysics, School of Physics
and Astronomy, Rochester Institute of Technology, 84 Lomb Memorial Drive, Rochester, NY 14623, USA
\and
Space Telescope Science Institute, 3700 San Martin Drive, Baltimore, MD 21218, USA
\and
Oxford Astrophysics, Denys Wilkinson Building, Keble Road, Oxford OX1 3RH, UK
\and
Purple Mountain Observatory, Chinese Academy of Sciences, 10
Yuanhua Road, Nanjing 210023, China
\and 
Center for Astrophysics | Harvard \& Smithsonian, 60 Garden St, Cambridge, MA 02138, USA
\and
Black Hole Initiative, Harvard University, 20 Garden St, Cambridge, MA 02138, USA
\and 
Jet Propulsion Laboratory, California Institute of Technology, 4800
Oak Grove Drive, Pasadena, CA 91109
\and
 Department of Astronomy and Astrophysics, University of California, Santa Cruz, 1156 High Street, Santa Cruz, CA 95064 USA
 \and
 Physics Department, Lancaster University, Lancaster, LA1 4YB, UK
 \and
 Kavli Institute for the Physics and Mathematics of the Universe (WPI), The University of Tokyo, Kashiwa, Chiba 277-8583, Japan
 \and 
 Institute for Physics, Laboratory for Galaxy Evolution and Spectral modelling, Ecole Polytechnique Federale de Lausanne,
Observatoire de Sauverny, Chemin Pegasi 51, 1290 Versoix, Switzerland
\and
Departments of Astronomy and Physics, Haverford College, 370 Lancaster Ave., Haverford, PA 19041, USA
\and 
Department of Phyiscs University of California Santa Barbara, CA, 93106, CA
\and 
Department of Computer Science, Aalto University, PO Box 15400, Espoo, FI-00 076, Finland
\and
Department of Physics, Faculty of Science, University of Helsinki, 00014-Helsinki, Finland
\and 
Department of Theoretical Physics and Astrophysics, Faculty of Science, Masaryk University, Kotlářská 2, Brno, 611 37, Czech Republic}

   \date{Received September 15, 1996; accepted March 16, 1997}

 
  \abstract
   {The first JWST deep surveys have opened a new window into understanding the morphological evolution of galaxies across cosmic time. The improved spatial resolution and near-infrared (NIR) coverage have revealed a population of morphologically evolved galaxies at very early epochs. However, all previous works are based on small-number statistics, preventing accurate probing of the morphological diversity at cosmic dawn.}
   {Leveraging the wide area coverage of the COSMOS-Web survey, we quantify the abundance of different morphological types from $z\sim7$ with unprecedented statistics and establish robust constraints on the epoch of emergence of the Hubble sequence.}
    {We measure the global (spheroids, disk-dominated, bulge-dominated, peculiar) and resolved (stellar bars) morphologies for about 400,000 galaxies down to $F150W=27$ using deep learning, representing a two-orders-of-magnitude increase over previous studies. 
   We then provide reference Stellar Mass Functions (SMFs) of different morphologies between $z\sim0.2$ and $z\sim7$ as well as best-fit parameters to inform models of galaxy formation. All catalogs and data are made publicly available.}
    {(a) At redshift \( z > 4.5 \), the massive galaxy population ($\log M_*/M_\odot>10$) is dominated by disturbed morphologies (\( \sim70\% \)) - even in the optical rest frame - and very compact objects (\( \sim30\% \)) with effective radii smaller than \( \sim500\,\text{pc} \). This confirms that a significant fraction of the star formation at cosmic dawn occurs in very dense regions, although the stellar mass for these systems could be overestimated. (b) Galaxies with Hubble-type morphologies—including bulge- and disk-dominated galaxies—arose rapidly around \( z \sim 4 \) and dominate the morphological diversity of massive galaxies as early as \( z \sim 3 \). (c) Using stellar bars as a proxy, we speculate that stellar disks in massive galaxies might have been common ($>50\%$) among the star-forming population since cosmic noon (\( z \sim 2\text{--}2.5 \)) and formed as early as $z\sim7$. (d) Massive quenched galaxies are predominantly bulge-dominated from \( z \sim 4 \) onward, suggesting that morphological transformations briefly precede or are simultaneous to quenching mechanisms at the high-mass end. (e) Low-mass ($\log M_*/M_\odot<10$) quenched galaxies are typically disk-dominated, pointing to different quenching routes in the two ends of the stellar mass spectrum from cosmic dawn. }
   {}

   \keywords{Galaxies: evolution, Galaxies: high-redshift, Galaxies: statistics, Galaxies: structure}
   \maketitle

\section{Introduction}
When and how galaxies acquire their present-day stellar morphological diversity are still debated questions. Since its launch three years ago, JWST has provided a new window into the rest-frame optical emission of the first galaxies. A variety of works have shown that Hubble-type morphologies seem to emerge in massive galaxies very early in the history of the universe (e.g.,~\citealp{2022ApJ...938L...2F,2022arXiv221014713K,2024A&A...685A..48H,2024MNRAS.531.4857C}) compared to what was previously known from HST imaging (e.g., ~\citealp{2016MNRAS.462.4495H}) including the presence of cold stellar disk signatures such as stellar bars as early as $z\sim3$~(e.g.,~\citealp{2023ApJ...945L..10G,2023Natur.623..499C,2023ApJ...958...36S,2024arXiv240401918A}). This is somehow surprising given the expected high gas fractions and merger rates at these very early epochs. Although the exact nature of these galaxies is difficult to establish without kinematic measurements~\citep{2024ApJ...961...51V}, these purely morphological measurements seem to suggest a very rapid mass assembly at cosmic dawn, also in agreement with the high abundances of UV bright (e.g.,~\citealp{2024ApJ...969L...2F}) and massive galaxies at cosmic dawn. 

However, all previous results are based on relatively small volumes from the first JWST deep field observations, which makes it difficult to properly quantify the number densities of different morphological types as a function of stellar mass and redshift and hence establish robust constraints on the epoch of emergence of the Hubble sequence.  


In this study, we take advantage of the wide area coverage of the COSMOS-Web survey to quantify the stellar mass functions of different morphological types from $z\sim0.5$ to $z\sim7$ with the best statistics up to date. We employ deep learning models from~\cite{2024A&A...685A..48H} to classify $\sim 400.000$ galaxies into four broad morphological classes and provide accurate measurements of the stellar mass functions over $95\%$ of cosmic history to inform models of galaxy formation. Based on these new measurements, we discuss the implications for the epoch of emergence of the Hubble sequence and the connection between star formation and morphology. In the second step, we use a large foundation neural network model trained on a variety of Galaxy Zoo (GZ) campaigns (e.g.,~\citealp{2023MNRAS.526.4768W}) and fine-tuned on available classifications in the CEERS survey~\citep{2024ApJ...969L...2F} to provide finer classifications of the internal structure of galaxies. We then use the abundance of stellar bars to establish additional constraints on the abundance and emergence of stellar disks as a function of cosmic time fueled by theoretical findings (e.g., ~\citealp{2010ApJ...719.1470V,2012ApJ...757...60K,2018MNRAS.477.1451F,2022MNRAS.512..160R, 2024arXiv240609453F,2023ApJ...947...80B,2024ApJ...968...86B}).

The paper proceeds as follows. In Section~\ref{sec:data}, we describe the dataset and the data products used. Section~\ref{sec:methods} details the methods employed for morphological classification. The main results are presented in Sections~\ref{sec:results} and~\ref{sec:bars} and discussed in Section~\ref{sec:disc}. All data are made publicly available with this publication. Throughout the work we employ the AB magnitude system and use a~\cite{2016A&A...594A..13P} cosmology.

\section{Data}
\label{sec:data}

\subsection{COSMOS-Web survey and  photometric data}

The main dataset supporting this work is the multi-band imaging survey from the \JWST\ Cycle 1 program COSMOS-Web \citep[GO\#1727, PI: Casey \& Kartaltepe]{2023ApJ...954...31C}. COSMOS-Web is a wide imaging survey ($0.54$ sq. deg. of which $0.43$ are used in this work) in four \NIRCAM{} (F115W, F150W, F277W, F444W) filters and one \MIRI{} (F770W) filter. The \NIRCAM{} (\MIRI{}) filters reach a 5$\sigma$ depth of AB mag $27.2 - 28.2$ ($\sim25.7$), measured in empty apertures of $0.15\arcsec$ ($0.3\arcsec$) radius \citep{2023ApJ...954...31C}. Data reduction was carried out using the version 1.10.0 of the \JWST{} Calibration Pipeline \citep{2023zndo...6984365B}, Calibration Reference Data System (CRDS) pmap-1075 and \NIRCAM{} instrument mapping imap-0252. Mosaics are created at \SI{30}{\mas} for short-wavelength filters and \SI{60}{\mas} for long-wavelength and MIRI filters. The \NIRCAM{} and \MIRI{} image processing and mosaic making will be described in detail in Franco et al. (in prep.) and Santosh et al. (in prep.).

In addition to this new JWST data we also make use of more than 30 photometric bands from ground-based observatories and from the Hubble Space Telescope (HST) gathered in the COSMOS field over the years \citep{2022ApJS..258...11W}. This is required for deriving physical properties of galaxies (see Subsection~\ref{sec:phys_props}). The imaging data and the resulting galaxy catalog are presented in detail in Shuntov et al. (in prep.). In this particular work, we use the same data as in~\cite{2024arXiv241008290S}. We hence refer the reader to the aforementioned work for full details.  

In summary, we use:
\begin{itemize}
\item $u$ band imaging from the CFHT Large Area $U$ band Deep Survey (CLAUDS; \citealt{2019MNRAS.489.5202S}).
\item Optical data, i.e. $g,r,i,z,y$ and three narrow bands, comes from The Hyper Suprime-Cam (HSC) Subaru Strategic Program (HSC-SSP; \citealt{2018PASJ...70S...4A}) DR3 \citep{2022PASJ...74..247A}. In addition, we include the reprocessed Subaru Suprime-Cam images with 12 medium bands in optical \citep{2007ApJS..172....9T,2015PASJ...67..104T}. We also include \hst{}/ACS F814W band \citep{2007ApJS..172..196K} data available in the COSMOS area.
\item NIR data - $Y,J,H,K_s$ bands and one narrow band (\SI{1.18}{\micro\meter}) - from the \UVISTA{} DR6 survey \citep{2012A&A...544A.156M}
\end{itemize}.

\begin{figure}[t]
  \centering
  \setlength{\abovecaptionskip}{-3mm}
    \includegraphics[width=1\columnwidth]{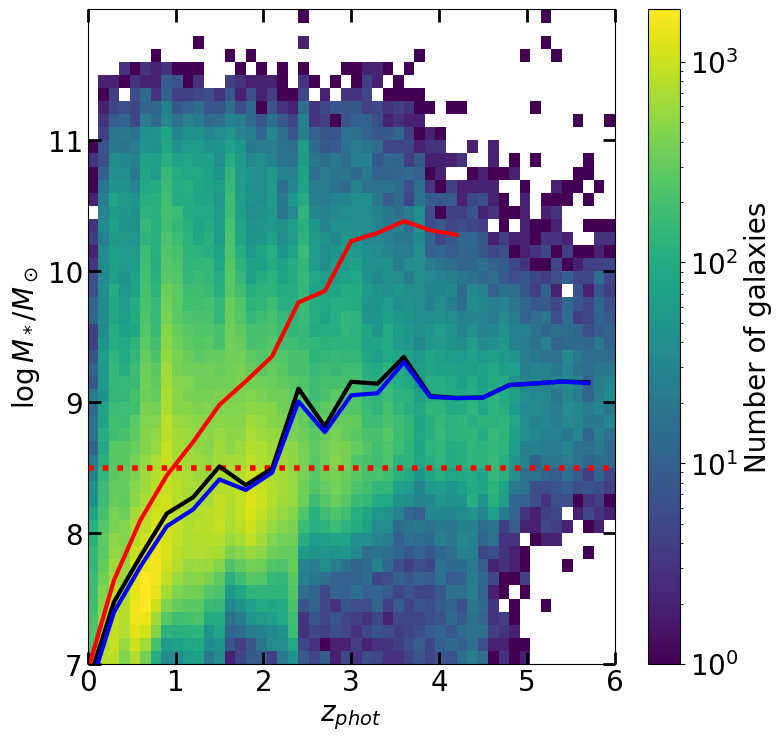}
    \vspace{0.2cm}
 \caption{Photo-$z$ vs. stellar mass diagram showing the completeness limits for the COSMOS-Web catalog. The stellar mass completeness limits are derived following~\cite{2010A&A...523A..13P} and are indicated by the solid black, red and blue lines for all, quiescent and star-forming galaxies, respectively. The horizontal dotted red line shows the stellar mass threshold adopted for the scientific analysis in this work.
 }
  \label{fig:photo-z_smass}
\end{figure}

\subsection{Galaxy physical properties}
\label{sec:phys_props}

We use a standard SED fitting approach (\Lephare;~\citealp{2002MNRAS.329..355A,2006A&A...457..841I}) to estimate photometric redshifts and physical properties of galaxies using the photometric data described in the previous section. As input for the SED fitting, we use model photometry obtained by fitting PSF convolved S\'ersic models with \textsc{SourceXtractor++}~\citep{2020ASPC..527..461B} in each filter for all sources detected in a positive-truncated $\sqrt{\chi^2}$ combination of the available \NIRCAM{} bands.  The structural parameters are constrained using the NIRCam bands while the flux is fitted independently for every filter (see~Shuntov et al. in prep. for more details). 

The SED fitting code \Lephare is run with BC03 template models~\citep{2003MNRAS.344.1000B} and a combination of 12 different star formation histories (SFH; exponentially declining and delayed) as described in~\cite{2015A&A...579A...2I}. Three different attenuation curves are used to correct for dust attenuation~\citep{2000ApJ...533..682C, 2013A&A...558A..67A, 2018ApJ...859...11S}. Emission lines are also taken into account by using the approach in~\citet{2020MNRAS.494..199S}. The IGM absorption is estimated using the analytic correction of~\citet{1995ApJ...441...18M}.  

\Lephare is run in two steps with the same configuration. The first one estimates the photometric redshift of each source. We adopt the median of the posterior distribution as an estimate of the photometric redshift.  The second step uses the derived photometric redshift to estimate the physical properties, namely stellar masses, star formation rates, and absolute magnitudes. 

For this work, we only make use of the stellar masses, which are also computed as the median of the marginalized probability density function provided by \Lephare. We also use absolute magnitudes to separate quenched and star-forming galaxies (see subsection~\ref{sec:sample}).

A detailed description of the methodology as well as a precise quantification of the the accuracy and reliability of the photometric redshifts and physical properties will be included in Shuntov et al. (in prep).

\subsection{Sample selection and completeness}
\label{sec:sample}
Our primary sample for this work is magnitude-selected in the F150W band. We select all galaxies brighter than 27. The magnitude cut is calibrated to ensure a SNR large enough for accurate morphological classifications (see~\citealp{2022arXiv221014713K, 2024A&A...685A..48H}). Although the COSMOS-Web survey is shallower than the  CEERS survey which was originally used to calibrate this SNR threshold, it is conservative enough to ensure reliable morphologies. This is quantified in more detail in Section~\ref{sec:methods}.

Additionally, we adopt the following color cuts to isolate quenched galaxies from the overall population~\citep{Ilbert2013}:
$$M_{NUV}-M_{R}>3\times(M_R-M_J)$$
$$M_{NUV}-M_R>3.1$$

 where $M_{NUV}$, $M_{R}$ and $M_{J}$ are the absolute magnitudes in the NUV, R, and J bands, respectively, estimated with SED fitting (see Subsection~\ref{sec:phys_props}).

Figure~\ref{fig:photo-z_smass} shows the photometric redshift - stellar mass plane for the selected sample. We estimate the completeness following the now standard~\cite{2010A&A...523A..13P} procedure. The $90\%$ stellar mass completeness goes from $\log M_*/M_\odot\sim8$ at $z<1$ to $\log M_*/M_\odot\sim9$ at $z\sim6$. To limit the effects of incompleteness we analyze only galaxies with stellar masses larger than $10^{8.5}$ solar masses throughout this work. We are hence excluding the local equivalent of dwarf galaxies.

\section{Methods}
\label{sec:methods}
\subsection{Galaxy morphologies}

The key ingredient of this work is galaxy morphology. We use two different approaches to obtain 1- broad global morphologies and 2 - detailed internal structure of galaxies. Throughout this work, we use the morphological classification in a different filter depending on redshift in order to keep a similar optical rest-frame band in all the considered redshift range, i.e., $F150W$ for galaxies at $z<1$, $F277W$ for galaxies at $1<z<3$ and $F444W$ for galaxies at $3<z<6$. 

\subsubsection{Global Morphologies}
\label{sec:broad_morph}

We use the neural network models from~\cite{2024A&A...685A..48H} to classify galaxies into four broad morphological classes: spheroids, bulge-dominated, disk-dominated, and peculiars.

As explained in~\cite{2024A&A...685A..48H}, the supervised machine learning model is trained with labels from the CANDELS survey~\citep{2011ApJS..197...35G,2011ApJS..197...36K} and domain-adapted to JWST using adversarial domain adaptation. We refer the reader to the aforementioned work for a detailed explanation of the architecture used and a quantification of the reliability of the classifications as compared to independent visual classifications in the CEERS survey. In brief, we use a Convolutional Neural Network (CNN) with standard adversarial domain adaptation to transfer labels from HST to JWST. For this work, we use the same architecture, input image size ($32\times32$ pixels), and normalization as in~\cite{2024A&A...685A..48H}, but retrain from scratch using COSMOS-Web images as the target domain.

We follow a methodology analogous to the one presented in~\cite{2024A&A...685A..48H} to estimate uncertainties in the classification. We perform an ensemble of 10 separate trainings for each of the three filters ($F150W$, $F277W$, and $F444W$) with different initial conditions and slightly different training sets. Each of these provides four probabilities of belonging to one of the four morphological classes such that $\sum_{i=1,4} p_{cl=i}=1$.

The final probability for a given morphological class and a given filter is computed as the average of the outputs from the 10 networks. The standard deviation of these 10 classifications, which we find to be typically below 0.1, is used to assess robustness. Thus, unless otherwise noted, we define classes based on the maximum of the average probability from the 10 classification networks. Excluding objects with uncertain classifications does not significantly impact the main findings.

Alongside the four prior classifications, we categorize unresolved galaxies into a distinct compact class. These galaxies have half-light radii smaller than the PSF half-width half-maximum (HWHM) in the $F444W$ filter, equivalent to an effective radius of approximately $\sim0.5$ kpc at $z>3$. This size threshold is consistently applied for defining compact galaxies in bluer filters. In addition, throughout this work, we will use the term "Early-Type galaxies" to refer to the combination of spheroids and bulge-dominated galaxies, and "Late-Type galaxies" for the combination of disk-dominated and peculiar galaxies. Finally, when referring to Hubble Types, we consider all galaxy types except peculiar galaxies.

In~\cite{2024A&A...685A..48H}, we showed that the model achieves an agreement of $\sim80-90\%$ compared to the visual classification in CEERS. Although the COSMOS-Web data is shallower than the CEERS data on which the deep learning model was originally applied, the conservative magnitude limit applied in our sample selection ensures a high enough SNR, and we therefore expect similar accuracy. To support this claim, we show in Appendix~\ref{app:morphs} some random example cutouts of the different morphological types, which present distinct features. We also compare the morphological fractions obtained in CEERS and COSMOS-Web, showing reasonable agreement. Finally, we compare our classifications with the S\'ersic index distributions from the parametric fits (see Shuntov et al. 2025, in preparation), which also exhibit the expected trends.

\subsubsection{Resolved Morphologies}
\label{sec:zoobot}

In addition to general morphology, we also explore resolved stellar structures. In particular, we focus on stellar bars as a proxy for thin, cold stellar disks (see Section~\ref{sec:bars} for more details). We use a fine-tuned version of the \textsc{Zoobot}\footnote{\url{https://github.com/mwalmsley/zoobot}} foundation model~\citep{2023MNRAS.526.4768W} to estimate the probability of a galaxy hosting a stellar bar. \textsc{Zoobot} is a deep learning foundation model trained on a combination of citizen science labels from multiple Galaxy Zoo (GZ) campaigns~\citep{2023MNRAS.526.4768W}. For this work, we use the pre-trained EfficientNetB0 model available on HuggingFace\footnote{\url{https://huggingface.co/collections/mwalmsley/zoobot-encoders-65fa14ae92911b173712b874}}. The model has 5.33M parameters and has been used and validated in~\cite{2022MNRAS.509.3966W,2023MNRAS.526.4768W}. It is trained on the GZ Evo dataset~\citep{2022mla..confE..29W}, which includes 820k images and over 100M volunteer votes drawn from every major Galaxy Zoo campaign: GZ2~\citep{2013MNRAS.435.2835W}, GZ UKIDSS~\citep{2024RNAAS...8..198M}, GZ Hubble~\citep{2017MNRAS.464.4176W}, GZ CANDELS~\citep{2017MNRAS.464.4420S}, GZ DECaLS/DESI~\citep{2022MNRAS.509.3966W,2023MNRAS.526.4768W}, and GZ Cosmic Dawn (unpublished).

For this work, the encoder is fine-tuned with Galaxy Zoo classifications performed on JWST CEERS images (G\'eron et al. 2025 in prep.) following a tree-like structure similar to GZ CANDELS. We adhere to the standard \textsc{Zoobot} fine-tuning strategy\footnote{\url{https://zoobot.readthedocs.io/en/latest/pretrained_models.html}}.

To match the image format expected by \textsc{Zoobot}, we preprocess the COSMOS-Web images. First, we generate cutouts with a size in pixels that depends on the galaxy's effective radius ($212\times0.04\times r_e\times\sqrt(b/a)$) to ensure a similar ratio between galaxy and background pixels for all objects. We then interpolate all cutouts to a common size of $424\times424$ to match the neural network model's input definition. As with the broadband morphologies (see Section~\ref{sec:broad_morph}), we perform three independent fine-tunings for each filter ($F150W$, $F277W$, and $F444W$). However, in this case, the labels remain unchanged, as the GZ CEERS classifications are performed on color images.

The \textsc{Zoobot} model is probabilistic, designed to emulate the classifications of GZ volunteers. It provides the parameters of a Dirichlet distribution, enabling the estimation of the number of classifiers $k$ out of a total of $N$ volunteers who would select a given morphological feature (see~\citealp{2023MNRAS.526.4768W} for more details).

To construct a morphological catalog of bars, we sample the fine-tuned model 100 times, assuming 100 volunteers per sample. Given the tree-like structure of the GZ classification scheme, where the bar-related question is only asked if the galaxy is classified as featured (as opposed to smooth) and face-on, we compute the following probabilities from the 100 samples:

\[
p^{i}_{featured} = \frac{k^{i}_{featured}}{100},
\]
\[
p^{i}_{edge-on} = \frac{k^{i}_{edge-on}}{k^{i}_{featured}},
\]
\[
p^{i}_{bar} = \frac{k^{i}_{bar}}{k^{i}_{faceon}},
\]

with $i = 1, \ldots, 100$ representing the samples and $k^{i}_{featured}$, $k^{i}_{faceon}$ being the estimated number of volunteers out of 100 who classified a galaxy as featured and face-on, respectively. Note that we do not divide by the total number of volunteers (100) to account for the tree-like structure of the GZ classifications.

The final assigned probabilities ($p_{featured}$, $p_{edge-on}$, and $p_{bar}$) are computed as the mean of the $p^{i}$ values from the 100 samples. For selecting candidates likely to host a bar or have spiral arms, we apply the following cuts:

\[
p_{featured} > 0.5, \quad p_{edge-on} < 0.5, \quad p_{bar} > 0.5.
\]

\begin{figure}[t]
  \centering
    \includegraphics[width=1\columnwidth]{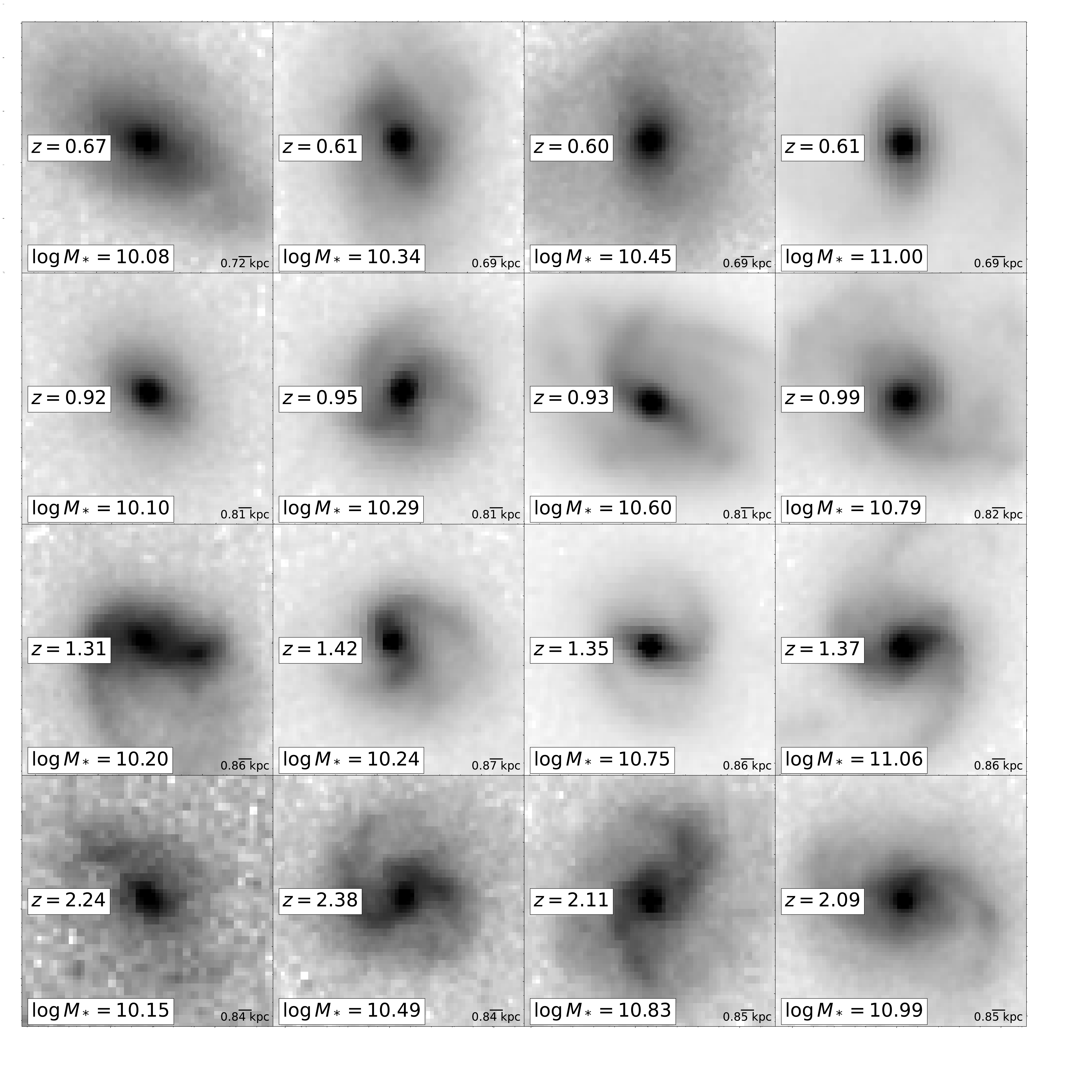}
 \caption{Random cutout examples of selected galaxies with bars with different stellar masses and redshifts. Images in filters $F150W$, $F277W$ and $F444W$ are shown for $z<1$, $1<z<3$ and $z>3$ respectively.}
  \label{fig:zoobot_bars}
\end{figure}

\begin{figure*}[t]
  \centering
    \includegraphics[width=0.3\textwidth]{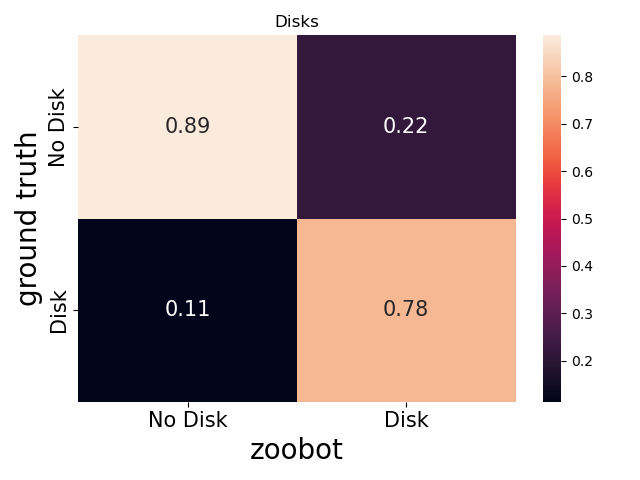}
     \includegraphics[width=0.3\textwidth]{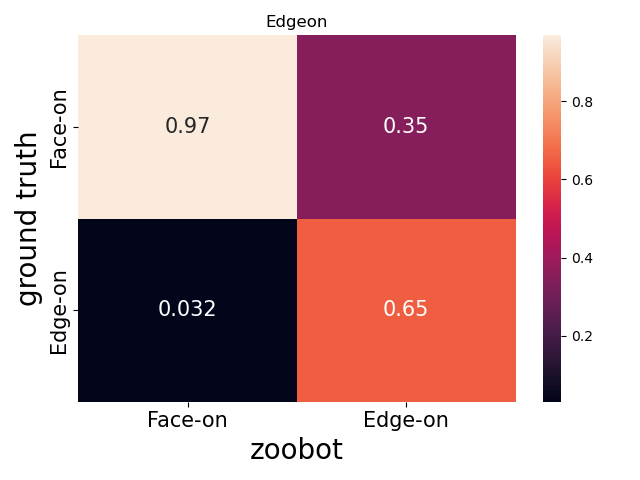}
      \includegraphics[width=0.3\textwidth]{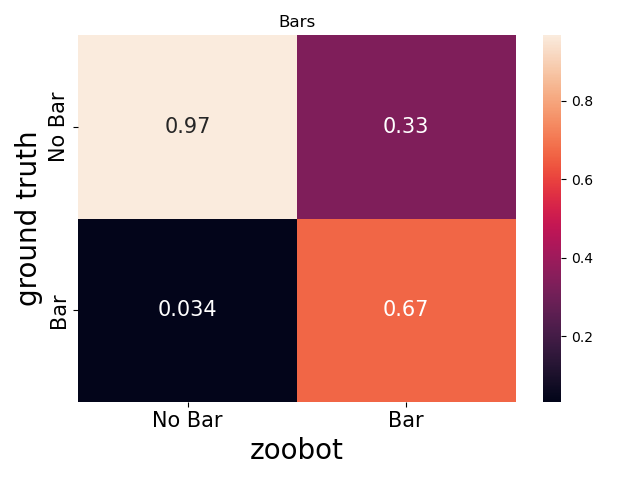}
 \caption{Confusion matrices of zoobot classifications. From left to right: featured, edge-on, bars.}
 \label{fig:conf_mat}
\end{figure*}

These thresholds account for the hierarchical structure of the Galaxy Zoo classifications. Different thresholds yield samples with varying purity and completeness. Figures~\ref{fig:zoobot_bars} and~\ref{fig:conf_mat} show random examples of selected barred galaxies and the confusion matrices for the three considered tasks in this work: featured, edge-on, and bar. The selected stamps clearly show bar signatures, supporting the sensitivity of the classification. The confusion matrices indicate satisfactory accuracies given the complexity of the task and the training set size.

For this study, the primary goal of morphological classification is not to obtain an absolute calibration of barred galaxy number densities but to use this feature as a proxy for the presence of stellar disks, as detailed in the following sections. From this perspective, the exact threshold values do not significantly affect the results, provided that the measurement biases are well calibrated and corrected. This is discussed further in Section~\ref{sec:bars}.

\subsection{Stellar Mass Functions}
\label{sec:SMFs}

The other main ingredient of this work is the stellar mass functions (SMFs). We use the same approach as in~\cite{2024arXiv241008290S}, with the only major difference being that galaxies are divided into four bins according to their morphological type, as discussed in Section~\ref{sec:broad_morph}. We refer the reader to the aforementioned work for further details. 

In summary, we use the $1/V_{max}$ estimator~\citep{1968ApJ...151..393S} in each redshift bin to account for the Malmquist bias~\citep{1922MeLuF.100....1M}. Each galaxy is weighted by the maximum volume in which it would be observed, given the redshift range of the sample and its magnitude:

\[
V_{\text{max},i} = \frac{4\pi \Omega_{\text{survey}}}{3 \Omega_{\text{sky}}} \left( d_c(z_{\text{max},i})^3 - d_c(z_{\text{min},i})^3 \right),
\]

where $\Omega_{\text{survey}}=1520$ arcmin$^2$ is the COSMOS-Web area, and $\Omega_{\text{sky}}=43\,331$ deg$^2$ is the area of the entire sky. $d_c$ refers to the comoving distance at a given redshift. $z_{\text{min}}$ is the lower redshift limit, and $z_{\text{max}} = \min(z_{bin,up},z_{lim})$, where $z_{bin,up}$ is the upper redshift limit of the bin, and $z_{lim}$ is the maximum redshift up to which a galaxy of a given magnitude can be observed, given the magnitude limit of the survey $m_{lim}$ in F150W.

The SMF is computed as a weighted sum of $\frac{1}{V_{\text{max},i}}$:

\[
\Phi(M_*) \Delta \log M_* = \sum_i^{N(M_*)} \frac{1}{V_{\text{max},i}},
\]

As described in~\cite{2024arXiv241008290S}, we include several sources of uncertainties in the SMF computation, namely: Poisson noise, SED fitting errors, and cosmic variance. Poisson noise accounts for statistical uncertainties in each stellar mass bin and is computed as $\sigma_{\text{Pois}} = \sqrt{\sum 1/V_{\text{max}}}$. Uncertainties from SED fitting translate into errors in the estimation of stellar masses. We estimate $\sigma_{\text{fit}}$ by sampling $10^3$ times the $PDF(M_*)$ produced by \Lephare, computing $\Phi(M_*)$ for each realization, and taking the standard deviation. Finally, cosmic variance measures the effect of observing a small patch of the sky. We use the same estimates of $\sigma_{\text{cv}}$ reported in Figure 3 of~\cite{2024arXiv241008290S}, based on the results from~\cite{2024arXiv240300050K} as discussed in the aforementioned work.

The final uncertainty on the SMF is then computed as:

\[
\sigma^2_{\Phi} = \sigma^2_{\text{Pois}} + \sigma^2_{\text{fit}} + \sigma^2_{\text{cv}}.
\]

More details on the relative contributions of each term as a function of stellar mass can be found in \cite{2024arXiv241008290S}.

Finally, we perform MCMC fits to the SMFs using three different models: a single Schechter (SSM) model, a double Schechter (DSM) model, and a double power-law model (DPLM):

\begin{align*}
\text{SSM:} \quad 
\Phi \, \mathrm{d}(\log M_*) &= \ln(10) \, \Phi^* \, e^{-10^{\log M_* - \log M^*}} \times \\
& \left( 10^{\log M_* - \log M^*} \right)^{\alpha + 1} \, \mathrm{d}(\log M_*), \\[1em]
\text{DSM:} \quad 
\Phi \, \mathrm{d}(\log M_*) &= \ln(10) \, e^{-10^{\log M_* - \log M^*}} \times \\
& \left[ \Phi^*_1 \left( 10^{\log M_* - \log M^*} \right)^{\alpha_1 + 1} + \right. \\ 
& \left. \Phi^*_2 \left( 10^{\log M_* - \log M^*} \right)^{\alpha_2 + 1} \right] \, \mathrm{d}(\log M_*), \\[1em]
\text{DPLM:} \quad 
\Phi \, \mathrm{d}(\log M_*) &= \Phi_* \Big/ \\
& \left[ 10^{-\left(\log M_*-\log M^*\right)\left(1+\alpha_1\right)} + \right. \\
& \left. 10^{-\left(\log M_*-\log M^*\right)\left(1+\alpha_2\right)} \right] \, \mathrm{d}(\log M_*),
\end{align*}

To account for incompleteness, we only fit the SMFs for galaxies more massive than $10^{8.5}$ solar masses. Even though the $V_{max}$ formalism should correct for incompleteness, Figure~\ref{fig:photo-z_smass} shows that the correction below this mass is large. Hence we decided not to use these values in the fitting procedure. We run $60,000$ iterations with a number of chains equal to eight times the number of free parameters in each model. We discard the first $10,000$ iterations to compute the summary statistics of the posterior distributions. To select the best model for each redshift and morphology bin, we apply the Corrected Akaike Information Criterion (AICc):

\[
\text{AICc} = 2k - 2\ln (L_{\text{max}}) + \frac{2k(k+1)}{n-k-1},
\]

where $k$ is the number of free parameters, $n$ is the number of observed data points in the SMF, and $L_{\text{max}}$ is the maximum likelihood from the MCMC chain. We use AICc instead of the Bayesian Information Criterion (BIC) because the former is more accurate when the ratio $n/k$ between observed data points and free parameters is small, as is the case here. The results of the best-fit models for all galaxies, as well as for star-forming and quenched ones, are reported in Appendix~\ref{app:smfs}.

\section{SMFs as a Function of Morphology and Star Formation from $z \sim 7$}
\label{sec:results}

The top row of Figure~\ref{fig:smf_4types} shows the stellar mass functions (SMFs) for the four broad morphological types considered in this work—spheroids, bulge-dominated, disk-dominated, and peculiars—along with the best-fit models (see Section~\ref{sec:SMFs}) across different redshift bins. In the middle and bottom rows of Figure~\ref{fig:smf_4types}, galaxies are further divided into star-forming and quiescent populations using the color selection criteria detailed in Section~\ref{sec:data}. Additionally, Figures~\ref{fig:smf_fits_all} through~\ref{fig:smf_fits_sf} show the best-fit models for all morphological types at various redshifts. In this work, we do not show global SMFs since they are presented and extensively described in~\cite{2024arXiv241008290S} using the same data.

\subsection{Overall Galaxy Population}

The SMFs of the overall population exhibit distinct shapes depending on the morphological type, suggesting that the morphological selection is capturing different galaxy populations. Late-Type systems, i.e., disk-dominated and peculiar galaxies, are generally well described by a single component, while Early-Type galaxies, i.e., spheroids and bulge-dominated systems, tend to display a double-peak shape. The best-fit models reported in Tables~\ref{tab:early} and~\ref{tab:late} confirm this trend. Interestingly,~\cite{2024MNRAS.531.4857C}, which also computed SMFs for different morphological types, did not identify these distinct behaviors for bulge- and disk-dominated galaxies. This discrepancy could be attributed to differences in classification schemes or to the significantly smaller sample size in their study compared to this work.

We also observe a deficit of low-mass spheroids and bulge-dominated galaxies at $z > 3$, which appears to vanish at lower redshifts, where the double-peak shape becomes more pronounced. This deficit can be attributed to the inclusion of most low-mass systems in the compact class, as detailed in Section~\ref{sec:broad_morph}, which encompasses all galaxies with physical sizes smaller than $0.5$ kpc. The figure also reveals that the normalization of Hubble-Type galaxies (spheroids, bulge-dominated, and disk-dominated) increases by three to four orders of magnitude below $z \sim 5$, whereas for peculiar galaxies, the increase is only a factor of $\sim 2$. This disparity in the growth rates of number densities is further illustrated in Figure~\ref{fig:smf_peculiar_hubble}, where we compare the SMFs of all Hubble Types (spheroids, bulge-dominated, and disk-dominated systems) with the SMFs of peculiar galaxies alone. 

At $z > 4.5$, the abundance of Hubble Types is almost negligible compared to peculiar galaxies. However, there is a rapid increase in the number density of Hubble Types, leading to similar densities at $z \sim 3$, particularly at the high-mass end. This trend reflects the rapid assembly of the Hubble Sequence, which will be discussed further in Section~\ref{sec:disc}.

The rightmost panel of Figure~\ref{fig:smf_peculiar_hubble} also shows the evolution of the SMFs of compact galaxies. At $z > 4.5$, the abundance of compact galaxies is comparable to that of peculiar galaxies and exceeds that of Hubble Types. This trend is also evident in Figure~\ref{fig:smf_fits_all}, where compact galaxies become less abundant than disk-dominated galaxies only at $z \sim 4$ and below.

\subsection{Quiescent and Star-Forming Galaxies}

\begin{figure*}[t]
  \centering
    \includegraphics[width=1\textwidth]{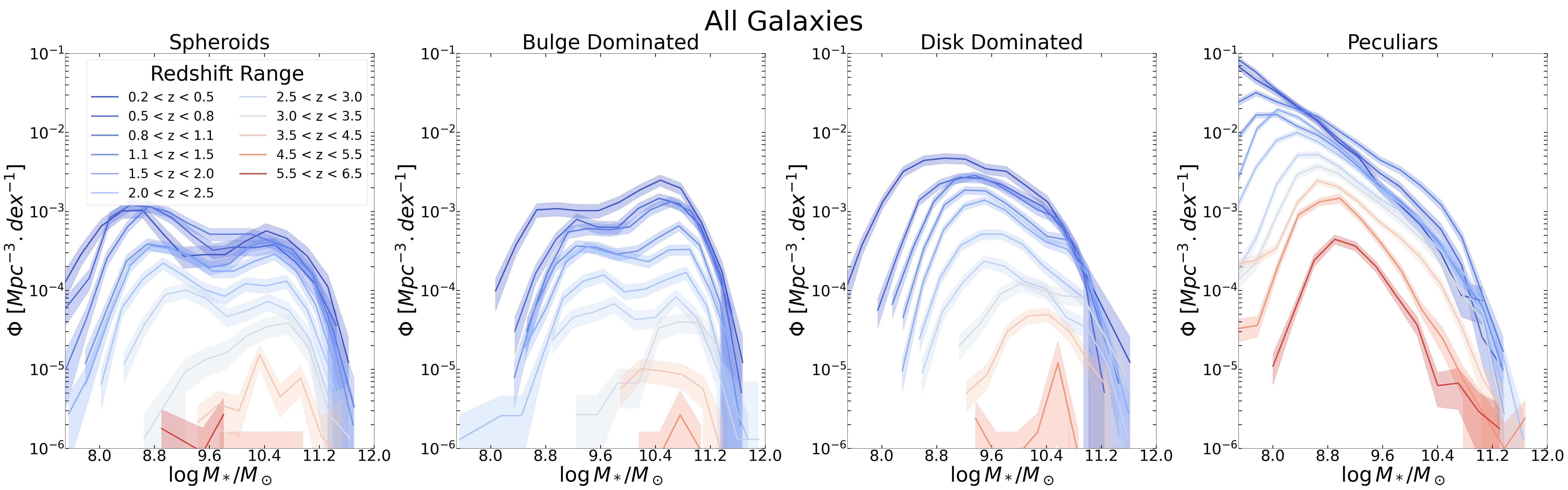}
    \includegraphics[width=1\textwidth]{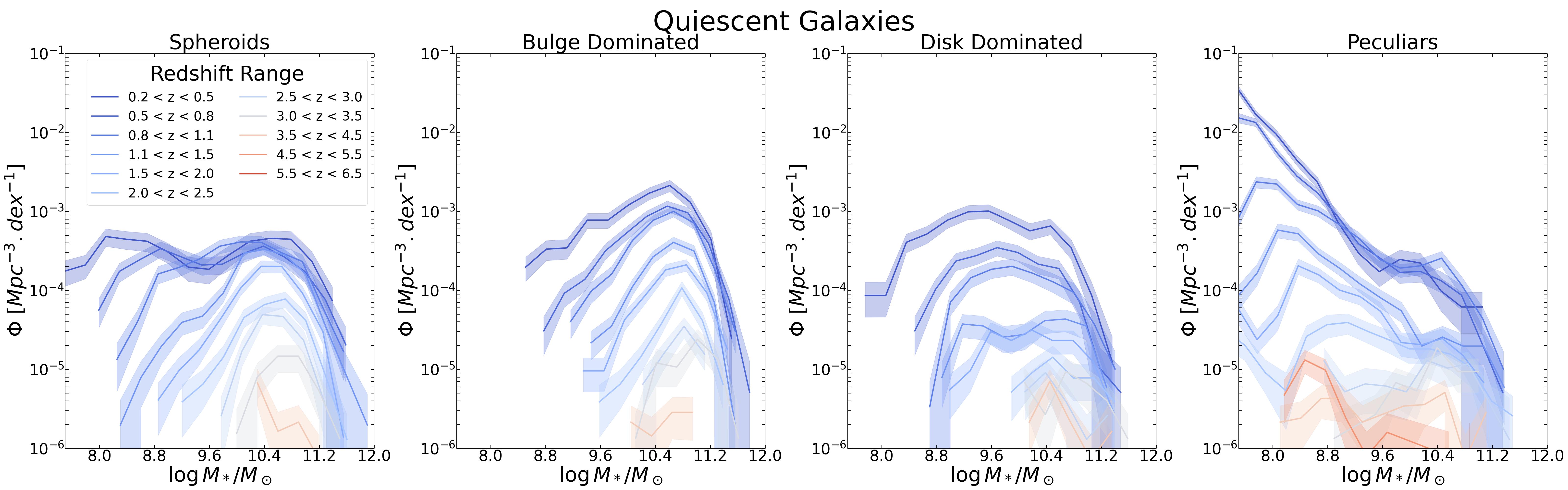}

    \includegraphics[width=1\textwidth]{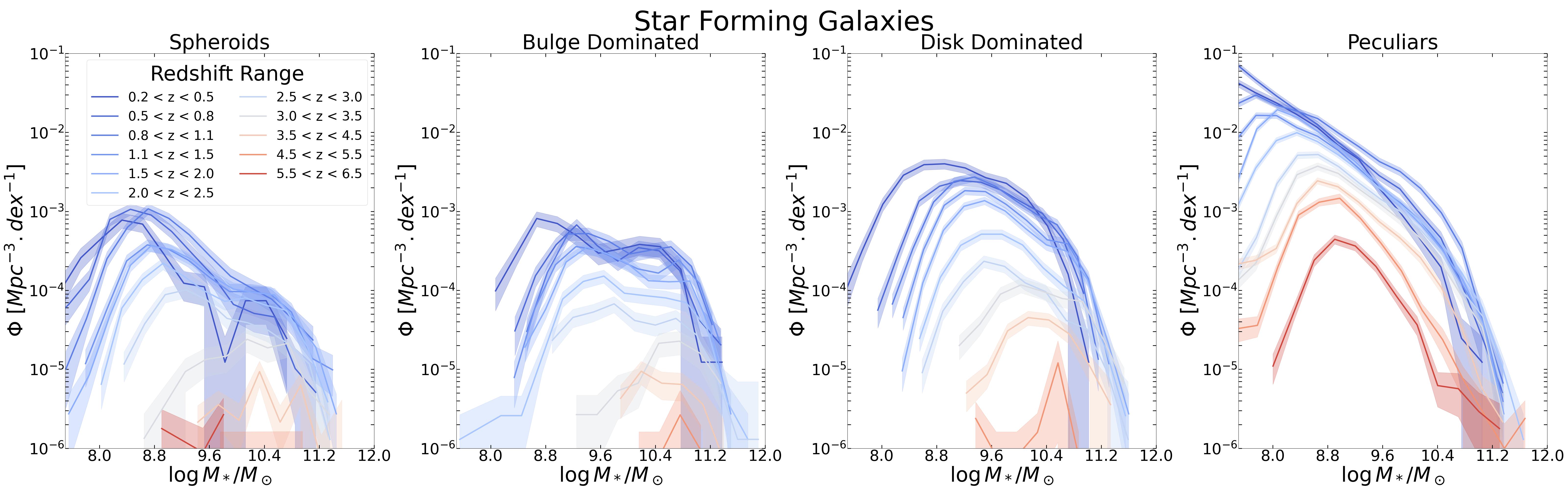}
 \caption{Evolution of the stellar mass functions of different Hubble types. From top to bottom: all galaxies, quiescent, star-forming galaxies. Each column shows a different morphological type, from left to right: spheroids, bulge-dominated, disk dominated and peculiar galaxies. The different colors correspond to different redshift as labeled. The shaded regions indicate the confidence intervals, combining Poisson errors and cosmic variance (see text for details). }
  \label{fig:smf_4types}
\end{figure*}

\begin{figure*}[t]
  \centering
    \includegraphics[width=1\textwidth]{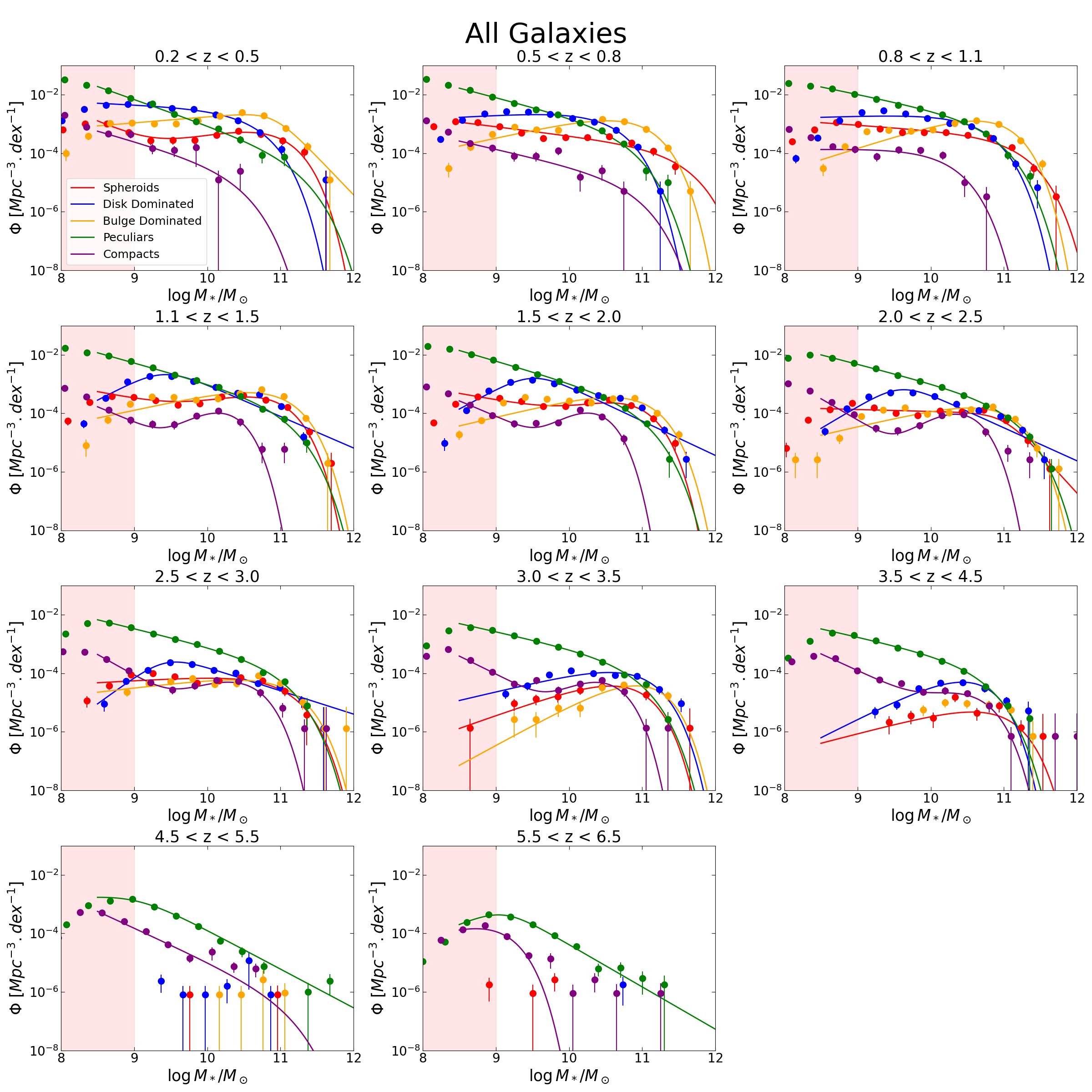}

 \caption{Evolution of the stellar mass functions of different morphological types. Each panel shows a different redshift bin. Filled colored dots indicate the observational measurements. The solid lines are the best fit models (see text for details). Only converged fits are shown. The vertical red filled region indicates the stellar mass range impacted by incompleteness }
  \label{fig:smf_fits_all}
\end{figure*}

\begin{figure*}[t]
  \centering
    \includegraphics[width=1\textwidth]{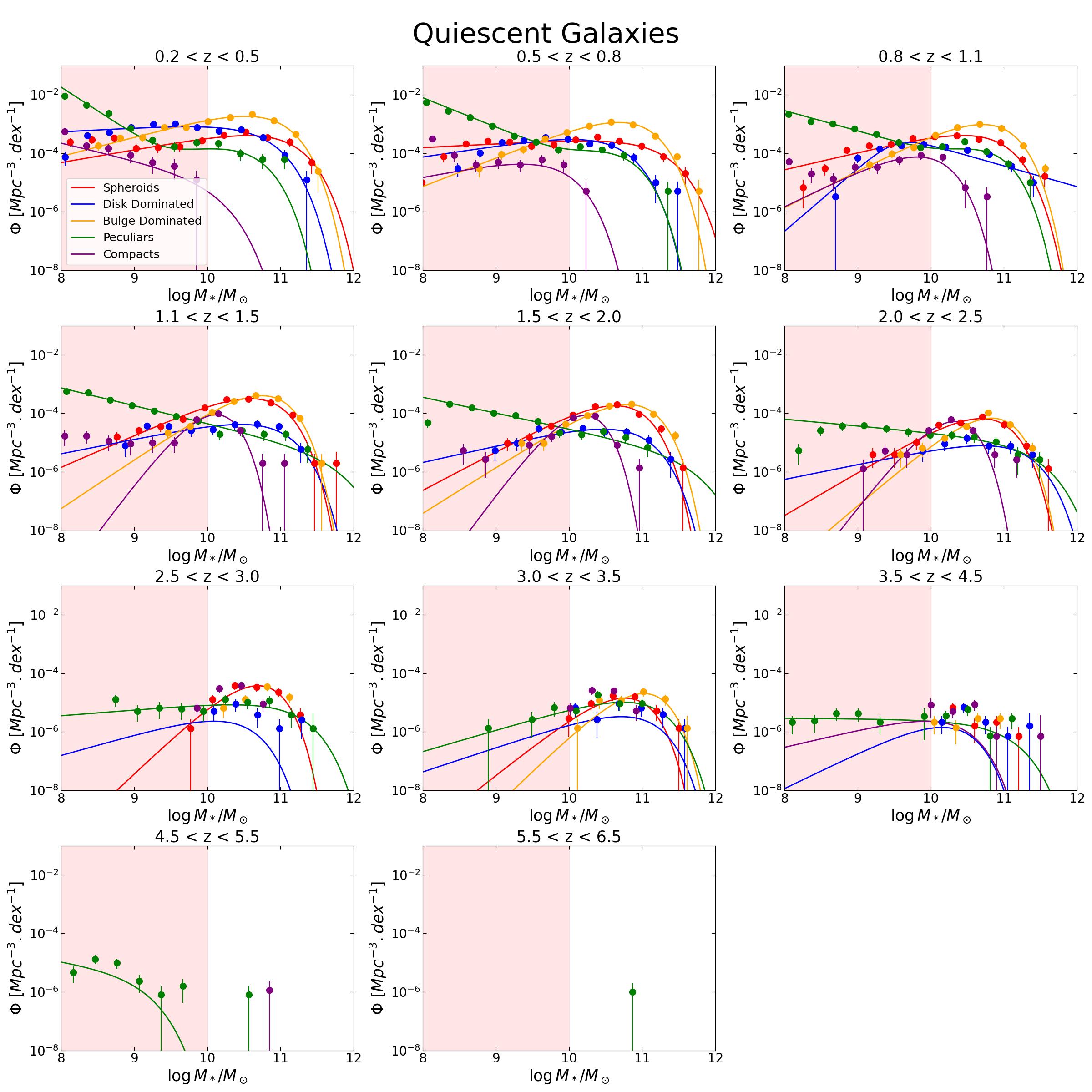}

 \caption{Evolution of the stellar mass functions of different morphological types for quiescent galaxies. Each panel shows a different redshift bin. Filled colored dots indicate the observational measurements. The solid lines are the best fit models (see text for details). The vertical red filled region indicates the stellar mass range impacted by incompleteness}
  \label{fig:smf_fits_q}
\end{figure*}

\begin{figure*}[t]
  \centering
    \includegraphics[width=1\textwidth]{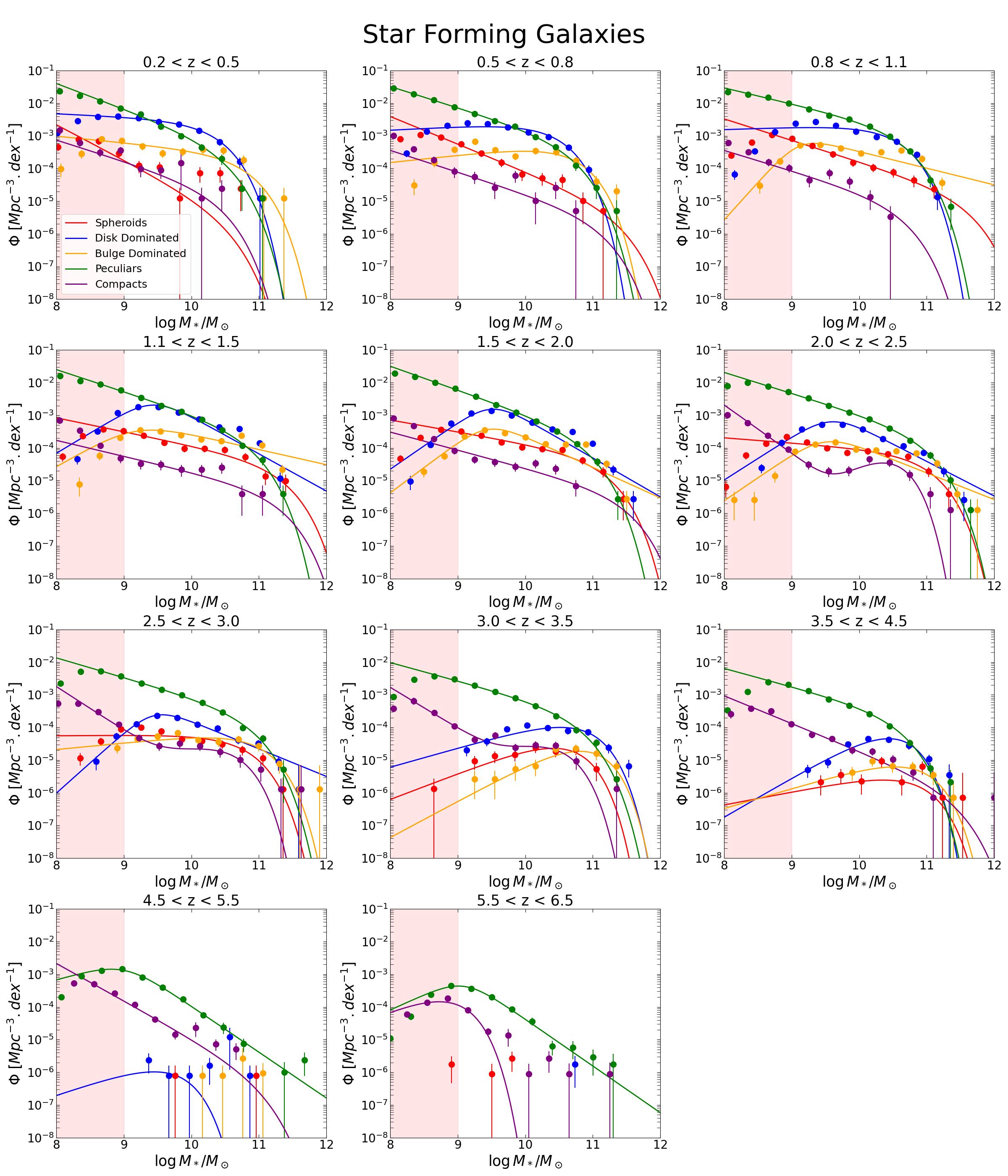}

 \caption{Evolution of the stellar mass functions of different morphological types for star forming galaxies. Each panel shows a different redshift bin. Filled colored dots indicate the observational measurements. The solid lines are the best fit models (see text for details). The vertical red filled region indicates the stellar mass range impacted by incompleteness.}
  \label{fig:smf_fits_sf}
\end{figure*}

\begin{figure*}[t]
  \centering
    \includegraphics[width=1\textwidth]{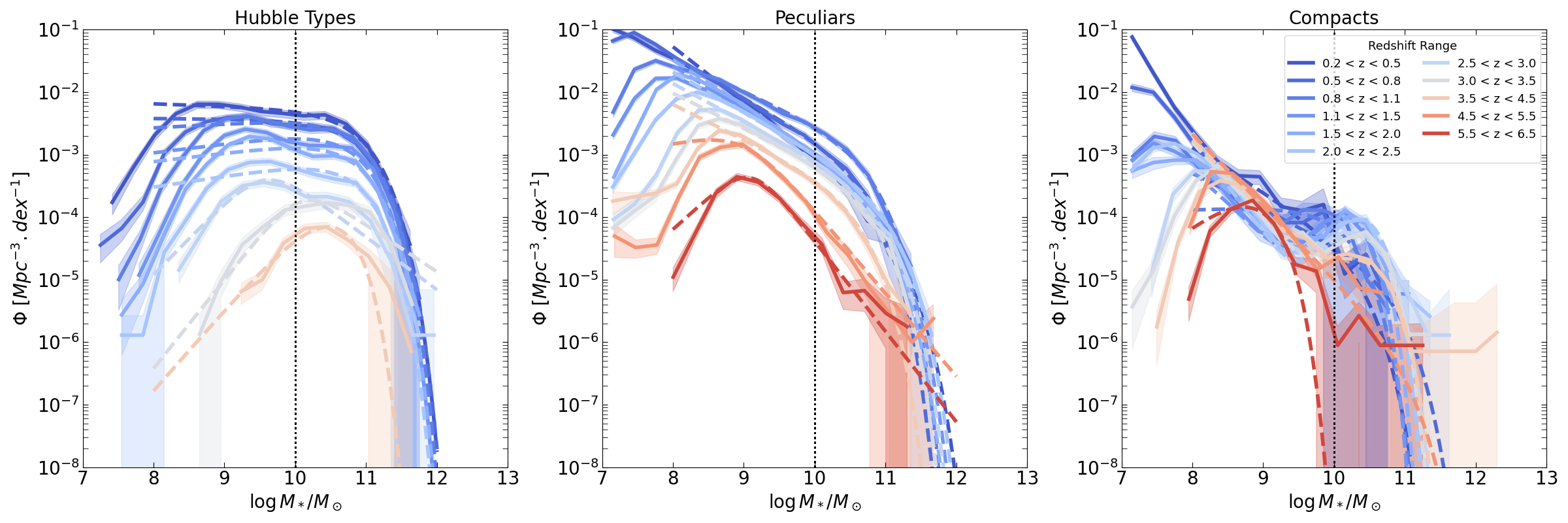}
 \caption{Stellar mass function of Hubble type galaxies (left panel), peculiar galaxies (middle panel) and compact galaxies (right panel). Each line shows a different redshift bin as labeled. The dashed lines indicate the best fit models. }
  \label{fig:smf_peculiar_hubble}
\end{figure*}

The middle and bottom rows of Figure~\ref{fig:smf_4types} present the SMFs of different morphologies for quiescent and star-forming galaxies separately.

We first observe that the abundance of quiescent galaxies is almost negligible at $z > 4$. Although some massive quenched galaxies have been reported at very high redshifts~\citep{2024Natur.628..277G,2024MNRAS.534..325C}, they remain extremely rare objects, as expected from state-of-the art models of galaxy formation. However, below $z \sim 3.5$, there is a rapid increase in the quiescent population, particularly at the high-mass end. Regarding the morphological mix, we detect a clear deficit of massive quiescent galaxies with late-type morphologies, even at the highest redshifts, when compared with the overall population. Conversely, the abundance of massive quiescent early-type galaxies is very similar to that of the overall population. Star-forming galaxies, on the other hand, are dominated by late-type morphologies across all cosmic epochs. This suggests that the connection between quenching and morphology at the high stellar-mass end has been established since the emergence of the first quenched galaxies. We will explore this further in Section~\ref{sec:disc}.

\begin{figure*}[t]
  \centering
    \includegraphics[width=1\textwidth]{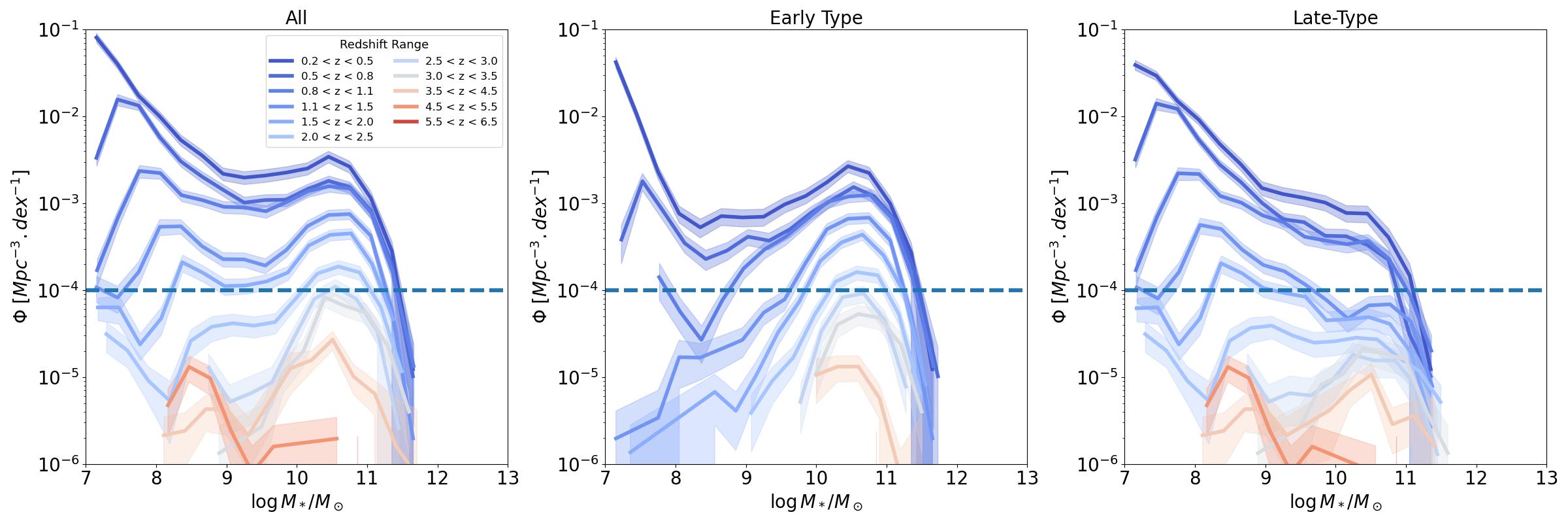}
 \caption{Stellar mass function of quenched galaxies. The left panel shows all quenched galaxies, the middle panel quenched galaxies with an early-type morphology and the right panel quenched galaxies with a late-type morphology. Different colors indicate different redshift bins as labeled. The vertical red filled region indicates the stellar mass range impacted by incompleteness }
  \label{fig:early_late_Q}
\end{figure*}

Interestingly, at the low-mass end, we measure a significant fraction of quiescent galaxies with peculiar and disk-like morphologies. As suggested in previous studies (e.g., ~\citealp{2016A&A...590A.103M}), this could indicate the presence of different quenching channels dominating at different stellar mass regimes. To further quantify this morphological dichotomy among quiescent galaxies, we plot in Figure~\ref{fig:early_late_Q} the SMFs of quenched galaxies divided into Late-Type and Early-Type morphologies. 

This analysis highlights that the massive end of the SMFs for Early-Type galaxies is nearly identical to the SMFs of all quiescent galaxies. However, at the low-mass end, the opposite trend is observed. The shape of the SMFs for all quiescent galaxies closely follows the SMFs of Late-Type galaxies across all redshifts.



\section{Stellar Bars from $z \sim 7$}
\label{sec:bars}

In this section, we investigate the abundance of stellar bars as a function of redshift in massive galaxies using the \textsc{Zoobot} classifications (see Section~\ref{sec:zoobot}). The primary purpose of this work is to use bars as a proxy for disk formation (e.g.,~\citealp{2003MNRAS.341.1179A,2018MNRAS.477.1451F}). Therefore, we are mainly interested in the relative evolution of the bar frequency with respect to a reference population.

\begin{figure}[t]
  \centering
    \includegraphics[width=1\columnwidth]{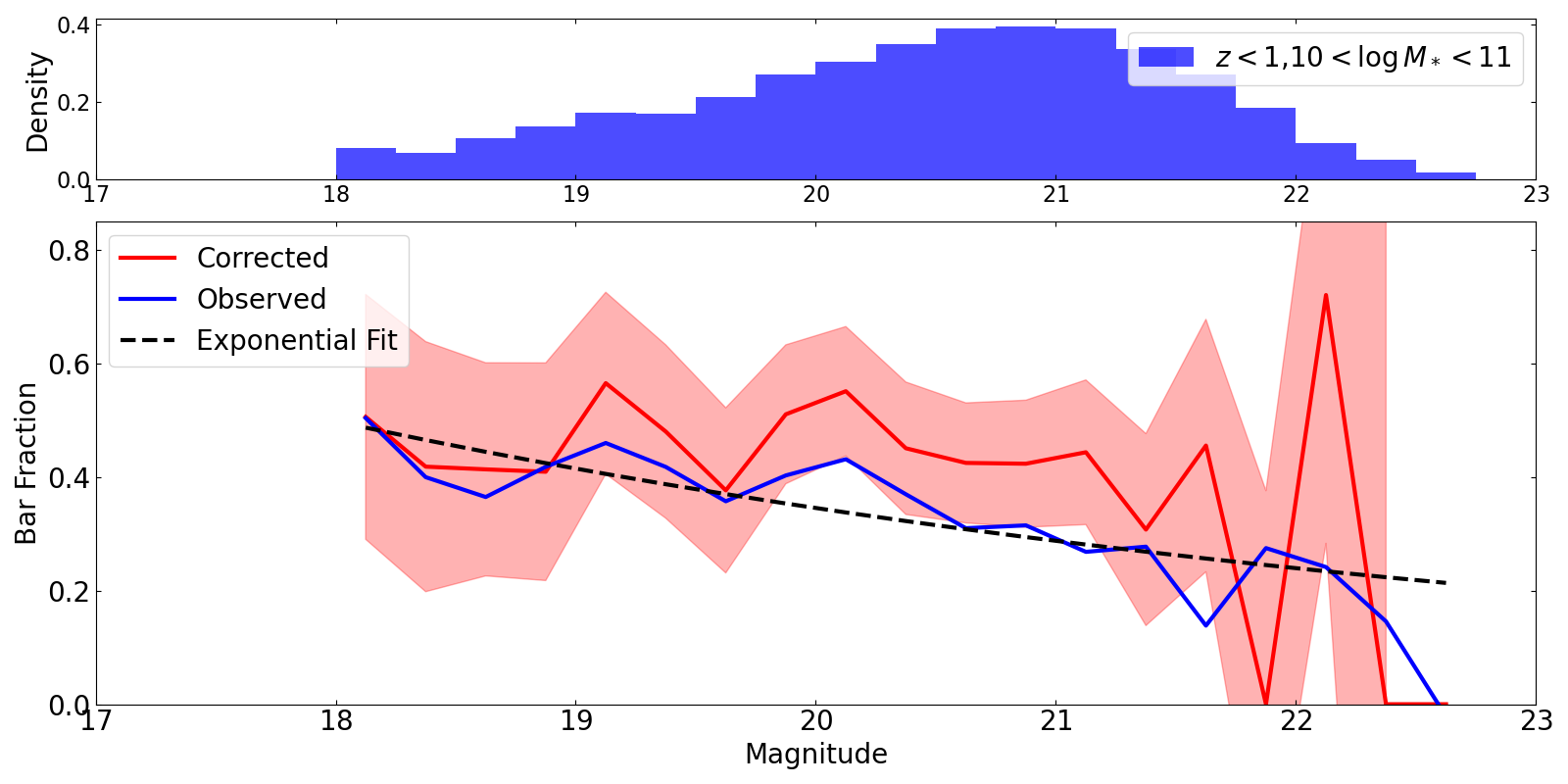}
    \includegraphics[width=1\columnwidth]{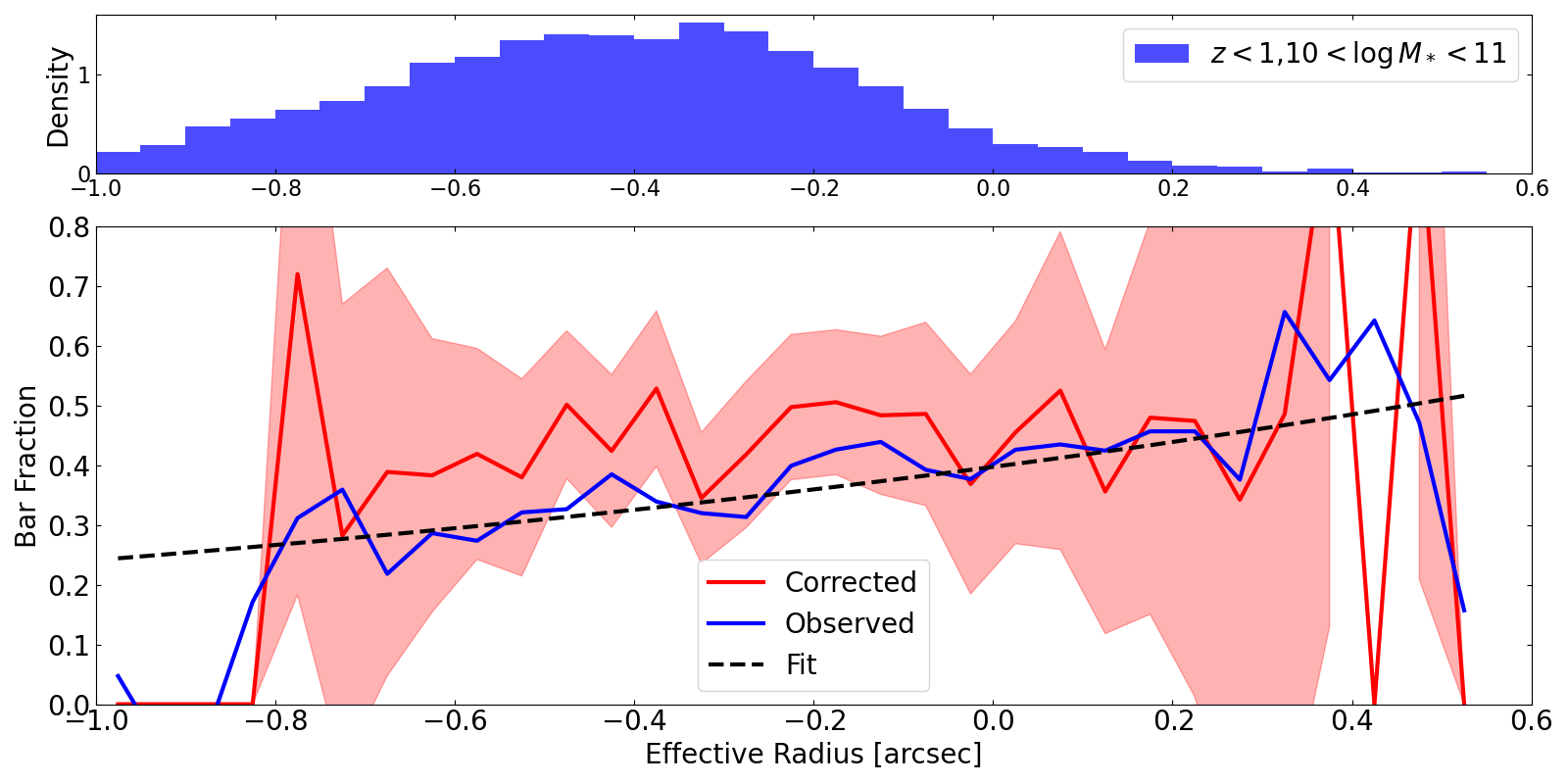}

 \caption{Bar fraction as a function of apparent $F150W$ magnitude (top panel) and effective radius (bottom panel) for a population of massive ($\log M_*/M_\odot > 10$) galaxies at $z < 1$. The blue solid line indicates the observed fraction, while the black dashed line shows the best fit (see text for details). The red solid line and shaded region show the corrected fractions and associated uncertainties after applying the weights.}
  \label{fig:bars_mag_rad}
\end{figure}

The main challenge lies in the biases introduced by observational effects (e.g.,~\citealp{2024arXiv240906100G}). By construction, identifying bars is more challenging in small and faint galaxies, which are more abundant at high redshifts, potentially leading to an artificially stronger decrease in the fraction of bars. To quantify this effect, we present Figure~\ref{fig:bars_mag_rad}, which shows the bar fraction for a sample of massive galaxies at $z < 1$ as a function of magnitude—a proxy for SNR—and apparent size. Assuming that this galaxy population is sufficiently homogeneous to share similar physical properties, any dependence of the bar fraction on apparent size or SNR can be attributed to detection biases. The figure indeed shows that the bar fraction is lower for small and faint galaxies, even if they share similar mass and redshift. While there remains a possibility that part of this trend is due to physical differences—e.g., small galaxies at fixed stellar mass hosting fewer bars—we assume it is entirely driven by the detection method.

To correct for this effect, we fit an exponential model to the observed bar fraction as a function of magnitude:

\[
f_{\text{bar}}(\text{mag}) = A \times e^{-B \times \text{mag}}
\]

Using the best-fit model, we assign a weight to every galaxy classified as hosting a bar as follows:

\[
w(\text{mag}) = \frac{f_{\text{bar}}(18)}{f_{\text{bar}}(\text{mag})}
\]

These weights are then used to compute the corrected bar fractions. This approach is analogous to the debiasing weights applied in some Galaxy Zoo catalogs (e.g.,~\citealp{2013MNRAS.435.2835W,2017ApJ...843...46S}). The top panel of Figure~\ref{fig:bars_mag_rad} shows the weighted bar fraction, which displays minimal dependence on magnitude, as expected. Given the correlation between magnitude and size, this correction also flattens the bar fraction as a function of size (bottom panel of Figure~\ref{fig:bars_mag_rad}). For simplicity, we apply only the magnitude-based correction.

It is important to note that this purely empirical correction might overestimate the true bar fraction, as we assume that there is no physical variation in bar fractions with magnitude or apparent size within the reference sample of massive galaxies at $z < 1$. Furthermore, the weights are extrapolated to fainter magnitudes ($F150W \sim 23$). Nevertheless, this method provides an upper limit to the abundance of stellar bars. In a recent work~\cite{2024A&A...688A.158L} explored the impact of observational effects on the detection of bars in simulated CEERS galaxies. They found that the decrease in the observed bar fraction due to these effects can be up to $\sim50-60\%$ at $z\sim3$ which is roughly consistent with our independent corrections.

\begin{figure*}[t]
  \centering
    \includegraphics[width=1\columnwidth]{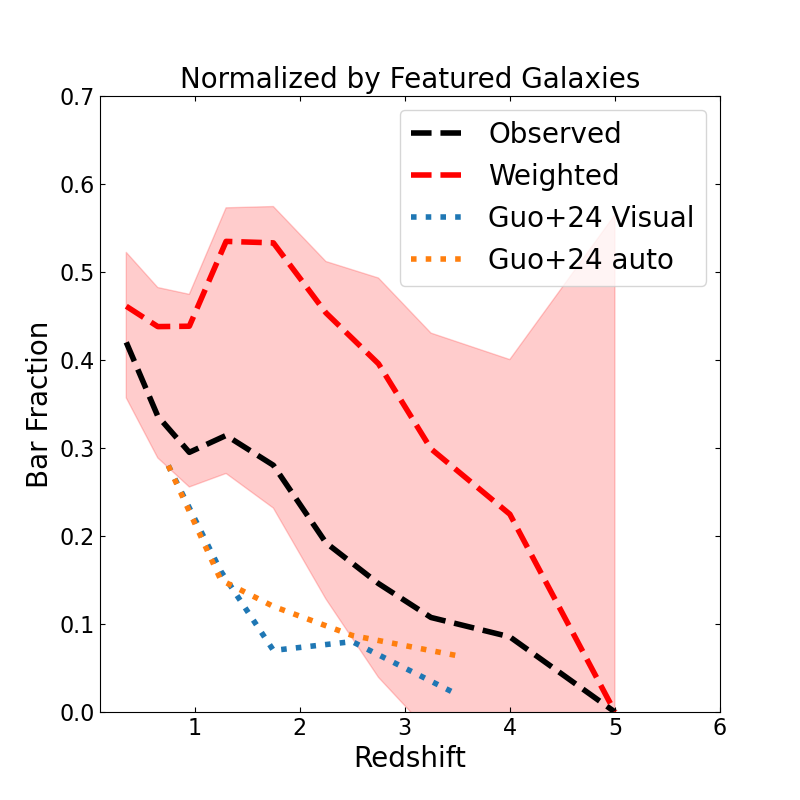}
    \includegraphics[width=1\columnwidth]{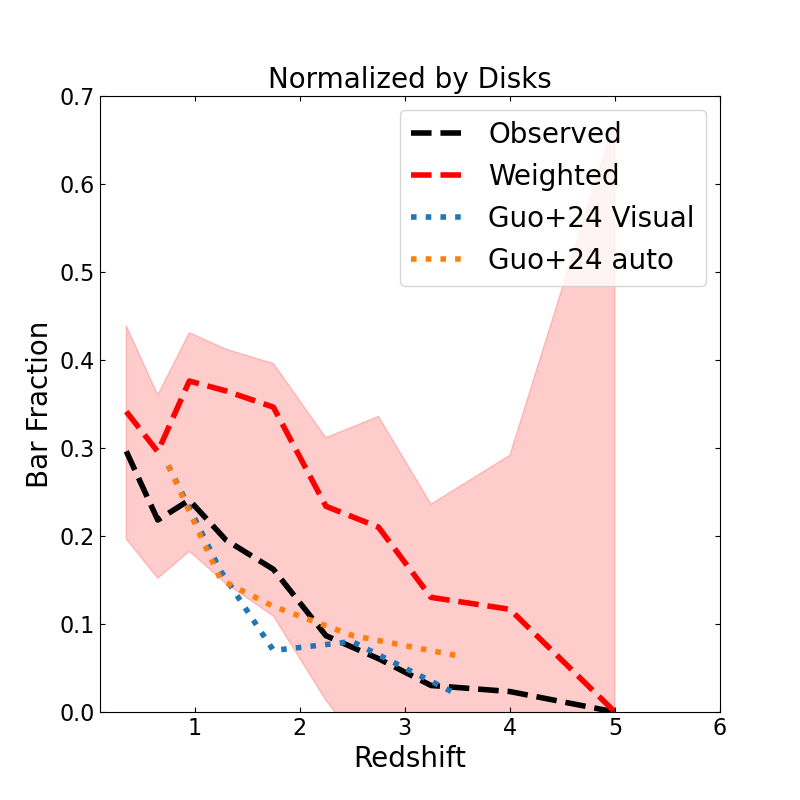}

 \caption{Bar fraction for massive galaxies as a function of redshift. The black dashed line shows the raw observed fraction while the red dashed line shows the fraction corrected from observational bias (see text for details). The orange and blue dotted lines indicate published results in CEERS~\citep{2024arXiv240906100G}. }
  \label{fig:bar_frac}
\end{figure*}

Using this correction, we present in Figure~\ref{fig:bar_frac} the evolution of the bar fraction as a function of redshift for massive galaxies, both before and after applying the correction. The bar fraction is computed in two ways:

\[
f^{\text{featured}}_{\text{bar}}(z) = \frac{N_{\text{bar}}(z)}{N_{\text{featured}}(z)}
\]
and
\[
f^{\text{disks}}_{\text{bar}}(z) = \frac{N_{\text{bar}}(z)}{N_{\text{disks}}(z)}
\]

where $N_{\text{bar}}$, $N_{\text{featured}}$, and $N_{\text{disks}}$ are the numbers of barred, featured, and disk-dominated galaxies, respectively, in a given redshift bin (see Section~\ref{sec:methods}). $N_{\text{bar}}$ and $N_{\text{featured}}$ are derived from the \textsc{Zoobot} classification, while $N_{\text{disks}}$ is based on the broad morphological classification. For the weighted bar fraction, $N_{\text{bar}}$ is computed using the estimated weights.

The abundance of bars decreases with redshift, as reported in numerous previous studies. When normalizing by the number of featured galaxies, we measure a higher overall bar fraction—even before correction—than recent results published using JWST data in the CEERS survey~\citep{2024arXiv240906100G}. However, the agreement improves substantially when normalizing by the number of disks, as done by~\cite{2024arXiv240906100G}. This discrepancy reflects that the featured classification commonly used in Galaxy Zoo does not completely overlap with the definition of disks, as smooth disks also exist~\citep{2017MNRAS.464.4420S}. Notably, the larger COSMOS-Web survey area allows us to probe higher redshifts than~\cite{2024arXiv240906100G}, revealing that the abundance of stellar bars is consistent with zero at $z \sim 4-5$. Consistent with high redshift stellar bars~\citealp{2024arXiv240401918A}. 

Ultimately, our primary focus is on the relative evolution of bars. From this perspective, the overall normalization is less critical as long as the measurements are internally consistent.

\section{Discussion}
\label{sec:disc}

\subsection{The Emergence of the Hubble Sequence}

\begin{figure*}[t]
  \centering
    \includegraphics[width=1\columnwidth]{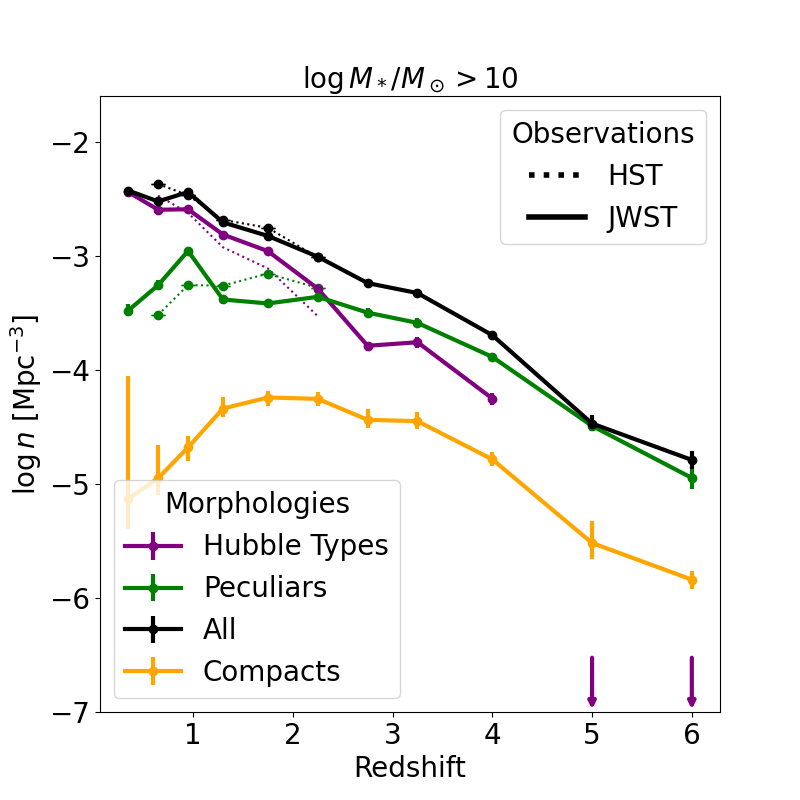}
    \includegraphics[width=1\columnwidth]{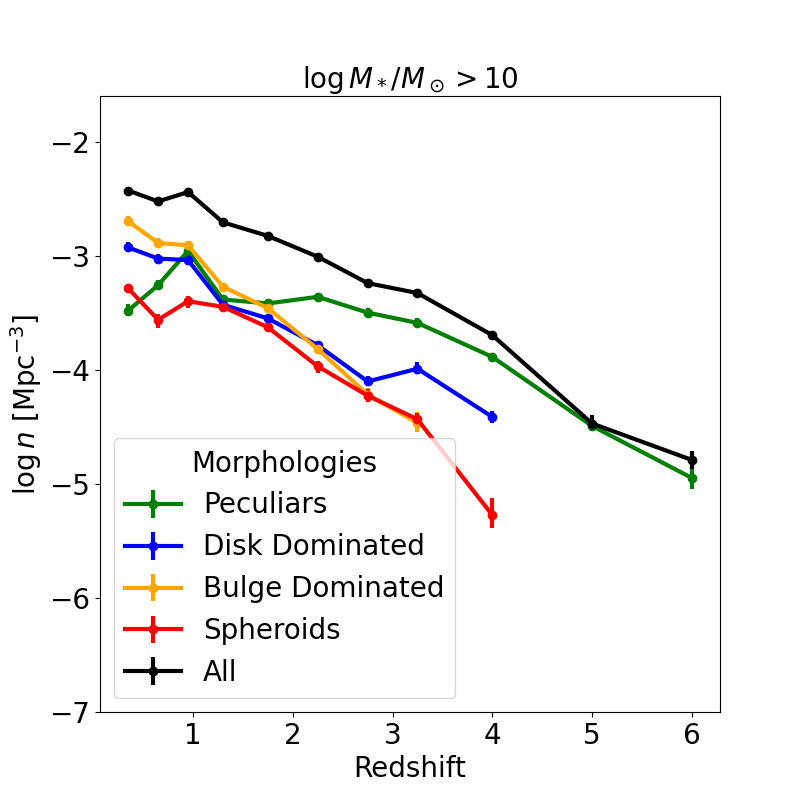}
 \caption{Evolution of the number densities of different morphological types with redshift. Left panel: Hubble Types (purple), Peculiars (green), compacts (orange) and all galaxies (black). The solid lines show JWST measurements from this work and the dotted lines indicate published HST measurements~\citep{2016MNRAS.462.4495H}. Right panel: Same as left panel but Hubble Types are split in disk-dominated (blue), bulge-dominated (orange) and spheroids (red).  }
  \label{fig:ndensity_peculiar_hubble}
\end{figure*}

An intriguing question is when the massive bulges and disks that constitute the present-day Hubble Sequence formed. Figure~\ref{fig:ndensity_peculiar_hubble} shows the evolution of the number densities of massive galaxies ($\log M_*/M_\odot > 10$) as a function of morphology. Number densities were obtained by integrating the best-fit model of the SMF (see Appendix~\ref{app:smfs} and Section~\ref{sec:methods}). We sample the posterior distribution of the SMF parameters 100 times and use the mean and scatter to estimate the number densities and associated uncertainties. To better probe the emergence of the Hubble Sequence, we group all Hubble types (spheroids, disk-dominated, and bulge-dominated systems) in the left panel of Figure~\ref{fig:ndensity_peculiar_hubble} and compare their number density evolution to that of peculiar and compact galaxies.

At $z > 4$, the abundance of Hubble types is very low. The galaxy population is dominated by peculiar galaxies (comprising $\sim70-80\%$) and compact/unresolved galaxies, which account for the remaining $\sim30\%$. This is consistent with previous findings of very compact galaxies at these early epochs, rapidly forming stars and possibly growing bulges in situ along with supermassive black holes at their centers (e.g.,~\citealp{2024arXiv240906772T,2024ApJ...963..129M,2024arXiv240403576K,2024ApJ...968....4P,2024arXiv241100279T,2024arXiv240918605D}). These results confirm that during cosmic dawn, a significant fraction of star formation activity occurs in extremely compact and dense regions, which might result in enhanced star formation efficiency, as discussed in several theoretical works (e.g.,~\citealp{2023MNRAS.523.3201D}).

At $z < 4$, we observe a rapid increase in Hubble sequence types, which become more abundant than peculiar galaxies at around $z \sim 3$—approximately 11 Gyr ago. The figure also includes pre-JWST HST-based measurements from~\cite{2016MNRAS.462.4495H}, showing reasonable agreement between HST and JWST results. However, as noted in several prior studies (e.g.,~\citealp{2022ApJ...938L...2F}), the abundance of regular Hubble-type morphologies is generally higher in JWST data than in HST data, particularly at $z > 1$. This discrepancy manifests in two key ways: (1) the transition time at which regular morphological types become more abundant than peculiar galaxies occurs $\sim1$ Gyr earlier in JWST data, and (2) the overall abundance of Hubble types at $z < 3$ is slightly higher in JWST measurements. As demonstrated in~\cite{2024A&A...685A..48H}, these differences are primarily driven by wavelength effects. The increase in Hubble types is accompanied by a flattening and subsequent decrease in the number densities of compact galaxies at $z < 2$, suggesting that some compact galaxies may evolve into the cores of massive Hubble-type galaxies (e.g.,~\citealp{2024arXiv241100279T}). It is important to highlight that the difference between peculiar and disk galaxies, especially at high redshifts, might be subtle. We discuss disk settling using stellar bars as a proxy in Section~\ref{sec:disksettling}. 

In the right panel of Figure~\ref{fig:ndensity_peculiar_hubble}, we examine in more detail which specific types—disk-dominated or bulge-dominated—drive the rapid increase in Hubble-type galaxies. This analysis reveals that the growth is driven by an increase in both populations, reflecting the simultaneous rapid buildup of bulges and disks from $z \sim 4-5$. Disks appear to grow slightly faster at higher redshifts, although their abundance is rapidly matched by bulges. In the following sections, we analyze in more detail the evolution of these two populations and their connections to quenching and disk settling.

\subsection{Disk Settling }
\label{sec:disksettling}
We now turn to analyzing the emergence of disk galaxies. The right panel of Figure~\ref{fig:ndensity_peculiar_hubble} shows a rapid increase in the abundance of disk dominated galaxies between $z \sim 4-5$, while the number density of peculiar galaxies tend to flatten out at $z \sim 3$. By $z \sim 3$, the abundance of disks increases by a factor of $\sim10$. This confirms two significant findings. First, galaxies with regular disk morphologies emerge as early as $z \sim 4$. Second, these galaxies become relatively common below $z \sim 3$, with their abundance only a factor of $\sim10$ lower than in the local universe (e.g.,~\citealp{2023ApJ...946L..13F}). However, it is important to note that the relative abundance of disks and peculiar galaxies is strongly mass-dependent, as reported in~\cite{2024A&A...685A..48H} and also reflected in Figure~\ref{fig:CEERS_CWEB}. At the very high stellar mass end ($\log M_*/M_\odot > 10.8$), symmetric disks dominate in numbers over peculiars even at $z > 3-6$. This is consistent with theoretical predictions of a critical mass for disk formation at all epochs~\citep{2020MNRAS.493.4126D}.

A critical caveat is that even though these galaxies are morphologically classified as disks, numerous studies have shown that it is challenging to definitively establish their nature without kinematic measurements (e.g.,~\citealp{2024ApJ...961...51V,2024ApJ...963...54P}). The morphological differences are sometimes subtle and hence SNR and spatial resolution can also affect the separation between disks and peculiars. In fact, the structural properties derived from Sersic fitting for peculiar and disk galaxies appear very similar (see Appendix~\ref{app:morphs}). Therefore, relying solely on morphology makes it difficult to accurately quantify the abundance of stellar rotating disks.

To provide additional constraints, we use the presence of stellar bars, as derived in Section~\ref{sec:bars}, as an additional proxy for the presence of stellar disks. Stellar bars are thought to reflect the dynamical status of stars in a galaxy and provide hints about the gas content. Early simulations of isolated galaxies indeed showed that high gas fraction tend to suppress the formation of bars (e.g.,\citealp{1986MNRAS.221..213A,2003MNRAS.341.1179A,2010ApJ...719.1470V}) pointing towards the need of dynamically cold, low turbulence, stellar disks for the formation of bars. Zoom-in cosmological simulations also tend to confirm that bar formation is suppressed when the disk is too dynamically hot (e.g.,~\citealp{2012ApJ...757...60K,2022MNRAS.512..160R}) suggesting that bars can be used as a proxy for cold disk formation. More recent cosmological simulations tend to suggest however that the ratio between dark and baryonic matter plays a major if not a dominant role in regulating bar formation (e.g.,~\citealp{2018MNRAS.477.1451F,2023ApJ...947...80B,2024arXiv240609453F}) being able to form bars even in gas-rich galaxies.~\cite{2024ApJ...968...86B} showed that if the fraction of baryonic mass in the disk is more than $\sim50\%$, a bar can be formed independently of the gas fraction.  In any case, the presence of a bar is a good tracer for the presence of a disk and the baryon dominance of central regions in galaxies, which complements the pure morphological classification.

We make the strong assumption that, at $z<0.5$, all massive morphologically classified galaxies disks are true stellar disks and use the fraction of bars in this population as a reference for bar frequency in a population of disks. Using the bar fraction evolution reported in Figure~\ref{fig:bar_frac}, we then compute the relative change in the bar fraction in massive galaxies as a function of redshift compared to the frequency of bars at $z < 0.5$:

\[
\Delta_{fb}(z) = \frac{f_{\text{bar}}(z) - f_{\text{bar}}(z < 0.5)}{f_{\text{bar}}(z < 0.5)},
\]

where $f_{\text{bar}}(z)$ denotes the bar fraction at redshift $z$, as shown in Figure~\ref{fig:bar_frac}. We employ normalization using disks, as is common in the literature, but since we rely on relative variations, the results are not significantly affected by the normalization choice.

\begin{figure}[t]
  \centering
    \includegraphics[width=\columnwidth]{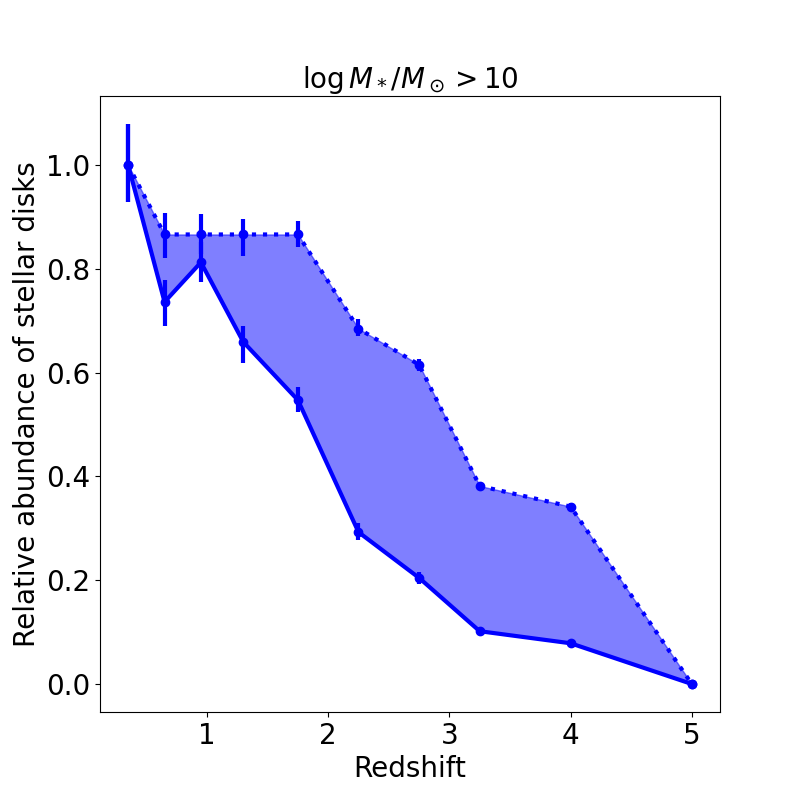}
 \caption{Evolution of the relative abundance of stellar disks based on the frequency of stellar bars (see text for details). }
  \label{fig:cold_disks}
\end{figure}

Figure~\ref{fig:cold_disks} translates this relative variation into a first-order estimate of the evolution of the fraction of baryon-dominated rotating disks based on the bar fraction. Up to $z \sim 2$, we find that the abundance of disks is as high as $\sim80\%$, but it rapidly decreases at higher redshifts. At $z \sim 3$, these mature disks represent a minority of the late-type population ($\sim20\%$), and by $z \sim 5$, our results suggest they constitute a negligible fraction of the massive galaxy population. Interestingly, this roughly corresponds to the highest redshift disk galaxies found by ALMA~\citep{2023ApJ...958..132N}. The abundance of stellar disks based on bars also roughly matches the evolution of the relative abundance between disk-dominated and peculiars reported in the right panel of Figure~\ref{fig:ndensity_peculiar_hubble}. The measurement of
Figure~\ref{fig:cold_disks} should be seen as a noisy first-order estimate. As previously mentioned, it is based on important assumptions. It also implicitly assumes that timescales for bar formation and survival are similar at different epochs, which is likely not true~\citep{2024arXiv240609453F}. Nevertheless, based on the results from the simulations reported by~\cite{2024ApJ...968...86B}, our results can help establish some constraints on the formation time of the first mature disks. The aforementioned work estimates that in order to observe stellar bars at $z\sim4-5$, galaxies should be baryon-dominated within two effective radii which pushes the formation of the first disks at least to $z\sim7$.~\cite{2020MNRAS.493.4126D} also predict that disks should be observed above a critical mass at all epochs.

\subsection{Bulge Growth and Quenching}

\begin{figure*}[t]
  \centering
    \includegraphics[width=1\columnwidth]{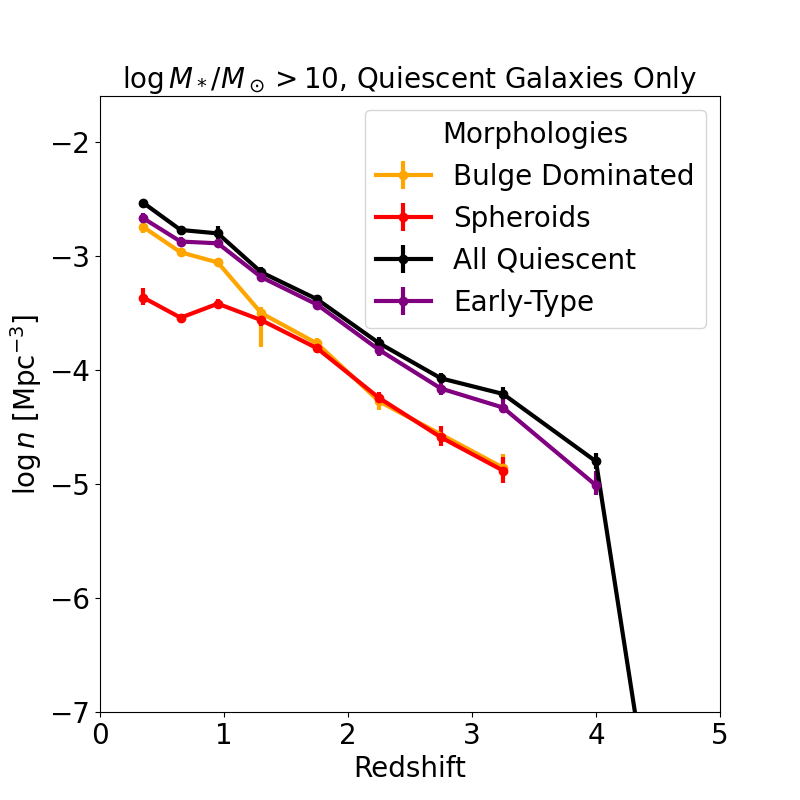}
    \includegraphics[width=1\columnwidth]{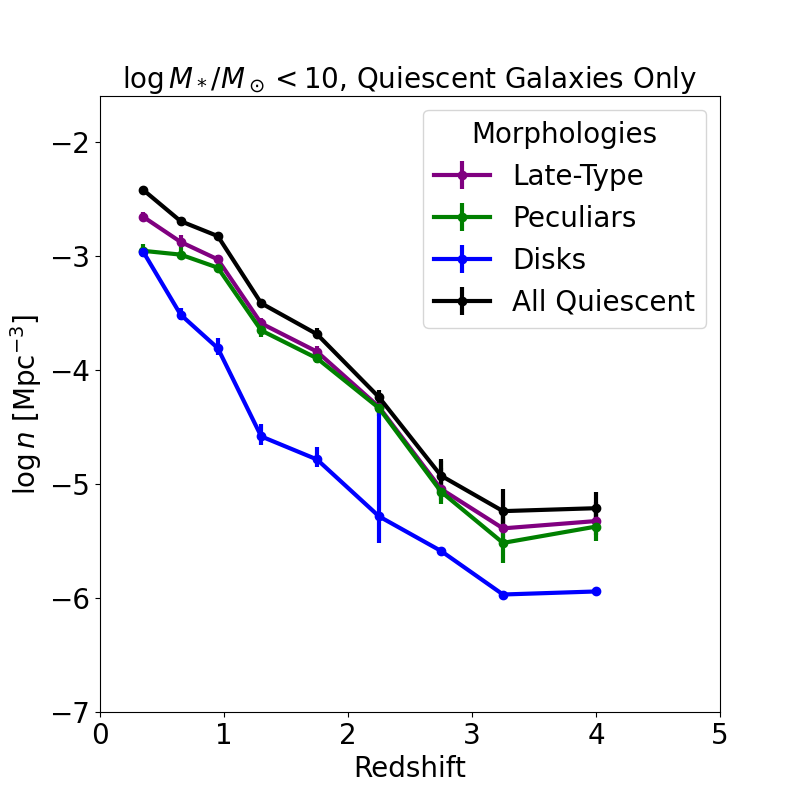}
 \caption{Evolution of the number densities of quiescent galaxies with redshift. Left panel: Massive galaxies ($M_*/M_\odot>10^{10}$). Black: all quiescent, purple: Early-Type, red: spheroids, orange: bulge dominated. Right panel: Low mass galaxies ($M_*/M_\odot<10^{10}$). Black: all, green: peculiars, purple: late-type, blue: disk dominated.}
  \label{fig:ndensity_Q}
\end{figure*}

Figure~\ref{fig:ndensity_Q} shows the evolution of the number densities of quenched galaxies, divided by morphology.

We observe a clear increase in the abundance of massive quiescent galaxies below $z \sim 4-5$ (see, e.g.,~\citealp{2023MNRAS.520.3974C, 2023ApJ...947...20V,2023AJ....165..248G, 2024ApJ...970...68L}), which is almost perfectly mirrored by the increase in bulge-dominated systems. This suggests, as previously highlighted in studies with smaller samples, that the distinct stellar structures of massive quenched and star-forming galaxies have been in place since the emergence of the first quenched galaxies. It strongly implies that morphological transformations are closely tied to the quenching of massive galaxies. Given the young age of the universe at $z \sim 4$, it is difficult to explain this correlation without invoking a causal connection, as proposed in some models (e.g.,~\citealp{2016ApJ...833....1L}). It also seems unlikely that bulge growth happens significantly earlier than quenching since this would imply a larger population of star-forming bulge dominated galaxies that we do not observe. This finding is consistent with the results of~\cite{2024ApJ...961..163B} and may indirectly point to AGN-driven quenching at the high-mass end, as black holes grow more efficiently in dense central cores (see also~\citealp{2020ApJ...897..102C}). Recent work by~\cite{2024arXiv240716578T} also finds evidence of rapid in-situ bulge formation in qualitative agreement with our results (see also~\citealp{2024A&A...688A..53L}). An indirect consequence of this connection is that central density could serve as a proxy for permanent quenching, as opposed to a temporary reduction in star formation rates due to more temporally dispersed burstiness at very high redshifts (e.g.,~\citealp{2024arXiv240903829W,2024ApJ...971...47F}). 

Interestingly, this correlation between quenching and morphology disappears at the low-mass end. The right panel of Figure~\ref{fig:ndensity_Q} shows the evolution of the number densities of quenched low-mass galaxies ($8.5 < \log M_*/M_\odot < 10$). The majority of these quenched galaxies exhibit peculiar morphologies, showing no significant differences compared to the star-forming population of similar stellar mass. This suggests that a different quenching mechanism operates in this population, one that does not require an increase in central mass density. A likely possibility is that quenching in these low-mass galaxies is driven by environmental effects (e.g.,~\citealp{2010ApJ...721..193P}), but this hypothesis requires further investigation through an analysis of the environments of these low-mass quenched galaxies, which is beyond the scope of this work. Another possible explanation could be related to merger timescales. If mergers are responsible for quenching at least some of these low-mass galaxies, relaxation times might be longer than for massive objects, which would increase the likelihood of finding peculiar morphologies.

\section{Summary and Conclusions}

In this work, we leveraged the wide area coverage of the COSMOS-Web survey to perform a comprehensive analysis of galaxy morphologies up to redshift \( z \sim 7 \). Utilizing deep learning techniques, we measured both global (spheroids, disk-dominated, bulge-dominated, peculiars) and internal morphologies (stellar bars) for approximately 400,000 galaxies. This represents a two-order-of-magnitude increase in sample size over previous studies. We constructed reference Stellar Mass Functions (SMFs) for different morphological types between \( z \sim 0.2 \) and \( z \sim 7 \), providing best-fit parameters to aid galaxy formation models. All catalogs and data products generated in this work are made publicly available to the scientific community.\\

Our main results are as follows.

\begin{itemize}
    \item \textbf{Clumpy and compact star formation at cosmic dawn:} At \( z > 4.5 \), the massive galaxy population (\( M_*/M_\odot > 10^{10} \)) is predominantly composed of galaxies with disturbed morphologies (\( \sim70\% \)) and very compact objects (\( \sim30\% \)) with effective radii smaller than \( \sim500\,\text{pc} \). This suggests that a significant fraction of star formation at cosmic dawn occurs in very dense regions.

    \item \textbf{Emergence of the Hubble Sequence:} Galaxies with Hubble-type morphologies, including bulge- and disk-dominated galaxies, rapidly emerge starting below \( z \sim 4 \). By \( z \sim 3 \), they dominate the morphological diversity of massive galaxies, indicating an early establishment of the Hubble sequence.

 \item \textbf{Early formation of stellar disks:} Using the presence of stellar bars as a proxy, we estimate that rotating stellar disks have been common (\( >50\% \)) among the star-forming galaxy population since cosmic noon (\( z \sim 2\text{--}2.5 \)). Massive stellar disks are observed as early as $z\sim4-5$, which would suggest a formation at even earlier epochs ($z\sim7$) based on simulations. 

    \item \textbf{Bulge growth and quenching connected from cosmic dawn:} Below \( z \sim 4 \) onward, massive quenched galaxies are predominantly bulge-dominated. This implies that morphological transformations are closely associated with quenching mechanisms at the high-mass end of the galaxy population.

    \item \textbf{Different Quenching Pathways for Low-Mass Galaxies:} Low-mass (\( M_*/M_\odot < 10^{10} \)) quenched galaxies are typically disk-dominated, pointing to different quenching processes operating at the low-mass end compared to the high-mass end of the stellar mass spectrum.

\end{itemize}

In future work, we plan to confront the results of this work with state-of-the-art forward modeled simulations. Investigating in more detail the class of peculiar galaxies and the morphology-quenching relation at low mass are additional promising research avenues.

\bibliographystyle{aa}
\bibliography{biblio.bib}

\begin{acknowledgements} 
MHC acknowledges support from the State Research Agency (AEIMCINN) of the Spanish Ministry of Science and Innovation under the grants “Galaxy Evolution with Artificial Intelligence" with reference
PGC2018-100852-A-I00 and "BASALT" with reference PID2021-126838NB-I00. This work was made possible by utilising the CANDIDE cluster at the Institut d’Astrophysique de Paris. The cluster was funded through grants from the PNCG, CNES, DIM-ACAV, the Euclid Consortium, and the Danish National Research Foundation Cosmic Dawn Center (DNRF140). It is maintained by Stephane Rouberol. The French contingent of the COSMOS team is partly supported by the Centre National d’Etudes Spatiales (CNES). We acknowledge the funding of the French Agence Nationale de la Recherche for the project iMAGE (grant ANR-22-CE31-0007). 
\end{acknowledgements}

\appendix

\section{Reliability of morphological classification}
\label{app:morphs}

\subsection{Cutouts}

\begin{figure*}[t]

    \includegraphics[width=\columnwidth]{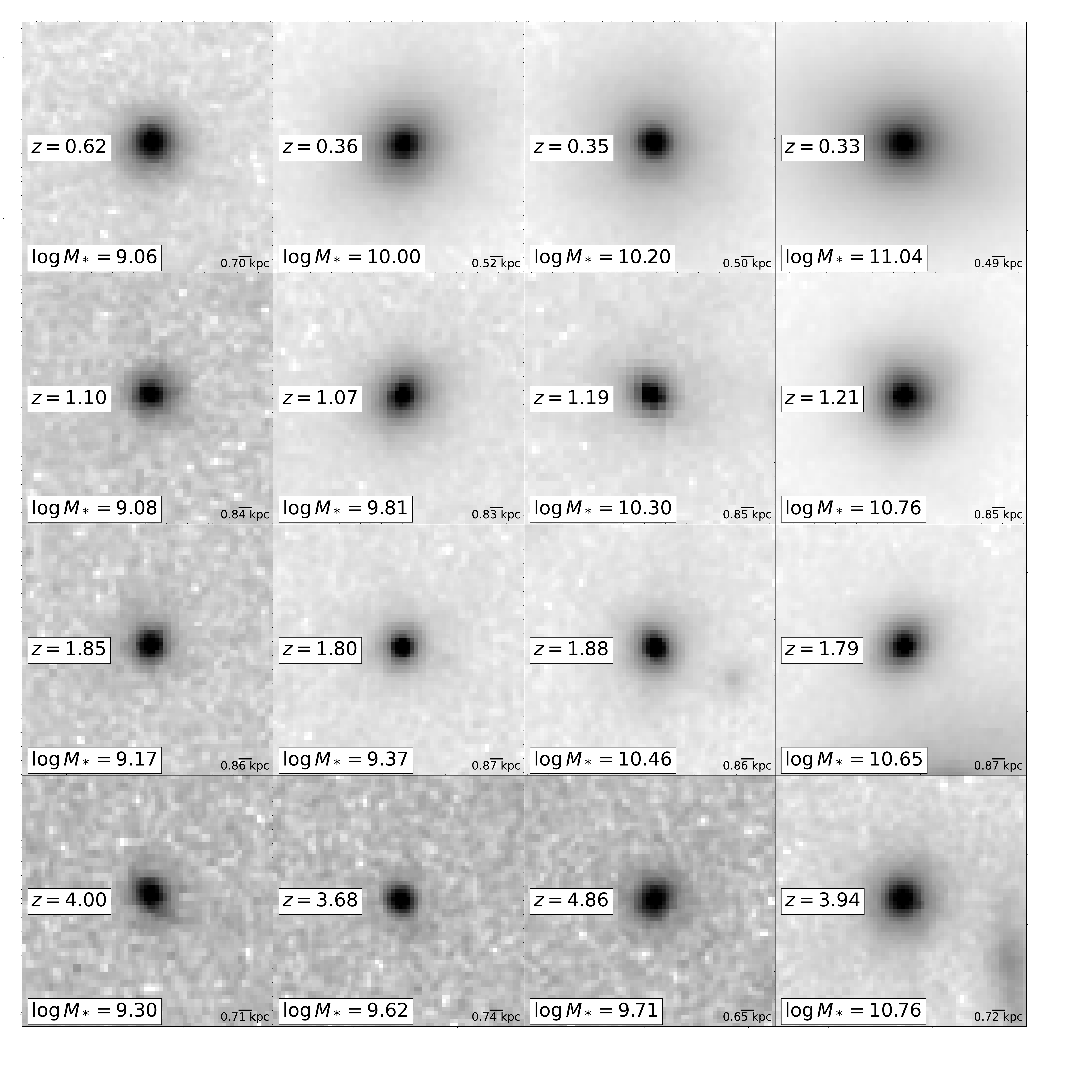}
    \includegraphics[width=\columnwidth]{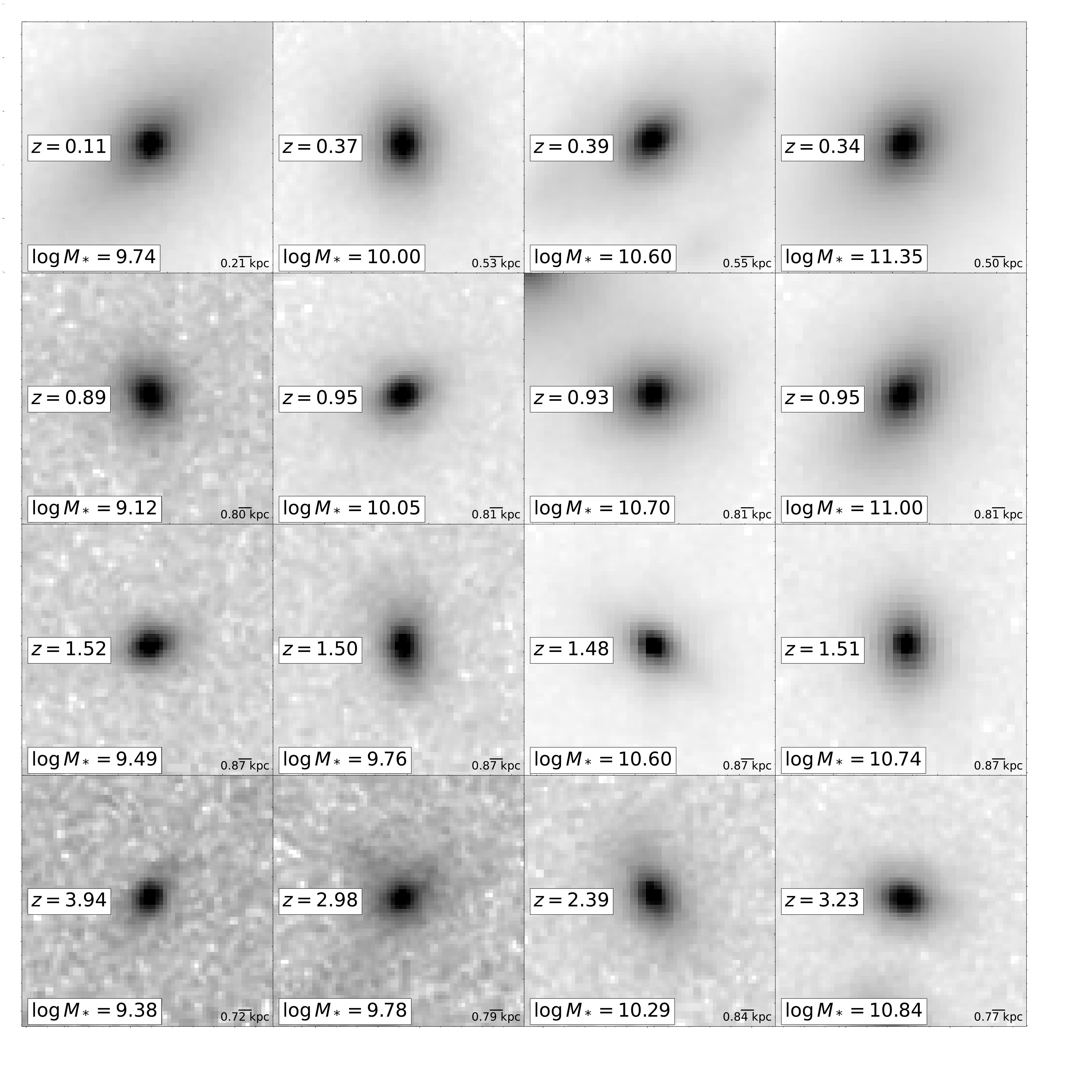}
    \includegraphics[width=\columnwidth]{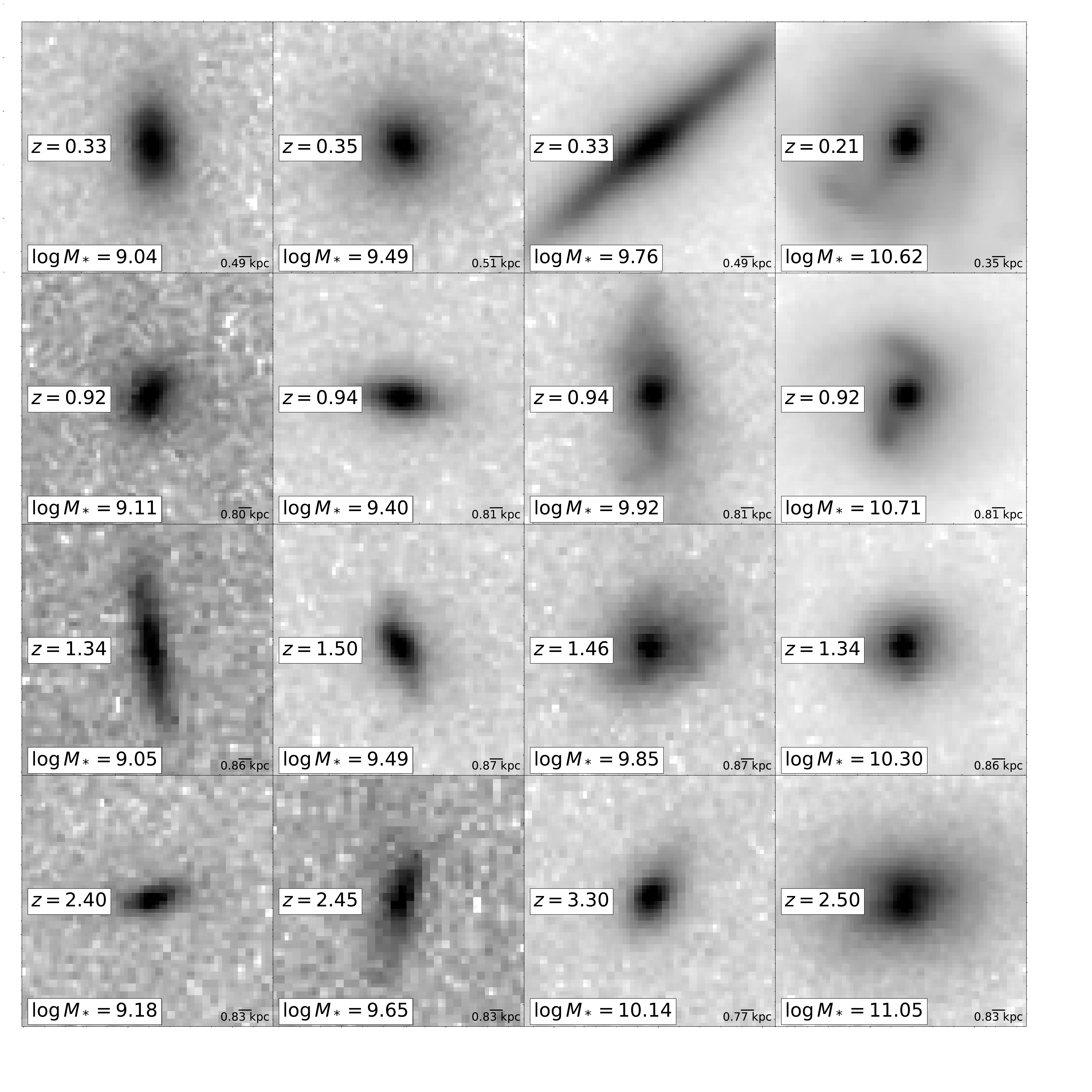}
    \includegraphics[width=\columnwidth]{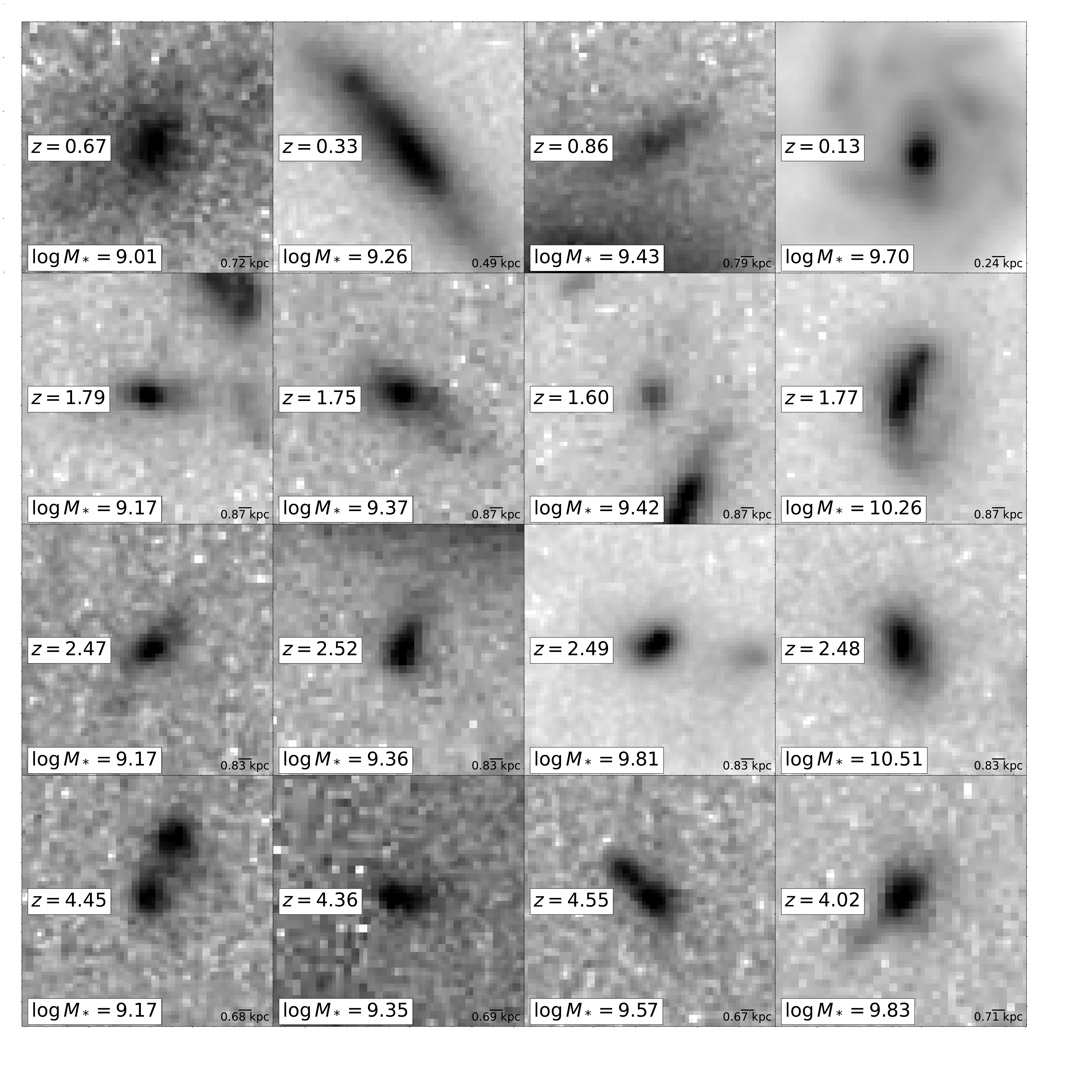}
    
 \caption{Example of cutouts from the COSMOS-Web survey of different morphological types. Top left: spheroids, top right: bulge dominated, bottom left: disks, bottom right: peculiars. The photometric redshift and stellar mass of each galaxy is indicated in the stamp. We show $F150W$, $F277W$ and $F444W$ images for galaxies at $z<1$, $1<z<3$, $z>3$ respectively. }
  \label{fig:cutouts_morph}
\end{figure*}

Figure~\ref{fig:cutouts_morph} shows some randomly selected example cutouts of galaxies with different morphological classifications, as detailed in the main text. We can clearly see that galaxies in different classes present different distinct features as expected. Spheroids are round and concentrated. Bulge-dominated galaxies present a clear central bulge but tend to be more elongated. A low surface brightness disk is sometimes detected. Disk dominated galaxies are predominantly less concentrated and appear with different orientations. Peculiar galaxies are a mix of several things, like merging, clumpy, or asymmetric systems. 

\subsection{Comparison with CEERS}

\begin{figure*}[t]

    \includegraphics[width=\textwidth]{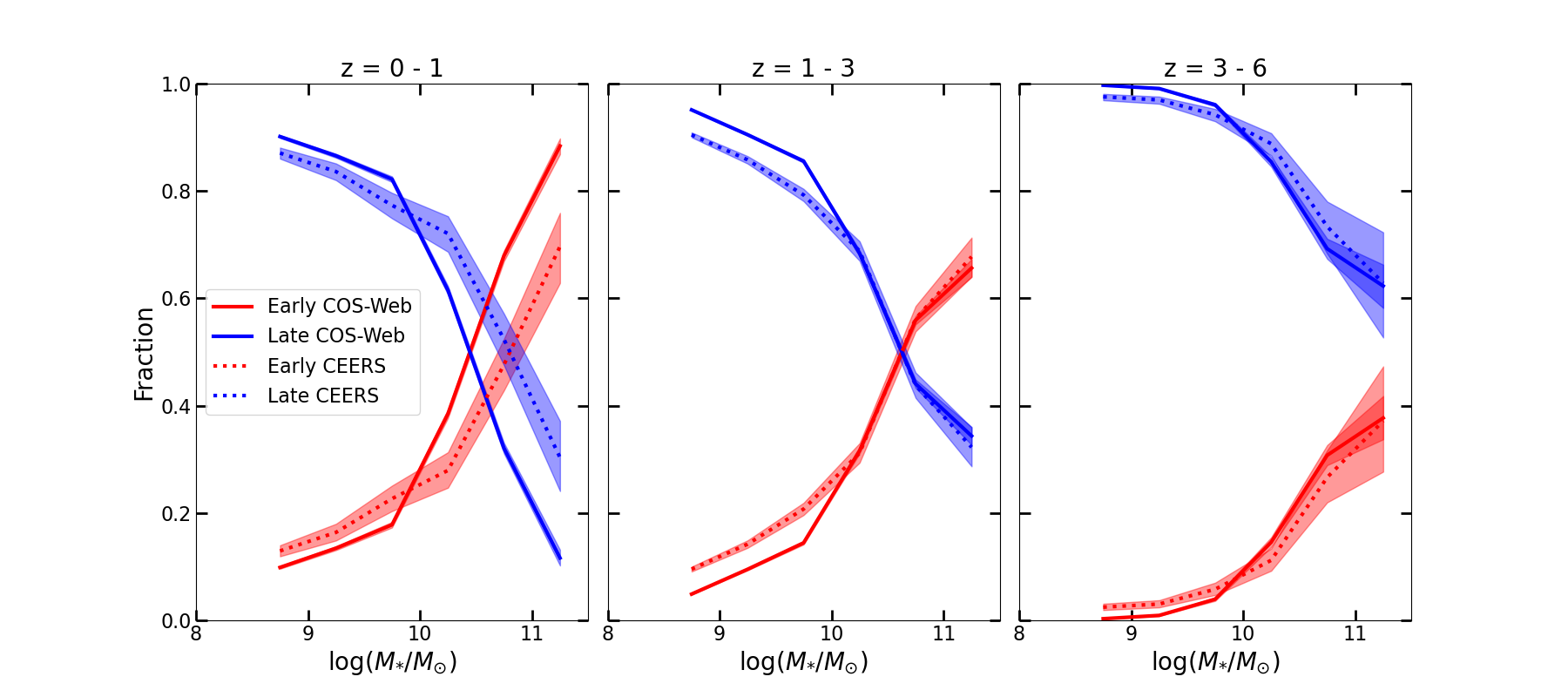}
    \includegraphics[width=\textwidth]{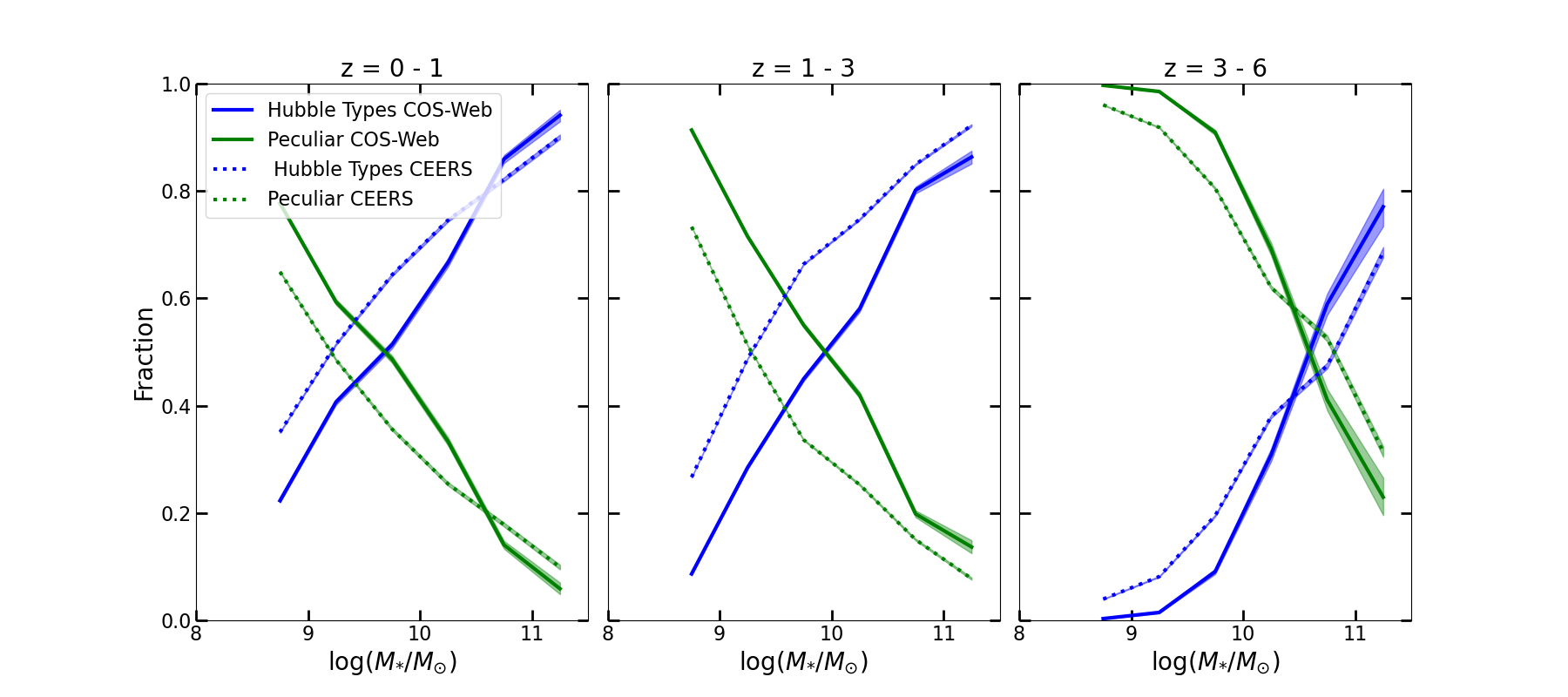}

 \caption{Comparison or morphological fractions of early vs. late (top panel) and Hubble Types vs. peculiar (bottom panel) between CEERS~\citealp{2024A&A...685A..48H} (dotted lines) and COSMOS-Web (solid lines). Optical rest-frame morphologies are plotted. At $z<1$, we use $F150W$, at $1<z<3$, $F356W$ for CEERS $F277W$ for COSMOS-Web and at $z>3$, $F444W$.   }
  \label{fig:CEERS_CWEB}
\end{figure*}

Figure~\ref{fig:CEERS_CWEB} comparison fraction published in the CEERS survey~\citep{2024A&A...685A..48H}
with those obtained in COSMOS-Web in this work. We compare both the fraction of early-type vs. late-type galaxies and the ones of peculiars vs. Hubble types. The figures show very similar trends which support the robustness of the morphological classifications presented in this work despite the fact that the COSMOS-Web data are shallower. The observed differences might be due to different stellar mass and photometric redshift estimators as well as the fact that slightly different filters - $F277W$ vs. $F356W$ for redshift bin $(1-3)$ - are used.

\subsection{Comparison with Sersic Fits}

\begin{figure*}[t]

    \includegraphics[width=\textwidth]{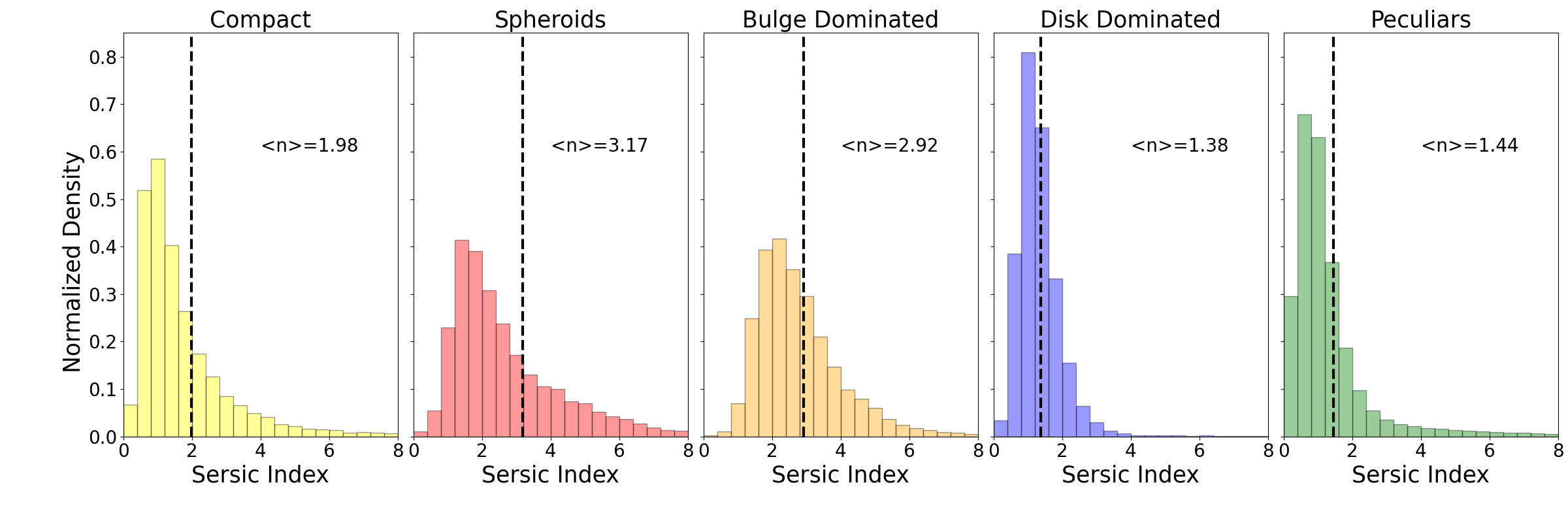}
    \includegraphics[width=\textwidth]{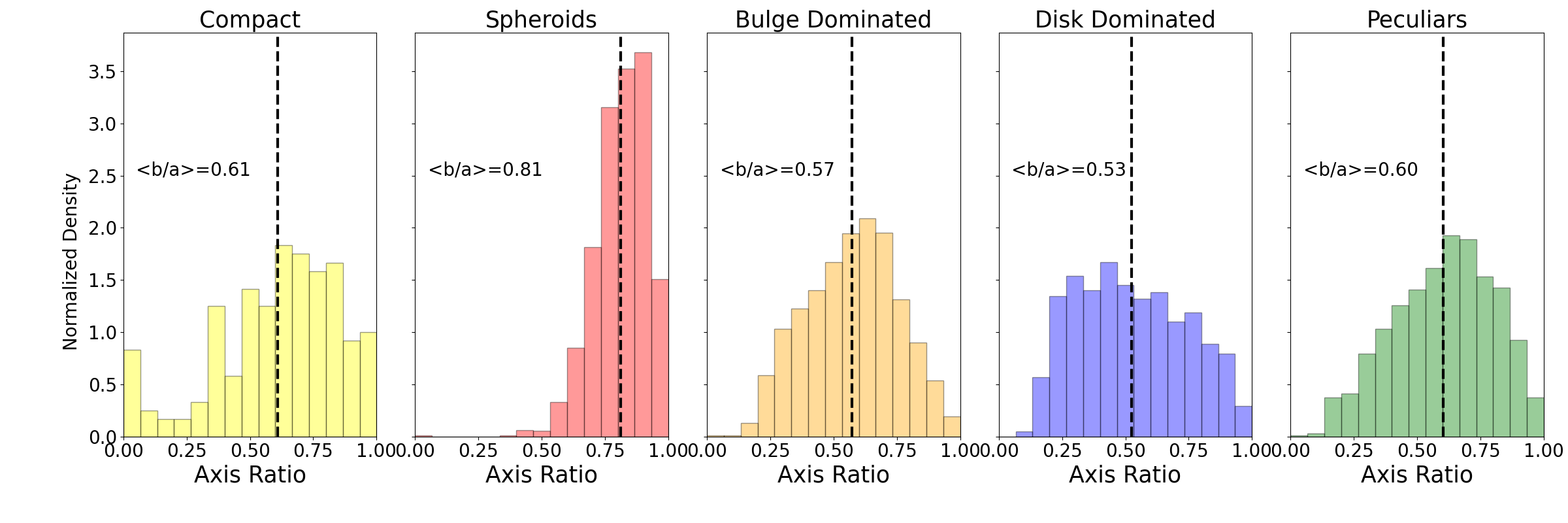}

 \caption{Sersic index (top row)  and axis ratios distributions (bottom row) for the different morphological types defined in this work. The dashed vertical lines indicate the mean of the distributions whose values are also written in each panel.  }
  \label{fig:Sersic}
\end{figure*}

We plot in figure~\ref{fig:Sersic} the Sersic index and axis ratio distributions for the different morphological types defined in this work. The

\section{Best fit parameters}
\label{app:smfs}
\subsection{All galaxies}

We report in tables~\ref{tab:early} to~\ref{tab:disk}, the best-fit parameters to the SMFs of early and late type galaxies (Tables~\ref{tab:early} and ~\ref{tab:late}), Hubble Types (Table~\ref{tab:reg}), spheroids (Table~\ref{tab:sph}) , bulge dominated (Table~\ref{tab:db}), disk dominated (Table~\ref{tab:disk}) and peculiars (Table~\ref{tab:irr}). 

\begin{table*}[ht]
\centering
\begin{tabular}{cccccccccc}
\hline
Redshift & Model & $\alpha$ & $\alpha_1$ & $\alpha_2$ & $logM\_star$ & $\log\phi_1$ & $\log\phi_2$ & $\log\phi_*$ & AICc \\
\hline
$0.2 \leq z < 0.5$ & Double & -- & $-1.72_{-0.51}^{+0.50}$ & $-0.41_{-0.28}^{+0.28}$ & $10.72_{-0.13}^{+0.10}$ & $-4.48_{-1.15}^{+0.93}$ & $-2.55_{-0.08}^{+0.10}$ & -- & $25.70$ \\[.3cm]
$0.5 \leq z < 0.8$ & Double & -- & $-1.33_{-0.35}^{+0.19}$ & $-0.18_{-0.31}^{+0.42}$ & $10.73_{-0.10}^{+0.09}$ & $-3.79_{-0.66}^{+0.38}$ & $-2.81_{-0.10}^{+0.07}$ & -- & $26.60$ \\[.3cm]
$0.8 \leq z < 1.1$ & DPL & -- & $-3.77_{-0.18}^{+2.83}$ & $-0.96_{-2.89}^{+0.05}$ & $11.01_{-0.04}^{+0.04}$ & -- & -- & $-2.77_{-0.05}^{+0.05}$ & $19.75$ \\[.3cm]
$1.1 \leq z < 1.5$ & Double & -- & $-1.10_{-0.16}^{+0.07}$ & $0.21_{-0.33}^{+0.34}$ & $10.60_{-0.07}^{+0.07}$ & $-3.75_{-0.30}^{+0.13}$ & $-3.01_{-0.06}^{+0.06}$ & -- & $25.39$ \\[.3cm]
$1.5 \leq z < 2$ & Double & -- & $-1.10_{-0.26}^{+0.08}$ & $-0.10_{-0.45}^{+0.43}$ & $10.70_{-0.10}^{+0.10}$ & $-3.78_{-0.44}^{+0.16}$ & $-3.27_{-0.09}^{+0.09}$ & -- & $23.99$ \\[.3cm]
$2 \leq z < 2.5$ & DPL & -- & $-3.82_{-0.13}^{+0.54}$ & $-1.02_{-0.10}^{+0.05}$ & $11.02_{-0.05}^{+0.05}$ & -- & -- & $-3.52_{-0.07}^{+0.06}$ & $22.05$ \\[.3cm]
$2.5 \leq z < 3$ & Double & -- & $-1.37_{-0.41}^{+1.63}$ & $-0.54_{-1.71}^{+0.62}$ & $10.71_{-0.31}^{+0.14}$ & $-4.58_{-1.02}^{+2.21}$ & $-3.78_{-0.15}^{+1.64}$ & -- & $20.44$ \\[.3cm]
$3 \leq z < 3.5$ & Double & -- & $-2.21_{-0.18}^{+0.25}$ & $-0.19_{-0.20}^{+0.25}$ & $10.73_{-0.13}^{+0.11}$ & $-6.48_{-0.36}^{+0.55}$ & $-3.92_{-0.09}^{+0.07}$ & -- & $26.49$ \\[.3cm]
$3.5 \leq z < 4.5$ & Double & -- & $-2.20_{-0.23}^{+0.33}$ & $-0.34_{-0.39}^{+0.45}$ & $10.56_{-0.18}^{+0.16}$ & $-6.12_{-0.54}^{+0.70}$ & $-4.37_{-0.12}^{+0.11}$ & -- & $21.49$ \\[.3cm]
$4.5 \leq z < 5.5$ & Single & $-2.13_{-0.08}^{+0.08}$ & -- & -- & $11.03_{-0.52}^{+0.33}$ & -- & -- & $-6.50_{-0.36}^{+0.65}$ & $21.47$ \\[.3cm]
$5.5 \leq z < 6.5$ & Single & $-0.10_{-0.57}^{+0.61}$ & -- & -- & $8.80_{-0.13}^{+0.21}$ & -- & -- & $-3.78_{-0.17}^{+0.07}$ & $21.87$ \\[.3cm]
\hline
\end{tabular}
\caption{Fit results for Early-Type galaxies}
\label{tab:early}
\end{table*}

\begin{table*}[ht]
\centering
\begin{tabular}{cccccccccc}
\hline
Redshift & Model & $\alpha$ & $\alpha_1$ & $\alpha_2$ & $logM\_star$ & $\log\phi_1$ & $\log\phi_2$ & $\log\phi_*$ & AICc \\
\hline
$0.2 \leq z < 0.5$ & Single & $-1.48_{-0.05}^{+0.06}$ & -- & -- & $10.74_{-0.09}^{+0.09}$ & -- & -- & $-3.11_{-0.11}^{+0.14}$ & $11.28$ \\[.3cm]
$0.5 \leq z < 0.8$ & Single & $-1.42_{-0.04}^{+0.04}$ & -- & -- & $10.65_{-0.06}^{+0.06}$ & -- & -- & $-3.02_{-0.09}^{+0.08}$ & $12.68$ \\[.3cm]
$0.8 \leq z < 1.1$ & Single & $-1.38_{-0.03}^{+0.04}$ & -- & -- & $10.72_{-0.05}^{+0.05}$ & -- & -- & $-2.92_{-0.07}^{+0.07}$ & $11.05$ \\[.3cm]
$1.1 \leq z < 1.5$ & Single & $-1.47_{-0.03}^{+0.03}$ & -- & -- & $10.85_{-0.06}^{+0.05}$ & -- & -- & $-3.38_{-0.07}^{+0.08}$ & $15.75$ \\[.3cm]
$1.5 \leq z < 2$ & Single & $-1.54_{-0.03}^{+0.03}$ & -- & -- & $10.91_{-0.05}^{+0.05}$ & -- & -- & $-3.55_{-0.07}^{+0.07}$ & $16.05$ \\[.3cm]
$2 \leq z < 2.5$ & Single & $-1.54_{-0.03}^{+0.03}$ & -- & -- & $10.96_{-0.07}^{+0.06}$ & -- & -- & $-3.70_{-0.08}^{+0.09}$ & $10.75$ \\[.3cm]
$2.5 \leq z < 3$ & Single & $-1.55_{-0.04}^{+0.05}$ & -- & -- & $10.91_{-0.12}^{+0.12}$ & -- & -- & $-3.87_{-0.14}^{+0.14}$ & $12.93$ \\[.3cm]
$3 \leq z < 3.5$ & Single & $-1.53_{-0.04}^{+0.04}$ & -- & -- & $11.03_{-0.09}^{+0.09}$ & -- & -- & $-4.00_{-0.11}^{+0.11}$ & $12.78$ \\[.3cm]
$3.5 \leq z < 4.5$ & Single & $-1.56_{-0.05}^{+0.04}$ & -- & -- & $10.73_{-0.11}^{+0.09}$ & -- & -- & $-4.08_{-0.12}^{+0.13}$ & $14.28$ \\[.3cm]
$4.5 \leq z < 5.5$ & DPL & -- & $-2.32_{-0.17}^{+0.19}$ & $-0.34_{-1.02}^{+1.04}$ & $8.95_{-0.13}^{+0.23}$ & -- & -- & $-2.56_{-0.15}^{+0.07}$ & $19.24$ \\[.3cm]
$5.5 \leq z < 6.5$ & DPL & -- & $-2.39_{-0.22}^{+2.42}$ & $-0.14_{-2.30}^{+0.63}$ & $9.08_{-0.12}^{+0.15}$ & -- & -- & $-3.07_{-0.08}^{+0.06}$ & $21.75$ \\[.3cm]
\hline
\end{tabular}
\caption{Fit results for Late Type}
\label{tab:late}
\end{table*}

\begin{table*}[ht]
\centering
\begin{tabular}{cccccccccc}
\hline
Redshift & Model & $\alpha$ & $\alpha_1$ & $\alpha_2$ & $logM\_star$ & $\log\phi_1$ & $\log\phi_2$ & $\log\phi_*$ & AICc \\
\hline
$0.2 \leq z < 0.5$ & Single & $-1.04_{-0.04}^{+0.04}$ & -- & -- & $10.90_{-0.05}^{+0.05}$ & -- & -- & $-2.67_{-0.07}^{+0.06}$ & $13.65$ \\[.3cm]
$0.5 \leq z < 0.8$ & Single & $-1.01_{-0.04}^{+0.04}$ & -- & -- & $10.92_{-0.05}^{+0.05}$ & -- & -- & $-2.81_{-0.06}^{+0.06}$ & $13.26$ \\[.3cm]
$0.8 \leq z < 1.1$ & Single & $-0.94_{-0.04}^{+0.03}$ & -- & -- & $10.86_{-0.04}^{+0.04}$ & -- & -- & $-2.75_{-0.05}^{+0.05}$ & $24.26$ \\[.3cm]
$1.1 \leq z < 1.5$ & Single & $-0.86_{-0.04}^{+0.03}$ & -- & -- & $10.82_{-0.04}^{+0.03}$ & -- & -- & $-2.93_{-0.05}^{+0.04}$ & $50.45$ \\[.3cm]
$1.5 \leq z < 2$ & Single & $-0.86_{-0.03}^{+0.03}$ & -- & -- & $10.84_{-0.04}^{+0.03}$ & -- & -- & $-3.08_{-0.04}^{+0.04}$ & $57.54$ \\[.3cm]
$2 \leq z < 2.5$ & Single & $-0.82_{-0.05}^{+0.04}$ & -- & -- & $10.82_{-0.06}^{+0.04}$ & -- & -- & $-3.39_{-0.06}^{+0.05}$ & $34.59$ \\[.3cm]
$2.5 \leq z < 3$ & DPL & -- & $-1.84_{-0.12}^{+0.10}$ & $0.15_{-0.31}^{+0.29}$ & $9.57_{-0.14}^{+0.17}$ & -- & -- & $-3.11_{-0.04}^{+0.04}$ & $42.70$ \\[.3cm]
$3 \leq z < 3.5$ & DPL & -- & $-1.75_{-0.23}^{+0.16}$ & $0.42_{-0.21}^{+0.25}$ & $10.10_{-0.19}^{+0.19}$ & -- & -- & $-3.45_{-0.08}^{+0.06}$ & $28.65$ \\[.3cm]
$3.5 \leq z < 4.5$ & Single & $0.29_{-0.21}^{+0.24}$ & -- & -- & $10.35_{-0.13}^{+0.12}$ & -- & -- & $-4.10_{-0.05}^{+0.05}$ & $19.25$ \\[.3cm]
$4.5 \leq z < 5.5$ & -- & -- & -- & -- & -- & -- & -- & -- & -- \\[.3cm]
$5.5 \leq z < 6.5$ & -- & -- & -- & -- & -- & -- & -- & -- & -- \\[.3cm]
\hline
\end{tabular}
\caption{Fit results for Hubble Types}
\label{tab:reg}
\end{table*}

\begin{table*}[ht]
\centering
\begin{tabular}{cccccccccc}
\hline
Redshift & Model & $\alpha$ & $\alpha_1$ & $\alpha_2$ & $logM^*$& $\log\phi_1$ & $\log\phi_2$ & $\log\phi_*$ & AICc \\
\hline
$0.2 \leq z < 0.5$ & Double & -- & $-2.08_{-0.36}^{+1.07}$ & $-0.42_{-1.00}^{+0.29}$ & $10.76_{-0.21}^{+0.14}$ & $-5.70_{-0.85}^{+3.68}$ & $-3.26_{-0.13}^{+0.18}$ & -- & $23.40$ \\[.3cm]
$0.5 \leq z < 0.8$ & Single & $-1.31_{-0.09}^{+0.08}$ & -- & -- & $11.41_{-0.23}^{+0.27}$ & -- & -- & $-4.20_{-0.24}^{+0.23}$ & $23.65$ \\[.3cm]
$0.8 \leq z < 1.1$ & Single & $-1.17_{-0.07}^{+0.06}$ & -- & -- & $11.06_{-0.10}^{+0.08}$ & -- & -- & $-3.76_{-0.11}^{+0.12}$ & $17.47$ \\[.3cm]
$1.1 \leq z < 1.5$ & Double & -- & $-1.50_{-0.34}^{+0.31}$ & $-0.19_{-0.36}^{+0.37}$ & $10.67_{-0.13}^{+0.10}$ & $-4.71_{-0.69}^{+0.66}$ & $-3.41_{-0.08}^{+0.09}$ & -- & $27.25$ \\[.3cm]
$1.5 \leq z < 2$ & Double & -- & $-1.45_{-0.24}^{+0.16}$ & $-0.02_{-0.40}^{+0.50}$ & $10.63_{-0.13}^{+0.12}$ & $-4.63_{-0.50}^{+0.32}$ & $-3.62_{-0.10}^{+0.07}$ & -- & $29.46$ \\[.3cm]
$2 \leq z < 2.5$ & DPL & -- & $-3.75_{-0.18}^{+0.35}$ & $-1.07_{-0.06}^{+0.06}$ & $11.01_{-0.06}^{+0.07}$ & -- & -- & $-4.00_{-0.10}^{+0.08}$ & $27.83$ \\[.3cm]
$2.5 \leq z < 3$ & Single & $-0.86_{-0.08}^{+0.08}$ & -- & -- & $10.86_{-0.11}^{+0.11}$ & -- & -- & $-4.35_{-0.10}^{+0.11}$ & $23.78$ \\[.3cm]
$3 \leq z < 3.5$ & Single & $-0.11_{-0.16}^{+0.20}$ & -- & -- & $10.62_{-0.11}^{+0.09}$ & -- & -- & $-4.37_{-0.10}^{+0.07}$ & $11.29$ \\[.3cm]
$3.5 \leq z < 4.5$ & Single & $-0.35_{-0.42}^{+0.58}$ & -- & -- & $10.75_{-0.30}^{+0.38}$ & -- & -- & $-5.28_{-0.33}^{+0.17}$ & $18.55$ \\[.3cm]
$4.5 \leq z < 5.5$ & -- & -- & -- & -- & -- & -- & -- & -- & -- \\[.3cm]
$5.5 \leq z < 6.5$ & -- & -- & -- & -- & -- & -- & -- & -- & -- \\[.3cm]
\hline
\end{tabular}
\caption{Fit results for Spheroids}
\label{tab:sph}
\end{table*}

\begin{table*}[ht]
\centering
\begin{tabular}{cccccccccc}
\hline
Redshift & Model & $\alpha$ & $\alpha_1$ & $\alpha_2$ & $logM^*$ & $\log\phi_1$ & $\log\phi_2$ & $\log\phi_*$ & AICc \\
\hline
$0.2 \leq z < 0.5$ & DPL & -- & $-3.61_{-0.28}^{+2.85}$ & $-0.79_{-2.97}^{+0.08}$ & $10.91_{-0.06}^{+0.05}$ & -- & -- & $-2.57_{-0.07}^{+0.06}$ & $21.34$ \\[.3cm]
$0.5 \leq z < 0.8$ & Single & $-0.45_{-0.06}^{+0.06}$ & -- & -- & $10.78_{-0.05}^{+0.04}$ & -- & -- & $-2.87_{-0.05}^{+0.05}$ & $23.79$ \\[.3cm]
$0.8 \leq z < 1.1$ & Single & $-0.21_{-0.05}^{+0.06}$ & -- & -- & $10.70_{-0.04}^{+0.04}$ & -- & -- & $-2.84_{-0.04}^{+0.04}$ & $44.84$ \\[.3cm]
$1.1 \leq z < 1.5$ & Single & $-0.46_{-0.05}^{+0.06}$ & -- & -- & $10.81_{-0.03}^{+0.03}$ & -- & -- & $-3.28_{-0.05}^{+0.04}$ & $42.18$ \\[.3cm]
$1.5 \leq z < 2$ & Single & $-0.40_{-0.04}^{+0.04}$ & -- & -- & $10.78_{-0.05}^{+0.04}$ & -- & -- & $-3.43_{-0.04}^{+0.04}$ & $66.01$ \\[.3cm]
$2 \leq z < 2.5$ & Single & $-0.42_{-0.07}^{+0.07}$ & -- & -- & $10.77_{-0.05}^{+0.05}$ & -- & -- & $-3.79_{-0.06}^{+0.06}$ & $42.00$ \\[.3cm]
$2.5 \leq z < 3$ & Single & $-0.71_{-0.10}^{+0.10}$ & -- & -- & $10.90_{-0.10}^{+0.10}$ & -- & -- & $-4.31_{-0.10}^{+0.10}$ & $19.03$ \\[.3cm]
$3 \leq z < 3.5$ & Single & $0.37_{-0.39}^{+0.44}$ & -- & -- & $10.78_{-0.16}^{+0.18}$ & -- & -- & $-4.38_{-0.10}^{+0.08}$ & $16.00$ \\[.3cm]
$3.5 \leq z < 4.5$ & -- & -- & -- & -- & -- & -- & -- & -- & -- \\[.3cm]
$4.5 \leq z < 5.5$ & -- & -- & -- & -- & -- & -- & -- & -- & -- \\[.3cm]
$5.5 \leq z < 6.5$ & -- & -- & -- & -- & -- & -- & -- & -- & -- \\[.3cm]
\hline
\end{tabular}
\caption{Fit results for Bulge dominated}
\label{tab:db}
\end{table*}

\begin{table*}[ht]
\centering
\begin{tabular}{cccccccccc}
\hline
Redshift & Model & $\alpha$ & $\alpha_1$ & $\alpha_2$ & $logM^*$ & $\log\phi_1$ & $\log\phi_2$ & $\log\phi_*$ & AICc \\
\hline
$0.2 \leq z < 0.5$ & Single & $-1.11_{-0.06}^{+0.06}$ & -- & -- & $10.49_{-0.09}^{+0.08}$ & -- & -- & $-2.87_{-0.10}^{+0.10}$ & $13.33$ \\[.3cm]
$0.5 \leq z < 0.8$ & Single & $-0.86_{-0.05}^{+0.06}$ & -- & -- & $10.41_{-0.06}^{+0.06}$ & -- & -- & $-2.86_{-0.06}^{+0.06}$ & $22.80$ \\[.3cm]
$0.8 \leq z < 1.1$ & Single & $-0.92_{-0.05}^{+0.06}$ & -- & -- & $10.52_{-0.07}^{+0.06}$ & -- & -- & $-2.97_{-0.06}^{+0.07}$ & $36.34$ \\[.3cm]
$1.1 \leq z < 1.5$ & DPL & -- & $-2.08_{-0.07}^{+0.06}$ & $0.30_{-0.19}^{+0.21}$ & $9.40_{-0.08}^{+0.09}$ & -- & -- & $-2.38_{-0.03}^{+0.03}$ & $59.09$ \\[.3cm]
$1.5 \leq z < 2$ & DPL & -- & $-2.18_{-0.07}^{+2.55}$ & $0.35_{-2.53}^{+0.20}$ & $9.50_{-0.07}^{+0.07}$ & -- & -- & $-2.50_{-0.03}^{+0.03}$ & $74.64$ \\[.3cm]
$2 \leq z < 2.5$ & DPL & -- & $-2.13_{-0.08}^{+0.08}$ & $0.49_{-0.17}^{+0.18}$ & $9.58_{-0.08}^{+0.08}$ & -- & -- & $-2.90_{-0.04}^{+0.03}$ & $43.12$ \\[.3cm]
$2.5 \leq z < 3$ & DPL & -- & $-1.80_{-0.11}^{+0.19}$ & $0.82_{-2.53}^{+0.35}$ & $9.43_{-0.09}^{+0.09}$ & -- & -- & $-3.35_{-0.06}^{+0.05}$ & $20.33$ \\[.3cm]
$3 \leq z < 3.5$ & Single & $-0.41_{-0.12}^{+0.12}$ & -- & -- & $10.77_{-0.13}^{+0.10}$ & -- & -- & $-3.94_{-0.08}^{+0.08}$ & $20.57$ \\[.3cm]
$3.5 \leq z < 4.5$ & Single & $0.26_{-0.21}^{+0.24}$ & -- & -- & $10.33_{-0.12}^{+0.10}$ & -- & -- & $-4.24_{-0.05}^{+0.05}$ & $17.32$ \\[.3cm]
$4.5 \leq z < 5.5$ & -- & -- & -- & -- & -- & -- & -- & -- & -- \\[.3cm]
$5.5 \leq z < 6.5$ & -- & -- & -- & -- & -- & -- & -- & -- & -- \\[.3cm]
\hline
\end{tabular}
\caption{Fit results for disk dominated}
\label{tab:disk}
\end{table*}

\begin{table*}[ht]
\centering
\begin{tabular}{cccccccccc}
\hline
Redshift & Model & $\alpha$ & $\alpha_1$ & $\alpha_2$ & $logM^*$ & $\log\phi_1$ & $\log\phi_2$ & $\log\phi_*$ & AICc \\
\hline
$0.2 \leq z < 0.5$ & Single & $-1.89_{-0.06}^{+0.07}$ & -- & -- & $11.10_{-0.35}^{+0.50}$ & -- & -- & $-4.38_{-0.56}^{+0.43}$ & $13.74$ \\[.3cm]
$0.5 \leq z < 0.8$ & Single & $-1.64_{-0.05}^{+0.05}$ & -- & -- & $10.67_{-0.09}^{+0.09}$ & -- & -- & $-3.53_{-0.14}^{+0.12}$ & $13.08$ \\[.3cm]
$0.8 \leq z < 1.1$ & Single & $-1.51_{-0.05}^{+0.04}$ & -- & -- & $10.72_{-0.07}^{+0.07}$ & -- & -- & $-3.23_{-0.10}^{+0.10}$ & $13.29$ \\[.3cm]
$1.1 \leq z < 1.5$ & Single & $-1.70_{-0.04}^{+0.04}$ & -- & -- & $10.92_{-0.11}^{+0.10}$ & -- & -- & $-3.98_{-0.15}^{+0.13}$ & $12.23$ \\[.3cm]
$1.5 \leq z < 2$ & Single & $-1.73_{-0.04}^{+0.03}$ & -- & -- & $10.84_{-0.07}^{+0.06}$ & -- & -- & $-3.91_{-0.10}^{+0.10}$ & $12.09$ \\[.3cm]
$2 \leq z < 2.5$ & Single & $-1.64_{-0.03}^{+0.03}$ & -- & -- & $10.98_{-0.08}^{+0.07}$ & -- & -- & $-3.96_{-0.10}^{+0.10}$ & $10.45$ \\[.3cm]
$2.5 \leq z < 3$ & Single & $-1.61_{-0.04}^{+0.04}$ & -- & -- & $10.87_{-0.11}^{+0.11}$ & -- & -- & $-3.97_{-0.14}^{+0.13}$ & $12.96$ \\[.3cm]
$3 \leq z < 3.5$ & Single & $-1.56_{-0.05}^{+0.04}$ & -- & -- & $10.76_{-0.09}^{+0.08}$ & -- & -- & $-3.91_{-0.11}^{+0.12}$ & $13.79$ \\[.3cm]
$3.5 \leq z < 4.5$ & Single & $-1.56_{-0.05}^{+0.05}$ & -- & -- & $10.56_{-0.08}^{+0.09}$ & -- & -- & $-3.99_{-0.11}^{+0.11}$ & $15.31$ \\[.3cm]
$4.5 \leq z < 5.5$ & DPL & -- & $-2.33_{-0.19}^{+1.98}$ & $-0.77_{-1.67}^{+1.01}$ & $9.01_{-0.15}^{+0.24}$ & -- & -- & $-2.58_{-0.18}^{+0.09}$ & $18.91$ \\[.3cm]
$5.5 \leq z < 6.5$ & DPL & -- & $-2.45_{-0.20}^{+0.20}$ & $0.03_{-0.53}^{+0.54}$ & $9.09_{-0.12}^{+0.16}$ & -- & -- & $-3.07_{-0.09}^{+0.06}$ & $21.20$ \\[.3cm]
\hline
\end{tabular}
\caption{Fit results for peculiars}
\label{tab:irr}
\end{table*}

\subsection{Quiescent galaxies}

We report in tables~\ref{tab:early_Q} to~\ref{tab:irr_Q}, the best fit parameters to the SMFs of quiescent early and late type galaxies (Tables~\ref{tab:early_Q} and ~\ref{tab:late_Q}), Hubble Types (Table~\ref{tab:reg_Q}), spheroids (Table~\ref{tab:sph_Q}), bulge dominated (Table~\ref{tab:db_Q}), disk dominated (Table~\ref{tab:disk_Q}) and peculiars (Table~\ref{tab:irr_Q}).

\begin{table*}[ht]
\centering
\begin{tabular}{cccccccccc}
\hline
Redshift & Model & $alpha$ & $alpha1$ & $alpha2$ & $logM^*$ & $logphi1$ & $logphi2$ & $logphi\_star$ & AICc \\
\hline
$0.2 \leq z < 0.5$ & Single & $-0.52_{-0.07}^{+0.07}$ & -- & -- & $10.80_{-0.06}^{+0.05}$ & -- & -- & $-2.68_{-0.06}^{+0.06}$ & $17.37$ \\[.3cm]
$0.5 \leq z < 0.8$ & Single & $-0.46_{-0.07}^{+0.07}$ & -- & -- & $10.85_{-0.06}^{+0.05}$ & -- & -- & $-2.88_{-0.06}^{+0.06}$ & $14.85$ \\[.3cm]
$0.8 \leq z < 1.1$ & Single & $-0.25_{-0.06}^{+0.06}$ & -- & -- & $10.73_{-0.05}^{+0.04}$ & -- & -- & $-2.83_{-0.05}^{+0.04}$ & $10.64$ \\[.3cm]
$1.1 \leq z < 1.5$ & Single & $0.17_{-0.09}^{+0.09}$ & -- & -- & $10.59_{-0.04}^{+0.04}$ & -- & -- & $-3.08_{-0.04}^{+0.04}$ & $14.02$ \\[.3cm]
$1.5 \leq z < 2$ & Single & $0.33_{-0.10}^{+0.10}$ & -- & -- & $10.57_{-0.06}^{+0.05}$ & -- & -- & $-3.32_{-0.04}^{+0.03}$ & $19.93$ \\[.3cm]
$2 \leq z < 2.5$ & Single & $0.40_{-0.17}^{+0.17}$ & -- & -- & $10.56_{-0.08}^{+0.07}$ & -- & -- & $-3.73_{-0.05}^{+0.05}$ & $13.73$ \\[.3cm]
$2.5 \leq z < 3$ & Single & $1.22_{-0.38}^{+0.44}$ & -- & -- & $10.31_{-0.12}^{+0.10}$ & -- & -- & $-4.18_{-0.13}^{+0.09}$ & $22.00$ \\[.3cm]
$3 \leq z < 3.5$ & Single & $0.47_{-0.36}^{+0.50}$ & -- & -- & $10.63_{-0.16}^{+0.13}$ & -- & -- & $-4.26_{-0.09}^{+0.07}$ & $20.09$ \\[.3cm]
$3.5 \leq z < 4.5$ & Single & $0.07_{-0.89}^{+1.18}$ & -- & -- & $10.39_{-0.25}^{+0.29}$ & -- & -- & $-4.93_{-0.38}^{+0.15}$ & $18.09$ \\[.3cm]
$4.5 \leq z < 5.5$ & -- & -- & -- & -- & -- & -- & -- & -- & -- \\[.3cm]
$5.5 \leq z < 6.5$ & -- & -- & -- & -- & -- & -- & -- & -- & -- \\[.3cm]
\hline
\end{tabular}
\caption{Fit results for quiescent Early Type}
\label{tab:early_Q}
\end{table*}

\begin{table*}[ht]
\centering
\begin{tabular}{cccccccccc}
\hline
Redshift & Model & $alpha$ & $alpha1$ & $alpha2$ & $logM^*$ & $logphi1$ & $logphi2$ & $logphi\_star$ & AICc \\
\hline
$0.2 \leq z < 0.5$ & Single & $-1.19_{-0.07}^{+0.07}$ & -- & -- & $10.82_{-0.09}^{+0.09}$ & -- & -- & $-3.49_{-0.13}^{+0.11}$ & $15.61$ \\[.3cm]
$0.5 \leq z < 0.8$ & Double & -- & $-1.68_{-0.50}^{+0.35}$ & $-0.52_{-0.43}^{+0.86}$ & $10.46_{-0.20}^{+0.16}$ & $-4.36_{-1.05}^{+0.59}$ & $-3.48_{-0.16}^{+0.16}$ & -- & $19.54$ \\[.3cm]
$0.8 \leq z < 1.1$ & Single & $-1.23_{-0.07}^{+0.07}$ & -- & -- & $11.03_{-0.11}^{+0.12}$ & -- & -- & $-3.97_{-0.14}^{+0.13}$ & $15.68$ \\[.3cm]
$1.1 \leq z < 1.5$ & Double & -- & $-1.45_{-0.16}^{+0.12}$ & $0.82_{-1.03}^{+1.00}$ & $10.59_{-0.19}^{+0.32}$ & $-4.77_{-0.40}^{+0.24}$ & $-4.37_{-0.29}^{+0.19}$ & -- & $17.96$ \\[.3cm]
$1.5 \leq z < 2$ & Single & $-1.26_{-0.09}^{+0.09}$ & -- & -- & $11.09_{-0.15}^{+0.17}$ & -- & -- & $-4.85_{-0.20}^{+0.17}$ & $15.53$ \\[.3cm]
$2 \leq z < 2.5$ & Single & $-1.09_{-0.10}^{+0.11}$ & -- & -- & $11.34_{-0.26}^{+0.31}$ & -- & -- & $-5.01_{-0.22}^{+0.21}$ & $9.45$ \\[.3cm]
$2.5 \leq z < 3$ & Single & $-0.56_{-0.25}^{+0.28}$ & -- & -- & $10.89_{-0.24}^{+0.27}$ & -- & -- & $-4.85_{-0.23}^{+0.17}$ & $13.95$ \\[.3cm]
$3 \leq z < 3.5$ & Single & $0.03_{-0.30}^{+0.39}$ & -- & -- & $10.65_{-0.20}^{+0.16}$ & -- & -- & $-4.64_{-0.12}^{+0.10}$ & $12.35$ \\[.3cm]
$3.5 \leq z < 4.5$ & Single & $-0.77_{-0.19}^{+0.35}$ & -- & -- & $11.04_{-0.49}^{+0.58}$ & -- & -- & $-5.44_{-0.28}^{+0.26}$ & $16.05$ \\[.3cm]
$4.5 \leq z < 5.5$ & Single & $-1.40_{-1.34}^{+2.41}$ & -- & -- & $8.97_{-0.66}^{+0.85}$ & -- & -- & $-5.68_{-0.87}^{+0.99}$ & $16.49$ \\[.3cm]
$5.5 \leq z < 6.5$ & -- & -- & -- & -- & -- & -- & -- & -- & -- \\[.3cm]
\hline
\end{tabular}
\caption{Fit results for quiescent Late Type}
\label{tab:late_Q}
\end{table*}

\begin{table*}[ht]
\centering
\begin{tabular}{cccccccccc}
\hline
Redshift & Model & $alpha$ & $alpha1$ & $alpha2$ & $logM^*$ & $logphi1$ & $logphi2$ & $logphi\_star$ & AICc \\
\hline
$0.2 \leq z < 0.5$ & Single & $-0.67_{-0.06}^{+0.06}$ & -- & -- & $10.82_{-0.06}^{+0.05}$ & -- & -- & $-2.63_{-0.06}^{+0.06}$ & $13.25$ \\[.3cm]
$0.5 \leq z < 0.8$ & Single & $-0.51_{-0.06}^{+0.05}$ & -- & -- & $10.82_{-0.05}^{+0.05}$ & -- & -- & $-2.82_{-0.06}^{+0.05}$ & $13.26$ \\[.3cm]
$0.8 \leq z < 1.1$ & Single & $-0.15_{-0.06}^{+0.06}$ & -- & -- & $10.66_{-0.05}^{+0.04}$ & -- & -- & $-2.75_{-0.04}^{+0.04}$ & $24.87$ \\[.3cm]
$1.1 \leq z < 1.5$ & Single & $0.12_{-0.10}^{+0.09}$ & -- & -- & $10.60_{-0.04}^{+0.03}$ & -- & -- & $-3.08_{-0.05}^{+0.04}$ & $14.39$ \\[.3cm]
$1.5 \leq z < 2$ & Single & $0.20_{-0.10}^{+0.10}$ & -- & -- & $10.63_{-0.06}^{+0.05}$ & -- & -- & $-3.34_{-0.04}^{+0.04}$ & $14.28$ \\[.3cm]
$2 \leq z < 2.5$ & Single & $0.59_{-0.19}^{+0.21}$ & -- & -- & $10.56_{-0.08}^{+0.06}$ & -- & -- & $-3.80_{-0.05}^{+0.05}$ & $13.83$ \\[.3cm]
$2.5 \leq z < 3$ & Single & $1.62_{-0.45}^{+0.56}$ & -- & -- & $10.30_{-0.14}^{+0.12}$ & -- & -- & $-4.43_{-0.22}^{+0.13}$ & $24.44$ \\[.3cm]
$3 \leq z < 3.5$ & Single & $0.44_{-0.37}^{+0.47}$ & -- & -- & $10.74_{-0.14}^{+0.13}$ & -- & -- & $-4.35_{-0.10}^{+0.07}$ & $18.30$ \\[.3cm]
$3.5 \leq z < 4.5$ & Single & $0.01_{-0.76}^{+1.19}$ & -- & -- & $10.61_{-0.39}^{+0.50}$ & -- & -- & $-5.14_{-0.38}^{+0.16}$ & $32.26$ \\[.3cm]
$4.5 \leq z < 5.5$ & -- & -- & -- & -- & -- & -- & -- & -- & -- \\[.3cm]
$5.5 \leq z < 6.5$ & -- & -- & -- & -- & -- & -- & -- & -- & -- \\[.3cm]
\hline
\end{tabular}
\caption{Fit results for quiescent Hubble Types}
\label{tab:reg_Q}
\end{table*}

\begin{table*}[ht]
\centering
\begin{tabular}{cccccccccc}
\hline
Redshift & Model & $alpha$ & $alpha1$ & $alpha2$ & $logM^*$ & $logphi1$ & $logphi2$ & $logphi\_star$ & AICc \\
\hline
$0.2 \leq z < 0.5$ & Single & $-0.56_{-0.13}^{+0.14}$ & -- & -- & $10.90_{-0.11}^{+0.12}$ & -- & -- & $-3.41_{-0.13}^{+0.11}$ & $17.79$ \\[.3cm]
$0.5 \leq z < 0.8$ & Single & $-0.87_{-0.07}^{+0.08}$ & -- & -- & $11.08_{-0.11}^{+0.11}$ & -- & -- & $-3.77_{-0.12}^{+0.10}$ & $15.35$ \\[.3cm]
$0.8 \leq z < 1.1$ & Single & $-0.41_{-0.08}^{+0.07}$ & -- & -- & $10.67_{-0.10}^{+0.07}$ & -- & -- & $-3.36_{-0.07}^{+0.06}$ & $21.78$ \\[.3cm]
$1.1 \leq z < 1.5$ & Single & $0.09_{-0.12}^{+0.12}$ & -- & -- & $10.55_{-0.07}^{+0.05}$ & -- & -- & $-3.43_{-0.05}^{+0.05}$ & $14.29$ \\[.3cm]
$1.5 \leq z < 2$ & Single & $0.32_{-0.15}^{+0.17}$ & -- & -- & $10.51_{-0.08}^{+0.07}$ & -- & -- & $-3.69_{-0.04}^{+0.04}$ & $15.15$ \\[.3cm]
$2 \leq z < 2.5$ & Single & $0.47_{-0.23}^{+0.27}$ & -- & -- & $10.53_{-0.10}^{+0.08}$ & -- & -- & $-4.15_{-0.07}^{+0.06}$ & $14.27$ \\[.3cm]
$2.5 \leq z < 3$ & Single & $1.39_{-0.47}^{+0.68}$ & -- & -- & $10.32_{-0.17}^{+0.13}$ & -- & -- & $-4.65_{-0.26}^{+0.13}$ & $22.06$ \\[.3cm]
$3 \leq z < 3.5$ & Single & $0.98_{-0.80}^{+1.13}$ & -- & -- & $10.47_{-0.23}^{+0.25}$ & -- & -- & $-4.94_{-0.39}^{+0.16}$ & $18.82$ \\[.3cm]
$3.5 \leq z < 4.5$ & -- & -- & -- & -- & -- & -- & -- & -- & -- \\[.3cm]
$4.5 \leq z < 5.5$ & -- & -- & -- & -- & -- & -- & -- & -- & -- \\[.3cm]
$5.5 \leq z < 6.5$ & -- & -- & -- & -- & -- & -- & -- & -- & -- \\[.3cm]
\hline
\end{tabular}
\caption{Fit results for quiescent spheroids}
\label{tab:sph_Q}
\end{table*}

\begin{table*}[ht]
\centering
\begin{tabular}{cccccccccc}
\hline
Redshift & Model & $alpha$ & $alpha1$ & $alpha2$ & $logM^*$ & $logphi1$ & $logphi2$ & $logphi\_star$ & AICc \\
\hline
$0.2 \leq z < 0.5$ & Single & $-0.34_{-0.08}^{+0.07}$ & -- & -- & $10.70_{-0.06}^{+0.05}$ & -- & -- & $-2.70_{-0.06}^{+0.05}$ & $13.31$ \\[.3cm]
$0.5 \leq z < 0.8$ & Single & $-0.02_{-0.09}^{+0.09}$ & -- & -- & $10.69_{-0.06}^{+0.05}$ & -- & -- & $-2.87_{-0.04}^{+0.04}$ & $12.86$ \\[.3cm]
$0.8 \leq z < 1.1$ & Single & $0.25_{-0.13}^{+0.13}$ & -- & -- & $10.61_{-0.06}^{+0.05}$ & -- & -- & $-2.96_{-0.04}^{+0.04}$ & $15.29$ \\[.3cm]
$1.1 \leq z < 1.5$ & Single & $0.67_{-0.19}^{+0.23}$ & -- & -- & $10.52_{-0.07}^{+0.06}$ & -- & -- & $-3.41_{-0.05}^{+0.05}$ & $13.29$ \\[.3cm]
$1.5 \leq z < 2$ & Single & $0.57_{-0.16}^{+0.21}$ & -- & -- & $10.61_{-0.08}^{+0.06}$ & -- & -- & $-3.69_{-0.05}^{+0.04}$ & $16.44$ \\[.3cm]
$2 \leq z < 2.5$ & Single & $1.14_{-0.45}^{+0.67}$ & -- & -- & $10.50_{-0.14}^{+0.10}$ & -- & -- & $-4.31_{-0.19}^{+0.09}$ & $18.78$ \\[.3cm]
$2.5 \leq z < 3$ & -- & -- & -- & -- & -- & -- & -- & -- & -- \\[.3cm]
$3 \leq z < 3.5$ & Single & $1.43_{-0.74}^{+1.18}$ & -- & -- & $10.64_{-0.19}^{+0.18}$ & -- & -- & $-4.93_{-0.48}^{+0.20}$ & $19.48$ \\[.3cm]
$3.5 \leq z < 4.5$ & -- & -- & -- & -- & -- & -- & -- & -- & -- \\[.3cm]
$4.5 \leq z < 5.5$ & -- & -- & -- & -- & -- & -- & -- & -- & -- \\[.3cm]
$5.5 \leq z < 6.5$ & -- & -- & -- & -- & -- & -- & -- & -- & -- \\[.3cm]
\hline
\end{tabular}
\caption{Fit results for quiescent bulge dominated}
\label{tab:db_Q}
\end{table*}

\begin{table*}[ht]
\centering
\begin{tabular}{cccccccccc}
\hline
Redshift & Model & $alpha$ & $alpha1$ & $alpha2$ & $logM^*$ & $logphi1$ & $logphi2$ & $logphi\_star$ & AICc \\
\hline
$0.2 \leq z < 0.5$ & Single & $-0.87_{-0.08}^{+0.08}$ & -- & -- & $10.62_{-0.11}^{+0.09}$ & -- & -- & $-3.28_{-0.10}^{+0.09}$ & $14.59$ \\[.3cm]
$0.5 \leq z < 0.8$ & Single & $-0.62_{-0.11}^{+0.11}$ & -- & -- & $10.45_{-0.13}^{+0.10}$ & -- & -- & $-3.57_{-0.10}^{+0.09}$ & $15.80$ \\[.3cm]
$0.8 \leq z < 1.1$ & DPL & -- & $-1.71_{-0.17}^{+0.14}$ & $1.15_{-0.54}^{+0.74}$ & $9.53_{-0.17}^{+0.19}$ & -- & -- & $-3.39_{-0.08}^{+0.07}$ & $28.00$ \\[.3cm]
$1.1 \leq z < 1.5$ & Single & $-0.50_{-0.15}^{+0.17}$ & -- & -- & $10.76_{-0.13}^{+0.12}$ & -- & -- & $-4.38_{-0.13}^{+0.11}$ & $17.71$ \\[.3cm]
$1.5 \leq z < 2$ & Single & $-0.44_{-0.15}^{+0.17}$ & -- & -- & $10.71_{-0.18}^{+0.14}$ & -- & -- & $-4.52_{-0.12}^{+0.10}$ & $15.13$ \\[.3cm]
$2 \leq z < 2.5$ & Single & $-0.49_{-0.42}^{+0.80}$ & -- & -- & $11.00_{-0.48}^{+0.52}$ & -- & -- & $-5.10_{-0.33}^{+0.20}$ & $19.28$ \\[.3cm]
$2.5 \leq z < 3$ & Single & $-0.29_{-1.33}^{+2.46}$ & -- & -- & $10.24_{-1.20}^{+1.10}$ & -- & -- & $-5.59_{-0.75}^{+0.83}$ & $32.07$ \\[.3cm]
$3 \leq z < 3.5$ & Single & $-0.17_{-1.02}^{+1.85}$ & -- & -- & $10.79_{-0.81}^{+0.62}$ & -- & -- & $-5.42_{-0.66}^{+0.35}$ & $20.60$ \\[.3cm]
$3.5 \leq z < 4.5$ & Single & $0.22_{-1.75}^{+2.38}$ & -- & -- & $10.07_{-1.22}^{+0.96}$ & -- & -- & $-5.79_{-0.70}^{+1.49}$ & $31.31$ \\[.3cm]
$4.5 \leq z < 5.5$ & -- & -- & -- & -- & -- & -- & -- & -- & -- \\[.3cm]
$5.5 \leq z < 6.5$ & -- & -- & -- & -- & -- & -- & -- & -- & -- \\[.3cm]
\hline
\end{tabular}
\caption{Fit results for quiescent disk dominated}
\label{tab:disk_Q}
\end{table*}

\begin{table*}[ht]
\centering
\begin{tabular}{cccccccccc}
\hline
Redshift & Model & $alpha$ & $alpha1$ & $alpha2$ & $logM\_star$ & $logphi1$ & $logphi2$ & $logphi\_star$ & AICc \\
\hline
$0.2 \leq z < 0.5$ & Double & -- & $-2.62_{-0.36}^{+0.34}$ & $-0.41_{-0.39}^{+0.96}$ & $10.34_{-0.42}^{+0.42}$ & $-5.89_{-0.74}^{+0.97}$ & $-3.83_{-0.26}^{+0.20}$ & -- & $21.00$ \\[.3cm]
$0.5 \leq z < 0.8$ & Double & -- & $-2.00_{-0.26}^{+0.25}$ & $-0.16_{-0.69}^{+1.03}$ & $10.45_{-0.32}^{+0.27}$ & $-4.90_{-0.66}^{+0.63}$ & $-3.92_{-0.20}^{+0.17}$ & -- & $18.64$ \\[.3cm]
$0.8 \leq z < 1.1$ & Double & -- & $-1.69_{-0.20}^{+0.15}$ & $0.13_{-0.73}^{+1.11}$ & $10.52_{-0.28}^{+0.28}$ & $-4.63_{-0.53}^{+0.38}$ & $-3.85_{-0.25}^{+0.15}$ & -- & $21.43$ \\[.3cm]
$1.1 \leq z < 1.5$ & Single & $-1.61_{-0.08}^{+0.07}$ & -- & -- & $11.65_{-0.40}^{+0.25}$ & -- & -- & $-5.71_{-0.24}^{+0.30}$ & $13.85$ \\[.3cm]
$1.5 \leq z < 2$ & Single & $-1.54_{-0.09}^{+0.08}$ & -- & -- & $11.56_{-0.42}^{+0.30}$ & -- & -- & $-5.74_{-0.27}^{+0.31}$ & $13.64$ \\[.3cm]
$2 \leq z < 2.5$ & Single & $-1.23_{-0.11}^{+0.12}$ & -- & -- & $11.28_{-0.34}^{+0.37}$ & -- & -- & $-5.32_{-0.29}^{+0.27}$ & $9.16$ \\[.3cm]
$2.5 \leq z < 3$ & Single & $-0.80_{-0.23}^{+0.32}$ & -- & -- & $11.04_{-0.35}^{+0.46}$ & -- & -- & $-5.21_{-0.33}^{+0.27}$ & $14.70$ \\[.3cm]
$3 \leq z < 3.5$ & Single & $-0.27_{-0.30}^{+0.65}$ & -- & -- & $10.88_{-0.45}^{+0.65}$ & -- & -- & $-4.93_{-0.20}^{+0.14}$ & $16.79$ \\[.3cm]
$3.5 \leq z < 4.5$ & Single & $-1.02_{-0.22}^{+0.46}$ & -- & -- & $10.89_{-0.66}^{+0.73}$ & -- & -- & $-5.96_{-0.44}^{+0.46}$ & $16.56$ \\[.3cm]
$4.5 \leq z < 5.5$ & Single & $-1.38_{-1.31}^{+2.40}$ & -- & -- & $8.95_{-0.65}^{+0.89}$ & -- & -- & $-5.65_{-0.88}^{+0.99}$ & $16.50$ \\[.3cm]
$5.5 \leq z < 6.5$ & -- & -- & -- & -- & -- & -- & -- & -- & -- \\[.3cm]
\hline
\end{tabular}
\caption{Fit results for quiescent peculiars}
\label{tab:irr_Q}
\end{table*}

\subsection{Star Forming galaxies}

We report in tables~\ref{tab:early_SF} to~\ref{tab:irr_SF}, the best fit parameters to the SMFs of star-forming early and late type galaxies (Tables~\ref{tab:early_SF} and ~\ref{tab:late_SF}), Hubble Types (Table~\ref{tab:reg_SF}), spheroids (Table~\ref{tab:sph_SF}), bulge dominated (Table~\ref{tab:db_SF}), disk dominated (Table~\ref{tab:disk_SF}) and peculiars (Table~\ref{tab:irr_SF}). 

\begin{table*}[ht]
\centering
\begin{tabular}{cccccccccc}
\hline
Redshift & Model & $alpha$ & $alpha1$ & $alpha2$ & $logM^*$ & $logphi1$ & $logphi2$ & $logphi\_star$ & AICc \\
\hline
$0.2 \leq z < 0.5$ & Double & -- & $-1.59_{-0.31}^{+0.24}$ & $0.47_{-1.01}^{+1.10}$ & $10.19_{-0.17}^{+0.17}$ & $-3.98_{-0.58}^{+0.35}$ & $-3.40_{-0.29}^{+0.16}$ & -- & $18.35$ \\[.3cm]
$0.5 \leq z < 0.8$ & Single & $-1.33_{-0.07}^{+0.06}$ & -- & -- & $10.98_{-0.14}^{+0.15}$ & -- & -- & $-3.99_{-0.16}^{+0.15}$ & $16.75$ \\[.3cm]
$0.8 \leq z < 1.1$ & Double & -- & $-1.13_{-0.10}^{+1.75}$ & $1.29_{-3.29}^{+1.09}$ & $10.34_{-0.15}^{+0.43}$ & $-3.46_{-0.17}^{+1.09}$ & $-3.74_{-0.43}^{+2.12}$ & -- & $19.32$ \\[.3cm]
$1.1 \leq z < 1.5$ & Single & $-1.09_{-0.05}^{+0.05}$ & -- & -- & $10.95_{-0.07}^{+0.07}$ & -- & -- & $-3.79_{-0.08}^{+0.08}$ & $18.86$ \\[.3cm]
$1.5 \leq z < 2$ & Single & $-1.10_{-0.04}^{+0.04}$ & -- & -- & $10.86_{-0.05}^{+0.05}$ & -- & -- & $-3.80_{-0.07}^{+0.06}$ & $17.68$ \\[.3cm]
$2 \leq z < 2.5$ & Single & $-1.13_{-0.05}^{+0.05}$ & -- & -- & $11.01_{-0.08}^{+0.07}$ & -- & -- & $-4.09_{-0.09}^{+0.09}$ & $10.62$ \\[.3cm]
$2.5 \leq z < 3$ & Single & $-1.25_{-0.06}^{+0.06}$ & -- & -- & $11.14_{-0.14}^{+0.15}$ & -- & -- & $-4.53_{-0.14}^{+0.13}$ & $10.26$ \\[.3cm]
$3 \leq z < 3.5$ & Double & -- & $-2.19_{-0.20}^{+0.28}$ & $-0.40_{-0.23}^{+0.34}$ & $10.70_{-0.16}^{+0.18}$ & $-6.45_{-0.39}^{+0.60}$ & $-4.21_{-0.15}^{+0.11}$ & -- & $21.64$ \\[.3cm]
$3.5 \leq z < 4.5$ & Double & -- & $-2.18_{-0.22}^{+0.28}$ & $-0.55_{-0.31}^{+0.45}$ & $10.63_{-0.21}^{+0.19}$ & $-6.20_{-0.51}^{+0.64}$ & $-4.61_{-0.15}^{+0.14}$ & -- & $18.69$ \\[.3cm]
$4.5 \leq z < 5.5$ & Single & $-2.14_{-0.09}^{+0.08}$ & -- & -- & $11.01_{-0.58}^{+0.34}$ & -- & -- & $-6.48_{-0.38}^{+0.73}$ & $19.45$ \\[.3cm]
$5.5 \leq z < 6.5$ & DPL & -- & $-3.01_{-0.45}^{+0.29}$ & $0.52_{-0.94}^{+1.55}$ & $8.82_{-0.13}^{+0.17}$ & -- & -- & $-3.43_{-0.15}^{+0.09}$ & $20.39$ \\[.3cm]
\hline
\end{tabular}
\caption{Fit results for star forming Early Type}
\label{tab:early_SF}
\end{table*}

\begin{table*}[ht]
\centering
\begin{tabular}{cccccccccc}
\hline
Redshift & Model & $alpha$ & $alpha1$ & $alpha2$ & $logM^*$ & $logphi1$ & $logphi2$ & $logphi\_star$ & AICc \\
\hline
$0.2 \leq z < 0.5$ & Single & $-1.48_{-0.07}^{+0.08}$ & -- & -- & $10.43_{-0.10}^{+0.10}$ & -- & -- & $-3.03_{-0.15}^{+0.14}$ & $9.63$ \\[.3cm]
$0.5 \leq z < 0.8$ & Single & $-1.43_{-0.05}^{+0.05}$ & -- & -- & $10.58_{-0.07}^{+0.06}$ & -- & -- & $-3.05_{-0.09}^{+0.09}$ & $8.98$ \\[.3cm]
$0.8 \leq z < 1.1$ & Single & $-1.37_{-0.04}^{+0.04}$ & -- & -- & $10.61_{-0.05}^{+0.05}$ & -- & -- & $-2.89_{-0.07}^{+0.07}$ & $8.72$ \\[.3cm]
$1.1 \leq z < 1.5$ & Single & $-1.47_{-0.03}^{+0.03}$ & -- & -- & $10.78_{-0.05}^{+0.05}$ & -- & -- & $-3.36_{-0.07}^{+0.07}$ & $13.76$ \\[.3cm]
$1.5 \leq z < 2$ & Single & $-1.55_{-0.03}^{+0.03}$ & -- & -- & $10.89_{-0.05}^{+0.05}$ & -- & -- & $-3.56_{-0.08}^{+0.07}$ & $14.49$ \\[.3cm]
$2 \leq z < 2.5$ & Single & $-1.54_{-0.04}^{+0.03}$ & -- & -- & $10.91_{-0.08}^{+0.06}$ & -- & -- & $-3.68_{-0.09}^{+0.10}$ & $9.24$ \\[.3cm]
$2.5 \leq z < 3$ & Single & $-1.56_{-0.05}^{+0.04}$ & -- & -- & $10.86_{-0.11}^{+0.10}$ & -- & -- & $-3.84_{-0.13}^{+0.13}$ & $12.12$ \\[.3cm]
$3 \leq z < 3.5$ & Single & $-1.54_{-0.04}^{+0.04}$ & -- & -- & $11.02_{-0.10}^{+0.10}$ & -- & -- & $-4.01_{-0.12}^{+0.11}$ & $11.63$ \\[.3cm]
$3.5 \leq z < 4.5$ & Single & $-1.56_{-0.04}^{+0.04}$ & -- & -- & $10.71_{-0.08}^{+0.08}$ & -- & -- & $-4.07_{-0.11}^{+0.10}$ & $12.49$ \\[.3cm]
$4.5 \leq z < 5.5$ & DPL & -- & $-2.33_{-0.17}^{+0.14}$ & $-0.30_{-0.81}^{+1.04}$ & $8.96_{-0.14}^{+0.23}$ & -- & -- & $-2.56_{-0.16}^{+0.08}$ & $15.21$ \\[.3cm]
$5.5 \leq z < 6.5$ & DPL & -- & $-2.47_{-0.19}^{+0.15}$ & $0.05_{-0.44}^{+0.54}$ & $9.09_{-0.12}^{+0.15}$ & -- & -- & $-3.07_{-0.08}^{+0.06}$ & $19.15$ \\[.3cm]
\hline
\end{tabular}
\caption{Fit results for star forming Late Type}
\label{tab:late_SF}
\end{table*}

\begin{table*}[ht]
\centering
\begin{tabular}{cccccccccc}
\hline
Redshift & Model & $alpha$ & $alpha1$ & $alpha2$ & $logM^*$ & $logphi1$ & $logphi2$ & $logphi\_star$ & AICc \\
\hline
$0.2 \leq z < 0.5$ & Single & $-1.20_{-0.06}^{+0.06}$ & -- & -- & $10.49_{-0.08}^{+0.08}$ & -- & -- & $-2.99_{-0.10}^{+0.09}$ & $9.53$ \\[.3cm]
$0.5 \leq z < 0.8$ & Single & $-1.11_{-0.06}^{+0.06}$ & -- & -- & $10.58_{-0.08}^{+0.07}$ & -- & -- & $-3.02_{-0.08}^{+0.10}$ & $12.63$ \\[.3cm]
$0.8 \leq z < 1.1$ & Single & $-1.06_{-0.04}^{+0.04}$ & -- & -- & $10.67_{-0.05}^{+0.04}$ & -- & -- & $-3.02_{-0.06}^{+0.05}$ & $29.70$ \\[.3cm]
$1.1 \leq z < 1.5$ & DPL & -- & $-2.09_{-0.11}^{+0.17}$ & $-0.40_{-0.39}^{+0.22}$ & $9.62_{-0.14}^{+0.16}$ & -- & -- & $-2.35_{-0.07}^{+0.05}$ & $58.50$ \\[.3cm]
$1.5 \leq z < 2$ & Single & $-0.95_{-0.04}^{+0.03}$ & -- & -- & $10.75_{-0.06}^{+0.04}$ & -- & -- & $-3.24_{-0.05}^{+0.06}$ & $70.94$ \\[.3cm]
$2 \leq z < 2.5$ & Single & $-0.88_{-0.05}^{+0.05}$ & -- & -- & $10.75_{-0.08}^{+0.06}$ & -- & -- & $-3.51_{-0.07}^{+0.06}$ & $41.98$ \\[.3cm]
$2.5 \leq z < 3$ & DPL & -- & $-1.82_{-0.11}^{+2.04}$ & $0.26_{-2.08}^{+0.33}$ & $9.47_{-0.12}^{+0.12}$ & -- & -- & $-3.12_{-0.05}^{+0.04}$ & $26.72$ \\[.3cm]
$3 \leq z < 3.5$ & DPL & -- & $-1.93_{-0.22}^{+2.16}$ & $0.34_{-2.23}^{+0.25}$ & $10.10_{-0.15}^{+0.16}$ & -- & -- & $-3.47_{-0.06}^{+0.05}$ & $30.57$ \\[.3cm]
$3.5 \leq z < 4.5$ & Single & $0.19_{-0.22}^{+0.26}$ & -- & -- & $10.37_{-0.16}^{+0.13}$ & -- & -- & $-4.17_{-0.06}^{+0.06}$ & $19.20$ \\[.3cm]
$4.5 \leq z < 5.5$ & Single & $-0.29_{-1.03}^{+2.07}$ & -- & -- & $10.45_{-1.56}^{+1.08}$ & -- & -- & $-5.73_{-0.72}^{+0.97}$ & $20.19$ \\[.3cm]
$5.5 \leq z < 6.5$ & -- & -- & -- & -- & -- & -- & -- & -- & -- \\[.3cm]
\hline
\end{tabular}
\caption{Fit results for star forming Hubble Types}
\label{tab:reg_SF}
\end{table*}

\begin{table*}[ht]
\centering
\begin{tabular}{cccccccccc}
\hline
Redshift & Model & $alpha$ & $alpha1$ & $alpha2$ & $logM^*$ & $logphi1$ & $logphi2$ & $logphi\_star$ & AICc \\
\hline
$0.2 \leq z < 0.5$ & Single & $-2.08_{-0.23}^{+0.28}$ & -- & -- & $10.43_{-0.85}^{+0.80}$ & -- & -- & $-5.65_{-0.93}^{+1.29}$ & $18.99$ \\[.3cm]
$0.5 \leq z < 0.8$ & Single & $-1.85_{-0.10}^{+0.09}$ & -- & -- & $11.30_{-0.55}^{+0.47}$ & -- & -- & $-5.56_{-0.49}^{+0.60}$ & $10.44$ \\[.3cm]
$0.8 \leq z < 1.1$ & Single & $-1.66_{-0.06}^{+0.06}$ & -- & -- & $11.52_{-0.41}^{+0.32}$ & -- & -- & $-5.16_{-0.28}^{+0.36}$ & $10.35$ \\[.3cm]
$1.1 \leq z < 1.5$ & Single & $-1.43_{-0.08}^{+0.08}$ & -- & -- & $11.25_{-0.27}^{+0.32}$ & -- & -- & $-4.83_{-0.29}^{+0.26}$ & $14.73$ \\[.3cm]
$1.5 \leq z < 2$ & Single & $-1.37_{-0.07}^{+0.06}$ & -- & -- & $11.08_{-0.12}^{+0.12}$ & -- & -- & $-4.63_{-0.15}^{+0.14}$ & $13.79$ \\[.3cm]
$2 \leq z < 2.5$ & Single & $-1.16_{-0.07}^{+0.07}$ & -- & -- & $10.92_{-0.11}^{+0.12}$ & -- & -- & $-4.51_{-0.13}^{+0.14}$ & $18.78$ \\[.3cm]
$2.5 \leq z < 3$ & Single & $-0.98_{-0.09}^{+0.08}$ & -- & -- & $10.73_{-0.21}^{+0.14}$ & -- & -- & $-4.57_{-0.13}^{+0.15}$ & $23.19$ \\[.3cm]
$3 \leq z < 3.5$ & Single & $-0.21_{-0.20}^{+0.27}$ & -- & -- & $10.51_{-0.17}^{+0.16}$ & -- & -- & $-4.58_{-0.11}^{+0.10}$ & $14.42$ \\[.3cm]
$3.5 \leq z < 4.5$ & Single & $-0.60_{-0.45}^{+0.94}$ & -- & -- & $10.75_{-0.60}^{+0.58}$ & -- & -- & $-5.64_{-0.49}^{+0.27}$ & $18.85$ \\[.3cm]
$4.5 \leq z < 5.5$ & -- & -- & -- & -- & -- & -- & -- & -- & -- \\[.3cm]
$5.5 \leq z < 6.5$ & -- & -- & -- & -- & -- & -- & -- & -- & -- \\[.3cm]
\hline
\end{tabular}
\caption{Fit results for star forming spheroids}
\label{tab:sph_SF}
\end{table*}

\begin{table*}[ht]
\centering
\begin{tabular}{cccccccccc}
\hline
Redshift & Model & $alpha$ & $alpha1$ & $alpha2$ & $logM^*$ & $logphi1$ & $logphi2$ & $logphi\_star$ & AICc \\
\hline
$0.2 \leq z < 0.5$ & Single & $-1.20_{-0.12}^{+0.13}$ & -- & -- & $10.72_{-0.15}^{+0.18}$ & -- & -- & $-3.92_{-0.25}^{+0.20}$ & $20.85$ \\[.3cm]
$0.5 \leq z < 0.8$ & Single & $-0.78_{-0.08}^{+0.08}$ & -- & -- & $10.62_{-0.09}^{+0.09}$ & -- & -- & $-3.59_{-0.10}^{+0.08}$ & $30.36$ \\[.3cm]
$0.8 \leq z < 1.1$ & DPL & -- & $-1.51_{-0.09}^{+3.02}$ & $1.31_{-2.84}^{+0.76}$ & $9.11_{-0.08}^{+0.09}$ & -- & -- & $-3.02_{-0.05}^{+0.05}$ & $37.41$ \\[.3cm]
$1.1 \leq z < 1.5$ & DPL & -- & $0.18_{-1.77}^{+1.02}$ & $-1.46_{-0.13}^{+2.61}$ & $9.17_{-0.10}^{+0.14}$ & -- & -- & $-3.19_{-0.05}^{+0.04}$ & $38.54$ \\[.3cm]
$1.5 \leq z < 2$ & DPL & -- & $-1.94_{-0.09}^{+2.30}$ & $0.55_{-2.44}^{+0.32}$ & $9.45_{-0.09}^{+0.10}$ & -- & -- & $-3.13_{-0.04}^{+0.04}$ & $51.96$ \\[.3cm]
$2 \leq z < 2.5$ & DPL & -- & $-1.82_{-0.13}^{+2.41}$ & $0.41_{-2.26}^{+0.58}$ & $9.48_{-0.12}^{+0.17}$ & -- & -- & $-3.52_{-0.06}^{+0.05}$ & $40.96$ \\[.3cm]
$2.5 \leq z < 3$ & Single & $-0.80_{-0.10}^{+0.10}$ & -- & -- & $10.89_{-0.13}^{+0.13}$ & -- & -- & $-4.45_{-0.12}^{+0.11}$ & $15.93$ \\[.3cm]
$3 \leq z < 3.5$ & Single & $0.11_{-0.37}^{+0.57}$ & -- & -- & $10.77_{-0.22}^{+0.23}$ & -- & -- & $-4.66_{-0.15}^{+0.11}$ & $13.89$ \\[.3cm]
$3.5 \leq z < 4.5$ & Single & $-0.39_{-0.51}^{+0.71}$ & -- & -- & $10.74_{-0.34}^{+0.42}$ & -- & -- & $-5.17_{-0.34}^{+0.16}$ & $19.13$ \\[.3cm]
$4.5 \leq z < 5.5$ & -- & -- & -- & -- & -- & -- & -- & -- & -- \\[.3cm]
$5.5 \leq z < 6.5$ & -- & -- & -- & -- & -- & -- & -- & -- & -- \\[.3cm]
\hline
\end{tabular}
\caption{Fit results for star forming bulge dominated}
\label{tab:db_SF}
\end{table*}

\begin{table*}[ht]
\centering
\begin{tabular}{cccccccccc}
\hline
Redshift & Model & $alpha$ & $alpha1$ & $alpha2$ & $logM^*$ & $logphi1$ & $logphi2$ & $logphi\_star$ & AICc \\
\hline
$0.2 \leq z < 0.5$ & Single & $-1.10_{-0.07}^{+0.08}$ & -- & -- & $10.27_{-0.08}^{+0.08}$ & -- & -- & $-2.90_{-0.10}^{+0.10}$ & $9.41$ \\[.3cm]
$0.5 \leq z < 0.8$ & Single & $-0.89_{-0.06}^{+0.06}$ & -- & -- & $10.36_{-0.07}^{+0.07}$ & -- & -- & $-2.93_{-0.07}^{+0.07}$ & $20.39$ \\[.3cm]
$0.8 \leq z < 1.1$ & Single & $-0.93_{-0.05}^{+0.05}$ & -- & -- & $10.44_{-0.06}^{+0.05}$ & -- & -- & $-3.00_{-0.06}^{+0.06}$ & $33.91$ \\[.3cm]
$1.1 \leq z < 1.5$ & DPL & -- & $-2.14_{-0.07}^{+0.07}$ & $0.24_{-0.19}^{+0.20}$ & $9.43_{-0.08}^{+0.09}$ & -- & -- & $-2.39_{-0.04}^{+0.03}$ & $57.40$ \\[.3cm]
$1.5 \leq z < 2$ & DPL & -- & $-2.19_{-0.06}^{+0.07}$ & $0.43_{-0.19}^{+0.17}$ & $9.50_{-0.07}^{+0.09}$ & -- & -- & $-2.51_{-0.04}^{+0.03}$ & $67.48$ \\[.3cm]
$2 \leq z < 2.5$ & DPL & -- & $-2.06_{-0.13}^{+2.62}$ & $0.32_{-2.50}^{+0.29}$ & $9.58_{-0.08}^{+0.08}$ & -- & -- & $-2.90_{-0.04}^{+0.03}$ & $38.72$ \\[.3cm]
$2.5 \leq z < 3$ & DPL & -- & $-1.84_{-0.10}^{+0.11}$ & $0.85_{-0.33}^{+0.33}$ & $9.44_{-0.08}^{+0.09}$ & -- & -- & $-3.34_{-0.05}^{+0.05}$ & $20.20$ \\[.3cm]
$3 \leq z < 3.5$ & Single & $-0.41_{-0.11}^{+0.11}$ & -- & -- & $10.75_{-0.11}^{+0.10}$ & -- & -- & $-3.97_{-0.08}^{+0.08}$ & $20.21$ \\[.3cm]
$3.5 \leq z < 4.5$ & Single & $0.20_{-0.22}^{+0.26}$ & -- & -- & $10.36_{-0.13}^{+0.12}$ & -- & -- & $-4.28_{-0.06}^{+0.06}$ & $17.76$ \\[.3cm]
$4.5 \leq z < 5.5$ & Single & $-0.29_{-1.85}^{+2.37}$ & -- & -- & $9.60_{-1.23}^{+1.60}$ & -- & -- & $-5.94_{-0.69}^{+1.84}$ & $20.24$ \\[.3cm]
$5.5 \leq z < 6.5$ & -- & -- & -- & -- & -- & -- & -- & -- & -- \\[.3cm]
\hline
\end{tabular}
\caption{Fit results for star forming Disk dominated}
\label{tab:disk_SF}
\end{table*}

\begin{table*}[ht]
\centering
\begin{tabular}{cccccccccc}
\hline
Redshift & Model & $alpha$ & $alpha1$ & $alpha2$ & $logM^*$ & $logphi1$ & $logphi2$ & $logphi\_star$ & AICc \\
\hline
$0.2 \leq z < 0.5$ & Single & $-1.78_{-0.09}^{+0.10}$ & -- & -- & $10.41_{-0.17}^{+0.19}$ & -- & -- & $-3.64_{-0.26}^{+0.26}$ & $11.12$ \\[.3cm]
$0.5 \leq z < 0.8$ & Single & $-1.65_{-0.05}^{+0.06}$ & -- & -- & $10.62_{-0.10}^{+0.10}$ & -- & -- & $-3.55_{-0.15}^{+0.14}$ & $9.81$ \\[.3cm]
$0.8 \leq z < 1.1$ & Single & $-1.49_{-0.05}^{+0.05}$ & -- & -- & $10.60_{-0.06}^{+0.06}$ & -- & -- & $-3.16_{-0.09}^{+0.09}$ & $11.29$ \\[.3cm]
$1.1 \leq z < 1.5$ & Single & $-1.68_{-0.04}^{+0.05}$ & -- & -- & $10.79_{-0.10}^{+0.08}$ & -- & -- & $-3.85_{-0.12}^{+0.13}$ & $10.08$ \\[.3cm]
$1.5 \leq z < 2$ & Single & $-1.73_{-0.03}^{+0.03}$ & -- & -- & $10.83_{-0.07}^{+0.06}$ & -- & -- & $-3.93_{-0.10}^{+0.10}$ & $9.47$ \\[.3cm]
$2 \leq z < 2.5$ & Single & $-1.64_{-0.03}^{+0.03}$ & -- & -- & $10.92_{-0.08}^{+0.07}$ & -- & -- & $-3.92_{-0.10}^{+0.10}$ & $9.15$ \\[.3cm]
$2.5 \leq z < 3$ & Single & $-1.60_{-0.04}^{+0.04}$ & -- & -- & $10.82_{-0.10}^{+0.10}$ & -- & -- & $-3.94_{-0.13}^{+0.13}$ & $11.42$ \\[.3cm]
$3 \leq z < 3.5$ & Single & $-1.56_{-0.05}^{+0.05}$ & -- & -- & $10.74_{-0.10}^{+0.08}$ & -- & -- & $-3.92_{-0.12}^{+0.13}$ & $11.75$ \\[.3cm]
$3.5 \leq z < 4.5$ & Single & $-1.56_{-0.05}^{+0.05}$ & -- & -- & $10.53_{-0.08}^{+0.08}$ & -- & -- & $-3.96_{-0.11}^{+0.13}$ & $12.95$ \\[.3cm]
$4.5 \leq z < 5.5$ & DPL & -- & $-2.41_{-0.17}^{+0.12}$ & $-0.42_{-0.59}^{+0.85}$ & $9.01_{-0.15}^{+0.23}$ & -- & -- & $-2.58_{-0.17}^{+0.09}$ & $14.86$ \\[.3cm]
$5.5 \leq z < 6.5$ & DPL & -- & $-2.43_{-0.20}^{+2.08}$ & $-0.06_{-2.26}^{+0.57}$ & $9.09_{-0.12}^{+0.15}$ & -- & -- & $-3.07_{-0.08}^{+0.06}$ & $18.61$ \\[.3cm]
\hline
\end{tabular}
\caption{Fit results for star forming peculiars}
\label{tab:irr_SF}
\end{table*}

\end{document}